\documentclass[longauth]{aa}

\usepackage{euclid}
\usepackage{graphicx}
\usepackage{natbib}
\usepackage{scalerel}
\usepackage{ulem}
\usepackage{empheq}
\usepackage{cancel}
\usepackage{booktabs}

\newcommand{\tj}[6]{ \begin{pmatrix}
  #1 & #2 & #3 \\
  #4 & #5 & #6 
\end{pmatrix}}

\def\ln{{\rm ln}}

\def\det{{\rm det}}

\def\be{\begin{equation}}
\def\ee{\end{equation}}
\def\bea{\begin{eqnarray}}
\def\eea{\end{eqnarray}}
\def\ba{\begin{align}}

\def\bi{\begin{itemize}}
\def\ei{\end{itemize}}

\newcommand{\nn}{ \nonumber }

\newcommand{\fnl}{f_{\rm NL}}

\def\bu{{\bf u}}
\def\bv{{\bf v}}
\def\bs{{\bf s}}
\def\br{{\bf r}}
\def\cH{{\mathcal H}}

\def\bn{{\bf n}}
\def\bx{{\bf x}}
\def\bk{{\bf k}}
\def\bq{{\bf q}}
\def\bp{{\bf p}}
\def\P{{\bf P}}
\def\B{{\bf B}}

\def\cG{{\mathcal G}}

\def\cM{{\mathcal M}}
\def\cL{{\mathcal L}}

\newcommand{\kv}{\mathbf{k}}

\renewcommand{\vec}{\mathbf}

\newcommand{\bdm}{\begin{displaymath}}
\newcommand{\edm}{\end{displaymath}}

\usepackage[table]{xcolor}

\newcommand{\immag}{\mathrm{i}}
\newcommand{\e}{\mathrm{e}}

\usepackage{txfonts}
\usepackage{glossaries}
\usepackage[pdfencoding=auto,psdextra]{hyperref}
\hypersetup{
    colorlinks=true,
    linkcolor=blue,
    filecolor=magenta,      
    urlcolor=blue,
    citecolor=blue
}
\makeatletter
\renewcommand*\aa@pageof{, page \thepage{} of \pageref*{LastPage}}
\makeatother

\usepackage[utf8]{inputenc}

\usepackage[switch, modulo]{lineno}
\renewcommand{\linenumbers}[0]{}
\usepackage{multirow}

\newacronym{hod}{HOD}{halo occupation distribution}
\newacronym{fob}{FoB}{figure of bias}
\newacronym{dic}{DIC}{Deviance Information Criterion}
\newacronym{spt}{SPT}{Standard Perturbation Theory}
\newacronym{eft}{EFTofLSS}{Effective Field Theory of Large Scale Structure}
\newacronym{bao}{BAO}{baryon-acoustic-oscillation}
\newacronym{mcmc}{MCMC}{Markov Chain Monte Carlo}

\usepackage{fontawesome5}
\usepackage{hyperref}

\makeatletter
\newcommand{\github}[1]{%
   \href{#1}{\faGithubSquare}%
}
\makeatother

\definecolor{MyDarkOrange}{rgb}{0.8, 0.4, 0.0}

\begin{document}
%
%
\title{\Euclid preparation}
\subtitle{Testing multi-field inflation with galaxy power spectrum and bispectrum}

\newcommand{\orcid}[1]{} 
\author{Euclid Collaboration: D.~Linde\orcid{0000-0001-7192-1067}\inst{\ref{aff1}}
\and A.~Moradinezhad~Dizgah\orcid{0000-0001-8841-9989}\thanks{\email{azadeh.moradinezhad@lapth.cnrs.fr}}\inst{\ref{aff2}}
\and G.~Parimbelli\orcid{0000-0002-2539-2472}\inst{\ref{aff3},\ref{aff4},\ref{aff5}}
\and K.~Pardede\orcid{0000-0002-7728-8220}\inst{\ref{aff1}}
\and E.~Sefusatti\orcid{0000-0003-0473-1567}\inst{\ref{aff6},\ref{aff7},\ref{aff8}}
\and M.~S.~Cagliari\orcid{0000-0002-2912-9233}\inst{\ref{aff2}}
\and G.~D'Amico\orcid{0000-0002-8183-1214}\inst{\ref{aff9},\ref{aff1}}
\and V.~Desjacques\orcid{0000-0003-2062-8172}\inst{\ref{aff10}}
\and A.~Eggemeier\orcid{0000-0002-1841-8910}\inst{\ref{aff11}}
\and M.~Biagetti\orcid{0000-0002-5097-479X}\inst{\ref{aff12}}
\and A.~Veropalumbo\orcid{0000-0003-2387-1194}\inst{\ref{aff13},\ref{aff14},\ref{aff15}}
\and B.~Camacho~Quevedo\orcid{0000-0002-8789-4232}\inst{\ref{aff7},\ref{aff5},\ref{aff6}}
\and A.~Chudaykin\inst{\ref{aff16}}
\and M.~Crocce\orcid{0000-0002-9745-6228}\inst{\ref{aff3},\ref{aff17}}
\and L.~Castiblanco\orcid{0000-0002-2324-7335}\inst{\ref{aff18}}
\and E.~Castorina\inst{\ref{aff19},\ref{aff20}}
\and A.~Farina\orcid{0009-0000-3420-929X}\inst{\ref{aff15},\ref{aff13},\ref{aff14}}
\and M.~Guidi\orcid{0000-0001-9408-1101}\inst{\ref{aff21},\ref{aff22}}
\and M.~K{\"a}rcher\orcid{0000-0001-5868-647X}\inst{\ref{aff19}}
\and A.~Pezzotta\orcid{0000-0003-0726-2268}\inst{\ref{aff13}}
\and A.~Pugno\inst{\ref{aff11}}
\and B.~Altieri\orcid{0000-0003-3936-0284}\inst{\ref{aff23}}
\and S.~Andreon\orcid{0000-0002-2041-8784}\inst{\ref{aff13}}
\and N.~Auricchio\orcid{0000-0003-4444-8651}\inst{\ref{aff22}}
\and C.~Baccigalupi\orcid{0000-0002-8211-1630}\inst{\ref{aff7},\ref{aff6},\ref{aff8},\ref{aff5}}
\and M.~Baldi\orcid{0000-0003-4145-1943}\inst{\ref{aff21},\ref{aff22},\ref{aff24}}
\and S.~Bardelli\orcid{0000-0002-8900-0298}\inst{\ref{aff22}}
\and P.~Battaglia\orcid{0000-0002-7337-5909}\inst{\ref{aff22}}
\and A.~Biviano\orcid{0000-0002-0857-0732}\inst{\ref{aff6},\ref{aff7}}
\and E.~Branchini\orcid{0000-0002-0808-6908}\inst{\ref{aff15},\ref{aff14},\ref{aff13}}
\and M.~Brescia\orcid{0000-0001-9506-5680}\inst{\ref{aff25},\ref{aff26}}
\and S.~Camera\orcid{0000-0003-3399-3574}\inst{\ref{aff27},\ref{aff28},\ref{aff29}}
\and G.~Ca\~nas-Herrera\orcid{0000-0003-2796-2149}\inst{\ref{aff30}}
\and V.~Capobianco\orcid{0000-0002-3309-7692}\inst{\ref{aff29}}
\and C.~Carbone\orcid{0000-0003-0125-3563}\inst{\ref{aff31}}
\and J.~Carretero\orcid{0000-0002-3130-0204}\inst{\ref{aff32},\ref{aff33}}
\and S.~Casas\orcid{0000-0002-4751-5138}\inst{\ref{aff34},\ref{aff35}}
\and M.~Castellano\orcid{0000-0001-9875-8263}\inst{\ref{aff36}}
\and G.~Castignani\orcid{0000-0001-6831-0687}\inst{\ref{aff22}}
\and S.~Cavuoti\orcid{0000-0002-3787-4196}\inst{\ref{aff26},\ref{aff37}}
\and K.~C.~Chambers\orcid{0000-0001-6965-7789}\inst{\ref{aff38}}
\and A.~Cimatti\inst{\ref{aff39}}
\and C.~Colodro-Conde\inst{\ref{aff40}}
\and G.~Congedo\orcid{0000-0003-2508-0046}\inst{\ref{aff41}}
\and L.~Conversi\orcid{0000-0002-6710-8476}\inst{\ref{aff42},\ref{aff23}}
\and Y.~Copin\orcid{0000-0002-5317-7518}\inst{\ref{aff43}}
\and F.~Courbin\orcid{0000-0003-0758-6510}\inst{\ref{aff44},\ref{aff45},\ref{aff46}}
\and H.~M.~Courtois\orcid{0000-0003-0509-1776}\inst{\ref{aff47}}
\and H.~Degaudenzi\orcid{0000-0002-5887-6799}\inst{\ref{aff48}}
\and S.~de~la~Torre\inst{\ref{aff49}}
\and G.~De~Lucia\orcid{0000-0002-6220-9104}\inst{\ref{aff6}}
\and H.~Dole\orcid{0000-0002-9767-3839}\inst{\ref{aff50}}
\and M.~Douspis\orcid{0000-0003-4203-3954}\inst{\ref{aff50}}
\and F.~Dubath\orcid{0000-0002-6533-2810}\inst{\ref{aff48}}
\and X.~Dupac\inst{\ref{aff23}}
\and S.~Escoffier\orcid{0000-0002-2847-7498}\inst{\ref{aff51}}
\and M.~Farina\orcid{0000-0002-3089-7846}\inst{\ref{aff52}}
\and R.~Farinelli\inst{\ref{aff22}}
\and S.~Ferriol\inst{\ref{aff43}}
\and F.~Finelli\orcid{0000-0002-6694-3269}\inst{\ref{aff22},\ref{aff53}}
\and P.~Fosalba\orcid{0000-0002-1510-5214}\inst{\ref{aff17},\ref{aff3}}
\and S.~Fotopoulou\orcid{0000-0002-9686-254X}\inst{\ref{aff54}}
\and M.~Frailis\orcid{0000-0002-7400-2135}\inst{\ref{aff6}}
\and M.~Fumana\orcid{0000-0001-6787-5950}\inst{\ref{aff31}}
\and S.~Galeotta\orcid{0000-0002-3748-5115}\inst{\ref{aff6}}
\and K.~George\orcid{0000-0002-1734-8455}\inst{\ref{aff55}}
\and B.~Gillis\orcid{0000-0002-4478-1270}\inst{\ref{aff41}}
\and C.~Giocoli\orcid{0000-0002-9590-7961}\inst{\ref{aff22},\ref{aff24}}
\and J.~Gracia-Carpio\orcid{0000-0003-4689-3134}\inst{\ref{aff56}}
\and A.~Grazian\orcid{0000-0002-5688-0663}\inst{\ref{aff57}}
\and F.~Grupp\inst{\ref{aff56},\ref{aff58}}
\and S.~V.~H.~Haugan\orcid{0000-0001-9648-7260}\inst{\ref{aff59}}
\and W.~Holmes\inst{\ref{aff60}}
\and F.~Hormuth\inst{\ref{aff61}}
\and A.~Hornstrup\orcid{0000-0002-3363-0936}\inst{\ref{aff62},\ref{aff63}}
\and K.~Jahnke\orcid{0000-0003-3804-2137}\inst{\ref{aff64}}
\and B.~Joachimi\orcid{0000-0001-7494-1303}\inst{\ref{aff65}}
\and S.~Kermiche\orcid{0000-0002-0302-5735}\inst{\ref{aff51}}
\and A.~Kiessling\orcid{0000-0002-2590-1273}\inst{\ref{aff60}}
\and B.~Kubik\orcid{0009-0006-5823-4880}\inst{\ref{aff43}}
\and M.~Kunz\orcid{0000-0002-3052-7394}\inst{\ref{aff16}}
\and H.~Kurki-Suonio\orcid{0000-0002-4618-3063}\inst{\ref{aff66},\ref{aff67}}
\and A.~M.~C.~Le~Brun\orcid{0000-0002-0936-4594}\inst{\ref{aff68}}
\and S.~Ligori\orcid{0000-0003-4172-4606}\inst{\ref{aff29}}
\and P.~B.~Lilje\orcid{0000-0003-4324-7794}\inst{\ref{aff59}}
\and V.~Lindholm\orcid{0000-0003-2317-5471}\inst{\ref{aff66},\ref{aff67}}
\and I.~Lloro\orcid{0000-0001-5966-1434}\inst{\ref{aff69}}
\and G.~Mainetti\orcid{0000-0003-2384-2377}\inst{\ref{aff70}}
\and O.~Mansutti\orcid{0000-0001-5758-4658}\inst{\ref{aff6}}
\and O.~Marggraf\orcid{0000-0001-7242-3852}\inst{\ref{aff11}}
\and M.~Martinelli\orcid{0000-0002-6943-7732}\inst{\ref{aff36},\ref{aff71}}
\and N.~Martinet\orcid{0000-0003-2786-7790}\inst{\ref{aff49}}
\and F.~Marulli\orcid{0000-0002-8850-0303}\inst{\ref{aff72},\ref{aff22},\ref{aff24}}
\and R.~J.~Massey\orcid{0000-0002-6085-3780}\inst{\ref{aff73}}
\and E.~Medinaceli\orcid{0000-0002-4040-7783}\inst{\ref{aff22}}
\and S.~Mei\orcid{0000-0002-2849-559X}\inst{\ref{aff74},\ref{aff75}}
\and M.~Meneghetti\orcid{0000-0003-1225-7084}\inst{\ref{aff22},\ref{aff24}}
\and E.~Merlin\orcid{0000-0001-6870-8900}\inst{\ref{aff36}}
\and G.~Meylan\inst{\ref{aff76}}
\and A.~Mora\orcid{0000-0002-1922-8529}\inst{\ref{aff77}}
\and M.~Moresco\orcid{0000-0002-7616-7136}\inst{\ref{aff72},\ref{aff22}}
\and L.~Moscardini\orcid{0000-0002-3473-6716}\inst{\ref{aff72},\ref{aff22},\ref{aff24}}
\and C.~Neissner\orcid{0000-0001-8524-4968}\inst{\ref{aff78},\ref{aff33}}
\and S.-M.~Niemi\orcid{0009-0005-0247-0086}\inst{\ref{aff79}}
\and J.~W.~Nightingale\orcid{0000-0002-8987-7401}\inst{\ref{aff80}}
\and C.~Padilla\orcid{0000-0001-7951-0166}\inst{\ref{aff78}}
\and S.~Paltani\orcid{0000-0002-8108-9179}\inst{\ref{aff48}}
\and F.~Pasian\orcid{0000-0002-4869-3227}\inst{\ref{aff6}}
\and K.~Pedersen\inst{\ref{aff81}}
\and W.~J.~Percival\orcid{0000-0002-0644-5727}\inst{\ref{aff82},\ref{aff83},\ref{aff84}}
\and V.~Pettorino\orcid{0000-0002-4203-9320}\inst{\ref{aff79}}
\and S.~Pires\orcid{0000-0002-0249-2104}\inst{\ref{aff85}}
\and G.~Polenta\orcid{0000-0003-4067-9196}\inst{\ref{aff86}}
\and M.~Poncet\inst{\ref{aff87}}
\and L.~A.~Popa\inst{\ref{aff88}}
\and F.~Raison\orcid{0000-0002-7819-6918}\inst{\ref{aff56}}
\and A.~Renzi\orcid{0000-0001-9856-1970}\inst{\ref{aff89},\ref{aff90},\ref{aff22}}
\and J.~Rhodes\orcid{0000-0002-4485-8549}\inst{\ref{aff60}}
\and G.~Riccio\inst{\ref{aff26}}
\and E.~Romelli\orcid{0000-0003-3069-9222}\inst{\ref{aff6}}
\and M.~Roncarelli\orcid{0000-0001-9587-7822}\inst{\ref{aff22}}
\and R.~Saglia\orcid{0000-0003-0378-7032}\inst{\ref{aff58},\ref{aff56}}
\and Z.~Sakr\orcid{0000-0002-4823-3757}\inst{\ref{aff91},\ref{aff92},\ref{aff93}}
\and A.~G.~S\'anchez\orcid{0000-0003-1198-831X}\inst{\ref{aff56}}
\and D.~Sapone\orcid{0000-0001-7089-4503}\inst{\ref{aff94}}
\and B.~Sartoris\orcid{0000-0003-1337-5269}\inst{\ref{aff58},\ref{aff6}}
\and A.~Secroun\orcid{0000-0003-0505-3710}\inst{\ref{aff51}}
\and G.~Seidel\orcid{0000-0003-2907-353X}\inst{\ref{aff64}}
\and E.~Sihvola\orcid{0000-0003-1804-7715}\inst{\ref{aff95}}
\and P.~Simon\inst{\ref{aff11}}
\and C.~Sirignano\orcid{0000-0002-0995-7146}\inst{\ref{aff89},\ref{aff90}}
\and G.~Sirri\orcid{0000-0003-2626-2853}\inst{\ref{aff24}}
\and A.~Spurio~Mancini\orcid{0000-0001-5698-0990}\inst{\ref{aff96}}
\and L.~Stanco\orcid{0000-0002-9706-5104}\inst{\ref{aff90}}
\and P.~Tallada-Cresp\'{i}\orcid{0000-0002-1336-8328}\inst{\ref{aff32},\ref{aff33}}
\and A.~N.~Taylor\inst{\ref{aff41}}
\and I.~Tereno\orcid{0000-0002-4537-6218}\inst{\ref{aff97},\ref{aff98}}
\and N.~Tessore\orcid{0000-0002-9696-7931}\inst{\ref{aff99}}
\and S.~Toft\orcid{0000-0003-3631-7176}\inst{\ref{aff100},\ref{aff101}}
\and R.~Toledo-Moreo\orcid{0000-0002-2997-4859}\inst{\ref{aff102}}
\and F.~Torradeflot\orcid{0000-0003-1160-1517}\inst{\ref{aff33},\ref{aff32}}
\and I.~Tutusaus\orcid{0000-0002-3199-0399}\inst{\ref{aff3},\ref{aff17},\ref{aff92}}
\and L.~Valenziano\orcid{0000-0002-1170-0104}\inst{\ref{aff22},\ref{aff53}}
\and J.~Valiviita\orcid{0000-0001-6225-3693}\inst{\ref{aff66},\ref{aff67}}
\and T.~Vassallo\orcid{0000-0001-6512-6358}\inst{\ref{aff6},\ref{aff55}}
\and G.~Verdoes~Kleijn\orcid{0000-0001-5803-2580}\inst{\ref{aff103}}
\and Y.~Wang\orcid{0000-0002-4749-2984}\inst{\ref{aff104}}
\and J.~Weller\orcid{0000-0002-8282-2010}\inst{\ref{aff58},\ref{aff56}}
\and G.~Zamorani\orcid{0000-0002-2318-301X}\inst{\ref{aff22}}
\and F.~M.~Zerbi\orcid{0000-0002-9996-973X}\inst{\ref{aff13}}
\and E.~Zucca\orcid{0000-0002-5845-8132}\inst{\ref{aff22}}
\and M.~Ballardini\orcid{0000-0003-4481-3559}\inst{\ref{aff105},\ref{aff106},\ref{aff22}}
\and E.~Bozzo\orcid{0000-0002-8201-1525}\inst{\ref{aff48}}
\and C.~Burigana\orcid{0000-0002-3005-5796}\inst{\ref{aff107},\ref{aff53}}
\and R.~Cabanac\orcid{0000-0001-6679-2600}\inst{\ref{aff92}}
\and M.~Calabrese\orcid{0000-0002-2637-2422}\inst{\ref{aff108},\ref{aff31}}
\and T.~Castro\orcid{0000-0002-6292-3228}\inst{\ref{aff6},\ref{aff8},\ref{aff7},\ref{aff109}}
\and J.~A.~Escartin~Vigo\inst{\ref{aff56}}
\and J.~Garc\'ia-Bellido\orcid{0000-0002-9370-8360}\inst{\ref{aff91}}
\and J.~Macias-Perez\orcid{0000-0002-5385-2763}\inst{\ref{aff110}}
\and R.~Maoli\orcid{0000-0002-6065-3025}\inst{\ref{aff111},\ref{aff36}}
\and J.~Mart\'{i}n-Fleitas\orcid{0000-0002-8594-569X}\inst{\ref{aff112}}
\and N.~Mauri\orcid{0000-0001-8196-1548}\inst{\ref{aff39},\ref{aff24}}
\and R.~B.~Metcalf\orcid{0000-0003-3167-2574}\inst{\ref{aff72},\ref{aff22}}
\and P.~Monaco\orcid{0000-0003-2083-7564}\inst{\ref{aff113},\ref{aff6},\ref{aff8},\ref{aff7}}
\and M.~P\"ontinen\orcid{0000-0001-5442-2530}\inst{\ref{aff66}}
\and I.~Risso\orcid{0000-0003-2525-7761}\inst{\ref{aff13},\ref{aff14}}
\and V.~Scottez\orcid{0009-0008-3864-940X}\inst{\ref{aff114},\ref{aff115}}
\and M.~Sereno\orcid{0000-0003-0302-0325}\inst{\ref{aff22},\ref{aff24}}
\and M.~Tenti\orcid{0000-0002-4254-5901}\inst{\ref{aff24}}
\and M.~Tucci\inst{\ref{aff48}}
\and M.~Viel\orcid{0000-0002-2642-5707}\inst{\ref{aff7},\ref{aff6},\ref{aff5},\ref{aff8},\ref{aff109}}
\and M.~Wiesmann\orcid{0009-0000-8199-5860}\inst{\ref{aff59}}
\and Y.~Akrami\orcid{0000-0002-2407-7956}\inst{\ref{aff91},\ref{aff116}}
\and I.~T.~Andika\orcid{0000-0001-6102-9526}\inst{\ref{aff55}}
\and G.~Angora\orcid{0000-0002-0316-6562}\inst{\ref{aff26},\ref{aff105}}
\and M.~Archidiacono\orcid{0000-0003-4952-9012}\inst{\ref{aff19},\ref{aff20}}
\and F.~Atrio-Barandela\orcid{0000-0002-2130-2513}\inst{\ref{aff117}}
\and S.~Avila\orcid{0000-0001-5043-3662}\inst{\ref{aff32}}
\and L.~Bazzanini\orcid{0000-0003-0727-0137}\inst{\ref{aff105},\ref{aff22}}
\and J.~Bel\inst{\ref{aff118}}
\and D.~Bertacca\orcid{0000-0002-2490-7139}\inst{\ref{aff89},\ref{aff57},\ref{aff90}}
\and M.~Bethermin\orcid{0000-0002-3915-2015}\inst{\ref{aff119}}
\and F.~Beutler\orcid{0000-0003-0467-5438}\inst{\ref{aff41}}
\and A.~Blanchard\orcid{0000-0001-8555-9003}\inst{\ref{aff92}}
\and L.~Blot\orcid{0000-0002-9622-7167}\inst{\ref{aff120},\ref{aff68}}
\and H.~B\"ohringer\orcid{0000-0001-8241-4204}\inst{\ref{aff56},\ref{aff55},\ref{aff121}}
\and M.~Bonici\orcid{0000-0002-8430-126X}\inst{\ref{aff82},\ref{aff31}}
\and S.~Borgani\orcid{0000-0001-6151-6439}\inst{\ref{aff113},\ref{aff7},\ref{aff6},\ref{aff8},\ref{aff109}}
\and M.~L.~Brown\orcid{0000-0002-0370-8077}\inst{\ref{aff122}}
\and S.~Bruton\orcid{0000-0002-6503-5218}\inst{\ref{aff123}}
\and A.~Calabro\orcid{0000-0003-2536-1614}\inst{\ref{aff36}}
\and F.~Caro\orcid{0009-0003-1053-0507}\inst{\ref{aff36}}
\and C.~S.~Carvalho\inst{\ref{aff98}}
\and F.~Cogato\orcid{0000-0003-4632-6113}\inst{\ref{aff72},\ref{aff22}}
\and A.~R.~Cooray\orcid{0000-0002-3892-0190}\inst{\ref{aff124}}
\and S.~Davini\orcid{0000-0003-3269-1718}\inst{\ref{aff14}}
\and G.~Desprez\orcid{0000-0001-8325-1742}\inst{\ref{aff103}}
\and A.~D\'iaz-S\'anchez\orcid{0000-0003-0748-4768}\inst{\ref{aff125}}
\and S.~Di~Domizio\orcid{0000-0003-2863-5895}\inst{\ref{aff15},\ref{aff14}}
\and J.~M.~Diego\orcid{0000-0001-9065-3926}\inst{\ref{aff126}}
\and V.~Duret\orcid{0009-0009-0383-4960}\inst{\ref{aff51}}
\and M.~Y.~Elkhashab\orcid{0000-0001-9306-2603}\inst{\ref{aff113},\ref{aff6},\ref{aff8},\ref{aff7}}
\and A.~Enia\orcid{0000-0002-0200-2857}\inst{\ref{aff22}}
\and Y.~Fang\orcid{0000-0002-0334-6950}\inst{\ref{aff58}}
\and A.~Finoguenov\orcid{0000-0002-4606-5403}\inst{\ref{aff66}}
\and A.~Franco\orcid{0000-0002-4761-366X}\inst{\ref{aff127},\ref{aff128},\ref{aff129}}
\and K.~Ganga\orcid{0000-0001-8159-8208}\inst{\ref{aff74}}
\and T.~Gasparetto\orcid{0000-0002-7913-4866}\inst{\ref{aff36}}
\and F.~Giacomini\orcid{0000-0002-3129-2814}\inst{\ref{aff24}}
\and F.~Gianotti\orcid{0000-0003-4666-119X}\inst{\ref{aff22}}
\and G.~Gozaliasl\orcid{0000-0002-0236-919X}\inst{\ref{aff130},\ref{aff66}}
\and A.~Gruppuso\orcid{0000-0001-9272-5292}\inst{\ref{aff22},\ref{aff24}}
\and C.~M.~Gutierrez\orcid{0000-0001-7854-783X}\inst{\ref{aff40},\ref{aff131}}
\and A.~Hall\orcid{0000-0002-3139-8651}\inst{\ref{aff41}}
\and C.~Hern\'andez-Monteagudo\orcid{0000-0001-5471-9166}\inst{\ref{aff131},\ref{aff40}}
\and H.~Hildebrandt\orcid{0000-0002-9814-3338}\inst{\ref{aff132}}
\and J.~Hjorth\orcid{0000-0002-4571-2306}\inst{\ref{aff81}}
\and J.~J.~E.~Kajava\orcid{0000-0002-3010-8333}\inst{\ref{aff133},\ref{aff134},\ref{aff135}}
\and Y.~Kang\orcid{0009-0000-8588-7250}\inst{\ref{aff48}}
\and V.~Kansal\orcid{0000-0002-4008-6078}\inst{\ref{aff136},\ref{aff137}}
\and D.~Karagiannis\orcid{0000-0002-4927-0816}\inst{\ref{aff105},\ref{aff138}}
\and K.~Kiiveri\inst{\ref{aff95}}
\and J.~Kim\orcid{0000-0003-2776-2761}\inst{\ref{aff139}}
\and C.~C.~Kirkpatrick\inst{\ref{aff95}}
\and S.~Kruk\orcid{0000-0001-8010-8879}\inst{\ref{aff23}}
\and M.~Lattanzi\orcid{0000-0003-1059-2532}\inst{\ref{aff106}}
\and L.~Legrand\orcid{0000-0003-0610-5252}\inst{\ref{aff140},\ref{aff141}}
\and M.~Lembo\orcid{0000-0002-5271-5070}\inst{\ref{aff142}}
\and F.~Lepori\orcid{0009-0000-5061-7138}\inst{\ref{aff143}}
\and G.~Leroy\orcid{0009-0004-2523-4425}\inst{\ref{aff144},\ref{aff73}}
\and G.~F.~Lesci\orcid{0000-0002-4607-2830}\inst{\ref{aff72},\ref{aff22}}
\and J.~Lesgourgues\orcid{0000-0001-7627-353X}\inst{\ref{aff34}}
\and T.~I.~Liaudat\orcid{0000-0002-9104-314X}\inst{\ref{aff145}}
\and S.~J.~Liu\orcid{0000-0001-7680-2139}\inst{\ref{aff52}}
\and G.~Maggio\orcid{0000-0003-4020-4836}\inst{\ref{aff6}}
\and M.~Magliocchetti\orcid{0000-0001-9158-4838}\inst{\ref{aff52}}
\and A.~Manj\'on-Garc\'ia\orcid{0000-0002-7413-8825}\inst{\ref{aff125}}
\and F.~Mannucci\orcid{0000-0002-4803-2381}\inst{\ref{aff146}}
\and C.~J.~A.~P.~Martins\orcid{0000-0002-4886-9261}\inst{\ref{aff147},\ref{aff148}}
\and L.~Maurin\orcid{0000-0002-8406-0857}\inst{\ref{aff50}}
\and C.~Moretti\orcid{0000-0003-3314-8936}\inst{\ref{aff6},\ref{aff7},\ref{aff8}}
\and G.~Morgante\inst{\ref{aff22}}
\and S.~Nadathur\orcid{0000-0001-9070-3102}\inst{\ref{aff149}}
\and K.~Naidoo\orcid{0000-0002-9182-1802}\inst{\ref{aff149},\ref{aff64}}
\and A.~Navarro-Alsina\orcid{0000-0002-3173-2592}\inst{\ref{aff11}}
\and S.~Nesseris\orcid{0000-0002-0567-0324}\inst{\ref{aff91}}
\and L.~Pagano\orcid{0000-0003-1820-5998}\inst{\ref{aff105},\ref{aff106}}
\and D.~Paoletti\orcid{0000-0003-4761-6147}\inst{\ref{aff22},\ref{aff53}}
\and F.~Passalacqua\orcid{0000-0002-8606-4093}\inst{\ref{aff89},\ref{aff90}}
\and K.~Paterson\orcid{0000-0001-8340-3486}\inst{\ref{aff64}}
\and L.~Patrizii\inst{\ref{aff24}}
\and C.~Pattison\orcid{0000-0003-3272-2617}\inst{\ref{aff149}}
\and A.~Pisani\orcid{0000-0002-6146-4437}\inst{\ref{aff51}}
\and D.~Potter\orcid{0000-0002-0757-5195}\inst{\ref{aff150}}
\and G.~W.~Pratt\inst{\ref{aff85}}
\and S.~Quai\orcid{0000-0002-0449-8163}\inst{\ref{aff72},\ref{aff22}}
\and M.~Radovich\orcid{0000-0002-3585-866X}\inst{\ref{aff57}}
\and K.~Rojas\orcid{0000-0003-1391-6854}\inst{\ref{aff151}}
\and W.~Roster\orcid{0000-0002-9149-6528}\inst{\ref{aff56}}
\and S.~Sacquegna\orcid{0000-0002-8433-6630}\inst{\ref{aff152}}
\and M.~Sahl\'en\orcid{0000-0003-0973-4804}\inst{\ref{aff153}}
\and D.~B.~Sanders\orcid{0000-0002-1233-9998}\inst{\ref{aff38}}
\and E.~Sarpa\orcid{0000-0002-1256-655X}\inst{\ref{aff5},\ref{aff109},\ref{aff6}}
\and A.~Schneider\orcid{0000-0001-7055-8104}\inst{\ref{aff150}}
\and M.~Schultheis\inst{\ref{aff154}}
\and D.~Sciotti\orcid{0009-0008-4519-2620}\inst{\ref{aff36},\ref{aff71}}
\and E.~Sellentin\inst{\ref{aff155},\ref{aff30}}
\and L.~C.~Smith\orcid{0000-0002-3259-2771}\inst{\ref{aff156}}
\and K.~Tanidis\orcid{0000-0001-9843-5130}\inst{\ref{aff157}}
\and F.~Tarsitano\orcid{0000-0002-5919-0238}\inst{\ref{aff158},\ref{aff48}}
\and G.~Testera\inst{\ref{aff14}}
\and R.~Teyssier\orcid{0000-0001-7689-0933}\inst{\ref{aff159}}
\and S.~Tosi\orcid{0000-0002-7275-9193}\inst{\ref{aff15},\ref{aff13},\ref{aff14}}
\and A.~Troja\orcid{0000-0003-0239-4595}\inst{\ref{aff6}}
\and A.~Venhola\orcid{0000-0001-6071-4564}\inst{\ref{aff160}}
\and D.~Vergani\orcid{0000-0003-0898-2216}\inst{\ref{aff22}}
\and F.~Vernizzi\orcid{0000-0003-3426-2802}\inst{\ref{aff161}}
\and G.~Verza\orcid{0000-0002-1886-8348}\inst{\ref{aff162},\ref{aff163}}
\and S.~Vinciguerra\orcid{0009-0005-4018-3184}\inst{\ref{aff49}}
\and N.~A.~Walton\orcid{0000-0003-3983-8778}\inst{\ref{aff156}}
\and A.~H.~Wright\orcid{0000-0001-7363-7932}\inst{\ref{aff132}}
\and H.~W.~Yeung\orcid{0000-0002-4993-9014}\inst{\ref{aff41}}}
										   
\institute{INFN Gruppo Collegato di Parma, Viale delle Scienze 7/A 43124 Parma, Italy\label{aff1}
\and
Laboratoire d'Annecy-le-Vieux de Physique Theorique, CNRS \& Universite Savoie Mont Blanc, 9 Chemin de Bellevue, BP 110, Annecy-le-Vieux, 74941 ANNECY Cedex, France\label{aff2}
\and
Institute of Space Sciences (ICE, CSIC), Campus UAB, Carrer de Can Magrans, s/n, 08193 Barcelona, Spain\label{aff3}
\and
Dipartimento di Fisica, Universit\`a degli studi di Genova, and INFN-Sezione di Genova, via Dodecaneso 33, 16146, Genova, Italy\label{aff4}
\and
SISSA, International School for Advanced Studies, Via Bonomea 265, 34136 Trieste TS, Italy\label{aff5}
\and
INAF-Osservatorio Astronomico di Trieste, Via G. B. Tiepolo 11, 34143 Trieste, Italy\label{aff6}
\and
IFPU, Institute for Fundamental Physics of the Universe, via Beirut 2, 34151 Trieste, Italy\label{aff7}
\and
INFN, Sezione di Trieste, Via Valerio 2, 34127 Trieste TS, Italy\label{aff8}
\and
Dipartimento di Scienze Matematiche, Fisiche e Informatiche, Universit\`a di Parma, Viale delle Scienze 7/A 43124 Parma, Italy\label{aff9}
\and
Technion Israel Institute of Technology, Israel\label{aff10}
\and
Universit\"at Bonn, Argelander-Institut f\"ur Astronomie, Auf dem H\"ugel 71, 53121 Bonn, Germany\label{aff11}
\and
Area Science Park, Localit\`a Padriciano 99, 34149, Trieste, Italy\label{aff12}
\and
INAF-Osservatorio Astronomico di Brera, Via Brera 28, 20122 Milano, Italy\label{aff13}
\and
INFN-Sezione di Genova, Via Dodecaneso 33, 16146, Genova, Italy\label{aff14}
\and
Dipartimento di Fisica, Universit\`a di Genova, Via Dodecaneso 33, 16146, Genova, Italy\label{aff15}
\and
Universit\'e de Gen\`eve, D\'epartement de Physique Th\'eorique and Centre for Astroparticle Physics, 24 quai Ernest-Ansermet, CH-1211 Gen\`eve 4, Switzerland\label{aff16}
\and
Institut d'Estudis Espacials de Catalunya (IEEC),  Edifici RDIT, Campus UPC, 08860 Castelldefels, Barcelona, Spain\label{aff17}
\and
Fakult\"at f\"ur Physik, Universit\"at Bielefeld, Postfach 100131, 33501 Bielefeld, Germany\label{aff18}
\and
Dipartimento di Fisica "Aldo Pontremoli", Universit\`a degli Studi di Milano, Via Celoria 16, 20133 Milano, Italy\label{aff19}
\and
INFN-Sezione di Milano, Via Celoria 16, 20133 Milano, Italy\label{aff20}
\and
Dipartimento di Fisica e Astronomia, Universit\`a di Bologna, Via Gobetti 93/2, 40129 Bologna, Italy\label{aff21}
\and
INAF-Osservatorio di Astrofisica e Scienza dello Spazio di Bologna, Via Piero Gobetti 93/3, 40129 Bologna, Italy\label{aff22}
\and
ESAC/ESA, Camino Bajo del Castillo, s/n., Urb. Villafranca del Castillo, 28692 Villanueva de la Ca\~nada, Madrid, Spain\label{aff23}
\and
INFN-Sezione di Bologna, Viale Berti Pichat 6/2, 40127 Bologna, Italy\label{aff24}
\and
Department of Physics "E. Pancini", University Federico II, Via Cinthia 6, 80126, Napoli, Italy\label{aff25}
\and
INAF-Osservatorio Astronomico di Capodimonte, Via Moiariello 16, 80131 Napoli, Italy\label{aff26}
\and
Dipartimento di Fisica, Universit\`a degli Studi di Torino, Via P. Giuria 1, 10125 Torino, Italy\label{aff27}
\and
INFN-Sezione di Torino, Via P. Giuria 1, 10125 Torino, Italy\label{aff28}
\and
INAF-Osservatorio Astrofisico di Torino, Via Osservatorio 20, 10025 Pino Torinese (TO), Italy\label{aff29}
\and
Leiden Observatory, Leiden University, Einsteinweg 55, 2333 CC Leiden, The Netherlands\label{aff30}
\and
INAF-IASF Milano, Via Alfonso Corti 12, 20133 Milano, Italy\label{aff31}
\and
Centro de Investigaciones Energ\'eticas, Medioambientales y Tecnol\'ogicas (CIEMAT), Avenida Complutense 40, 28040 Madrid, Spain\label{aff32}
\and
Port d'Informaci\'{o} Cient\'{i}fica, Campus UAB, C. Albareda s/n, 08193 Bellaterra (Barcelona), Spain\label{aff33}
\and
Institute for Theoretical Particle Physics and Cosmology (TTK), RWTH Aachen University, 52056 Aachen, Germany\label{aff34}
\and
Deutsches Zentrum f\"ur Luft- und Raumfahrt e. V. (DLR), Linder H\"ohe, 51147 K\"oln, Germany\label{aff35}
\and
INAF-Osservatorio Astronomico di Roma, Via Frascati 33, 00078 Monteporzio Catone, Italy\label{aff36}
\and
INFN section of Naples, Via Cinthia 6, 80126, Napoli, Italy\label{aff37}
\and
Institute for Astronomy, University of Hawaii, 2680 Woodlawn Drive, Honolulu, HI 96822, USA\label{aff38}
\and
Dipartimento di Fisica e Astronomia "Augusto Righi" - Alma Mater Studiorum Universit\`a di Bologna, Viale Berti Pichat 6/2, 40127 Bologna, Italy\label{aff39}
\and
Instituto de Astrof\'{\i}sica de Canarias, E-38205 La Laguna, Tenerife, Spain\label{aff40}
\and
Institute for Astronomy, University of Edinburgh, Royal Observatory, Blackford Hill, Edinburgh EH9 3HJ, UK\label{aff41}
\and
European Space Agency/ESRIN, Largo Galileo Galilei 1, 00044 Frascati, Roma, Italy\label{aff42}
\and
Universit\'e Claude Bernard Lyon 1, CNRS/IN2P3, IP2I Lyon, UMR 5822, Villeurbanne, F-69100, France\label{aff43}
\and
Institut de Ci\`{e}ncies del Cosmos (ICCUB), Universitat de Barcelona (IEEC-UB), Mart\'{i} i Franqu\`{e}s 1, 08028 Barcelona, Spain\label{aff44}
\and
Instituci\'o Catalana de Recerca i Estudis Avan\c{c}ats (ICREA), Passeig de Llu\'{\i}s Companys 23, 08010 Barcelona, Spain\label{aff45}
\and
Institut de Ciencies de l'Espai (IEEC-CSIC), Campus UAB, Carrer de Can Magrans, s/n Cerdanyola del Vall\'es, 08193 Barcelona, Spain\label{aff46}
\and
UCB Lyon 1, CNRS/IN2P3, IUF, IP2I Lyon, 4 rue Enrico Fermi, 69622 Villeurbanne, France\label{aff47}
\and
Department of Astronomy, University of Geneva, ch. d'Ecogia 16, 1290 Versoix, Switzerland\label{aff48}
\and
Aix-Marseille Universit\'e, CNRS, CNES, LAM, Marseille, France\label{aff49}
\and
Universit\'e Paris-Saclay, CNRS, Institut d'astrophysique spatiale, 91405, Orsay, France\label{aff50}
\and
Aix-Marseille Universit\'e, CNRS/IN2P3, CPPM, Marseille, France\label{aff51}
\and
INAF-Istituto di Astrofisica e Planetologia Spaziali, via del Fosso del Cavaliere, 100, 00100 Roma, Italy\label{aff52}
\and
INFN-Bologna, Via Irnerio 46, 40126 Bologna, Italy\label{aff53}
\and
School of Physics, HH Wills Physics Laboratory, University of Bristol, Tyndall Avenue, Bristol, BS8 1TL, UK\label{aff54}
\and
University Observatory, LMU Faculty of Physics, Scheinerstr.~1, 81679 Munich, Germany\label{aff55}
\and
Max Planck Institute for Extraterrestrial Physics, Giessenbachstr. 1, 85748 Garching, Germany\label{aff56}
\and
INAF-Osservatorio Astronomico di Padova, Via dell'Osservatorio 5, 35122 Padova, Italy\label{aff57}
\and
Universit\"ats-Sternwarte M\"unchen, Fakult\"at f\"ur Physik, Ludwig-Maximilians-Universit\"at M\"unchen, Scheinerstr.~1, 81679 M\"unchen, Germany\label{aff58}
\and
Institute of Theoretical Astrophysics, University of Oslo, P.O. Box 1029 Blindern, 0315 Oslo, Norway\label{aff59}
\and
Jet Propulsion Laboratory, California Institute of Technology, 4800 Oak Grove Drive, Pasadena, CA, 91109, USA\label{aff60}
\and
Felix Hormuth Engineering, Goethestr. 17, 69181 Leimen, Germany\label{aff61}
\and
Technical University of Denmark, Elektrovej 327, 2800 Kgs. Lyngby, Denmark\label{aff62}
\and
Cosmic Dawn Center (DAWN), Denmark\label{aff63}
\and
Max-Planck-Institut f\"ur Astronomie, K\"onigstuhl 17, 69117 Heidelberg, Germany\label{aff64}
\and
Department of Physics and Astronomy, University College London, Gower Street, London WC1E 6BT, UK\label{aff65}
\and
Department of Physics, P.O. Box 64, University of Helsinki, 00014 Helsinki, Finland\label{aff66}
\and
Helsinki Institute of Physics, Gustaf H{\"a}llstr{\"o}min katu 2, University of Helsinki, 00014 Helsinki, Finland\label{aff67}
\and
Laboratoire d'etude de l'Univers et des phenomenes eXtremes, Observatoire de Paris, Universit\'e PSL, Sorbonne Universit\'e, CNRS, 92190 Meudon, France\label{aff68}
\and
SKAO, Jodrell Bank, Lower Withington, Macclesfield SK11 9FT, UK\label{aff69}
\and
Centre de Calcul de l'IN2P3/CNRS, 21 avenue Pierre de Coubertin 69627 Villeurbanne Cedex, France\label{aff70}
\and
INFN-Sezione di Roma, Piazzale Aldo Moro, 2 - c/o Dipartimento di Fisica, Edificio G. Marconi, 00185 Roma, Italy\label{aff71}
\and
Dipartimento di Fisica e Astronomia "Augusto Righi" - Alma Mater Studiorum Universit\`a di Bologna, via Piero Gobetti 93/2, 40129 Bologna, Italy\label{aff72}
\and
Department of Physics, Institute for Computational Cosmology, Durham University, South Road, Durham, DH1 3LE, UK\label{aff73}
\and
Universit\'e Paris Cit\'e, CNRS, Astroparticule et Cosmologie, 75013 Paris, France\label{aff74}
\and
CNRS-UCB International Research Laboratory, Centre Pierre Bin\'etruy, IRL2007, CPB-IN2P3, Berkeley, USA\label{aff75}
\and
Institute of Physics, Laboratory of Astrophysics, Ecole Polytechnique F\'ed\'erale de Lausanne (EPFL), Observatoire de Sauverny, 1290 Versoix, Switzerland\label{aff76}
\and
Telespazio UK S.L. for European Space Agency (ESA), Camino bajo del Castillo, s/n, Urbanizacion Villafranca del Castillo, Villanueva de la Ca\~nada, 28692 Madrid, Spain\label{aff77}
\and
Institut de F\'{i}sica d'Altes Energies (IFAE), The Barcelona Institute of Science and Technology, Campus UAB, 08193 Bellaterra (Barcelona), Spain\label{aff78}
\and
European Space Agency/ESTEC, Keplerlaan 1, 2201 AZ Noordwijk, The Netherlands\label{aff79}
\and
School of Mathematics, Statistics and Physics, Newcastle University, Herschel Building, Newcastle-upon-Tyne, NE1 7RU, UK\label{aff80}
\and
DARK, Niels Bohr Institute, University of Copenhagen, Jagtvej 155, 2200 Copenhagen, Denmark\label{aff81}
\and
Waterloo Centre for Astrophysics, University of Waterloo, Waterloo, Ontario N2L 3G1, Canada\label{aff82}
\and
Department of Physics and Astronomy, University of Waterloo, Waterloo, Ontario N2L 3G1, Canada\label{aff83}
\and
Perimeter Institute for Theoretical Physics, Waterloo, Ontario N2L 2Y5, Canada\label{aff84}
\and
Universit\'e Paris-Saclay, Universit\'e Paris Cit\'e, CEA, CNRS, AIM, 91191, Gif-sur-Yvette, France\label{aff85}
\and
Space Science Data Center, Italian Space Agency, via del Politecnico snc, 00133 Roma, Italy\label{aff86}
\and
Centre National d'Etudes Spatiales -- Centre spatial de Toulouse, 18 avenue Edouard Belin, 31401 Toulouse Cedex 9, France\label{aff87}
\and
Institute of Space Science, Str. Atomistilor, nr. 409 M\u{a}gurele, Ilfov, 077125, Romania\label{aff88}
\and
Dipartimento di Fisica e Astronomia "G. Galilei", Universit\`a di Padova, Via Marzolo 8, 35131 Padova, Italy\label{aff89}
\and
INFN-Padova, Via Marzolo 8, 35131 Padova, Italy\label{aff90}
\and
Instituto de F\'isica Te\'orica UAM-CSIC, Campus de Cantoblanco, 28049 Madrid, Spain\label{aff91}
\and
Institut de Recherche en Astrophysique et Plan\'etologie (IRAP), Universit\'e de Toulouse, CNRS, UPS, CNES, 14 Av. Edouard Belin, 31400 Toulouse, France\label{aff92}
\and
Universit\'e St Joseph; Faculty of Sciences, Beirut, Lebanon\label{aff93}
\and
Departamento de F\'isica, FCFM, Universidad de Chile, Blanco Encalada 2008, Santiago, Chile\label{aff94}
\and
Department of Physics and Helsinki Institute of Physics, Gustaf H\"allstr\"omin katu 2, University of Helsinki, 00014 Helsinki, Finland\label{aff95}
\and
Department of Physics, Royal Holloway, University of London, Surrey TW20 0EX, UK\label{aff96}
\and
Departamento de F\'isica, Faculdade de Ci\^encias, Universidade de Lisboa, Edif\'icio C8, Campo Grande, PT1749-016 Lisboa, Portugal\label{aff97}
\and
Instituto de Astrof\'isica e Ci\^encias do Espa\c{c}o, Faculdade de Ci\^encias, Universidade de Lisboa, Tapada da Ajuda, 1349-018 Lisboa, Portugal\label{aff98}
\and
Mullard Space Science Laboratory, University College London, Holmbury St Mary, Dorking, Surrey RH5 6NT, UK\label{aff99}
\and
Cosmic Dawn Center (DAWN)\label{aff100}
\and
Niels Bohr Institute, University of Copenhagen, Jagtvej 128, 2200 Copenhagen, Denmark\label{aff101}
\and
Universidad Polit\'ecnica de Cartagena, Departamento de Electr\'onica y Tecnolog\'ia de Computadoras,  Plaza del Hospital 1, 30202 Cartagena, Spain\label{aff102}
\and
Kapteyn Astronomical Institute, University of Groningen, PO Box 800, 9700 AV Groningen, The Netherlands\label{aff103}
\and
Caltech/IPAC, 1200 E. California Blvd., Pasadena, CA 91125, USA\label{aff104}
\and
Dipartimento di Fisica e Scienze della Terra, Universit\`a degli Studi di Ferrara, Via Giuseppe Saragat 1, 44122 Ferrara, Italy\label{aff105}
\and
Istituto Nazionale di Fisica Nucleare, Sezione di Ferrara, Via Giuseppe Saragat 1, 44122 Ferrara, Italy\label{aff106}
\and
INAF, Istituto di Radioastronomia, Via Piero Gobetti 101, 40129 Bologna, Italy\label{aff107}
\and
Astronomical Observatory of the Autonomous Region of the Aosta Valley (OAVdA), Loc. Lignan 39, I-11020, Nus (Aosta Valley), Italy\label{aff108}
\and
ICSC - Centro Nazionale di Ricerca in High Performance Computing, Big Data e Quantum Computing, Via Magnanelli 2, Bologna, Italy\label{aff109}
\and
Univ. Grenoble Alpes, CNRS, Grenoble INP, LPSC-IN2P3, 53, Avenue des Martyrs, 38000, Grenoble, France\label{aff110}
\and
Dipartimento di Fisica, Sapienza Universit\`a di Roma, Piazzale Aldo Moro 2, 00185 Roma, Italy\label{aff111}
\and
Aurora Technology for European Space Agency (ESA), Camino bajo del Castillo, s/n, Urbanizacion Villafranca del Castillo, Villanueva de la Ca\~nada, 28692 Madrid, Spain\label{aff112}
\and
Dipartimento di Fisica - Sezione di Astronomia, Universit\`a di Trieste, Via Tiepolo 11, 34131 Trieste, Italy\label{aff113}
\and
Institut d'Astrophysique de Paris, 98bis Boulevard Arago, 75014, Paris, France\label{aff114}
\and
ICL, Junia, Universit\'e Catholique de Lille, LITL, 59000 Lille, France\label{aff115}
\and
CERCA/ISO, Department of Physics, Case Western Reserve University, 10900 Euclid Avenue, Cleveland, OH 44106, USA\label{aff116}
\and
Departamento de F{\'\i}sica Fundamental. Universidad de Salamanca. Plaza de la Merced s/n. 37008 Salamanca, Spain\label{aff117}
\and
Aix-Marseille Universit\'e, Universit\'e de Toulon, CNRS, CPT, Marseille, France\label{aff118}
\and
Universit\'e de Strasbourg, CNRS, Observatoire astronomique de Strasbourg, UMR 7550, 67000 Strasbourg, France\label{aff119}
\and
Center for Data-Driven Discovery, Kavli IPMU (WPI), UTIAS, The University of Tokyo, Kashiwa, Chiba 277-8583, Japan\label{aff120}
\and
Max-Planck-Institut f\"ur Physik, Boltzmannstr. 8, 85748 Garching, Germany\label{aff121}
\and
Jodrell Bank Centre for Astrophysics, Department of Physics and Astronomy, University of Manchester, Oxford Road, Manchester M13 9PL, UK\label{aff122}
\and
California Institute of Technology, 1200 E California Blvd, Pasadena, CA 91125, USA\label{aff123}
\and
Department of Physics \& Astronomy, University of California Irvine, Irvine CA 92697, USA\label{aff124}
\and
Departamento F\'isica Aplicada, Universidad Polit\'ecnica de Cartagena, Campus Muralla del Mar, 30202 Cartagena, Murcia, Spain\label{aff125}
\and
Instituto de F\'isica de Cantabria, Edificio Juan Jord\'a, Avenida de los Castros, 39005 Santander, Spain\label{aff126}
\and
INFN, Sezione di Lecce, Via per Arnesano, CP-193, 73100, Lecce, Italy\label{aff127}
\and
Department of Mathematics and Physics E. De Giorgi, University of Salento, Via per Arnesano, CP-I93, 73100, Lecce, Italy\label{aff128}
\and
INAF-Sezione di Lecce, c/o Dipartimento Matematica e Fisica, Via per Arnesano, 73100, Lecce, Italy\label{aff129}
\and
Department of Computer Science, Aalto University, PO Box 15400, Espoo, FI-00 076, Finland\label{aff130}
\and
Universidad de La Laguna, Dpto. Astrof\'\i sica, E-38206 La Laguna, Tenerife, Spain\label{aff131}
\and
Ruhr University Bochum, Faculty of Physics and Astronomy, Astronomical Institute (AIRUB), German Centre for Cosmological Lensing (GCCL), 44780 Bochum, Germany\label{aff132}
\and
Department of Physics and Astronomy, Vesilinnantie 5, University of Turku, 20014 Turku, Finland\label{aff133}
\and
Finnish Centre for Astronomy with ESO (FINCA), Quantum, Vesilinnantie 5, University of Turku, 20014 Turku, Finland\label{aff134}
\and
Serco for European Space Agency (ESA), Camino bajo del Castillo, s/n, Urbanizacion Villafranca del Castillo, Villanueva de la Ca\~nada, 28692 Madrid, Spain\label{aff135}
\and
ARC Centre of Excellence for Dark Matter Particle Physics, Melbourne, Australia\label{aff136}
\and
Centre for Astrophysics \& Supercomputing, Swinburne University of Technology,  Hawthorn, Victoria 3122, Australia\label{aff137}
\and
Department of Physics and Astronomy, University of the Western Cape, Bellville, Cape Town, 7535, South Africa\label{aff138}
\and
Department of Physics, Oxford University, Keble Road, Oxford OX1 3RH, UK\label{aff139}
\and
DAMTP, Centre for Mathematical Sciences, Wilberforce Road, Cambridge CB3 0WA, UK\label{aff140}
\and
Kavli Institute for Cosmology Cambridge, Madingley Road, Cambridge, CB3 0HA, UK\label{aff141}
\and
Institut d'Astrophysique de Paris, UMR 7095, CNRS, and Sorbonne Universit\'e, 98 bis boulevard Arago, 75014 Paris, France\label{aff142}
\and
Departement of Theoretical Physics, University of Geneva, Switzerland\label{aff143}
\and
Department of Physics, Centre for Extragalactic Astronomy, Durham University, South Road, Durham, DH1 3LE, UK\label{aff144}
\and
IRFU, CEA, Universit\'e Paris-Saclay 91191 Gif-sur-Yvette Cedex, France\label{aff145}
\and
INAF-Osservatorio Astrofisico di Arcetri, Largo E. Fermi 5, 50125, Firenze, Italy\label{aff146}
\and
Centro de Astrof\'{\i}sica da Universidade do Porto, Rua das Estrelas, 4150-762 Porto, Portugal\label{aff147}
\and
Instituto de Astrof\'isica e Ci\^encias do Espa\c{c}o, Universidade do Porto, CAUP, Rua das Estrelas, PT4150-762 Porto, Portugal\label{aff148}
\and
Institute of Cosmology and Gravitation, University of Portsmouth, Portsmouth PO1 3FX, UK\label{aff149}
\and
Department of Astrophysics, University of Zurich, Winterthurerstrasse 190, 8057 Zurich, Switzerland\label{aff150}
\and
University of Applied Sciences and Arts of Northwestern Switzerland, School of Computer Science, 5210 Windisch, Switzerland\label{aff151}
\and
INAF - Osservatorio Astronomico d'Abruzzo, Via Maggini, 64100, Teramo, Italy\label{aff152}
\and
Theoretical astrophysics, Department of Physics and Astronomy, Uppsala University, Box 516, 751 37 Uppsala, Sweden\label{aff153}
\and
Universit\'e C\^{o}te d'Azur, Observatoire de la C\^{o}te d'Azur, CNRS, Laboratoire Lagrange, Bd de l'Observatoire, CS 34229, 06304 Nice cedex 4, France\label{aff154}
\and
Mathematical Institute, University of Leiden, Einsteinweg 55, 2333 CA Leiden, The Netherlands\label{aff155}
\and
Institute of Astronomy, University of Cambridge, Madingley Road, Cambridge CB3 0HA, UK\label{aff156}
\and
Center for Astrophysics and Cosmology, University of Nova Gorica, Nova Gorica, Slovenia\label{aff157}
\and
Institute for Particle Physics and Astrophysics, Dept. of Physics, ETH Zurich, Wolfgang-Pauli-Strasse 27, 8093 Zurich, Switzerland\label{aff158}
\and
Department of Astrophysical Sciences, Peyton Hall, Princeton University, Princeton, NJ 08544, USA\label{aff159}
\and
Space physics and astronomy research unit, University of Oulu, Pentti Kaiteran katu 1, FI-90014 Oulu, Finland\label{aff160}
\and
Institut de Physique Th\'eorique, CEA, CNRS, Universit\'e Paris-Saclay 91191 Gif-sur-Yvette Cedex, France\label{aff161}
\and
International Centre for Theoretical Physics (ICTP), Strada Costiera 11, 34151 Trieste, Italy\label{aff162}
\and
Center for Computational Astrophysics, Flatiron Institute, 162 5th Avenue, 10010, New York, NY, USA\label{aff163}}    
%
\abstract{Primordial non-Gaussianity (PNG) is a powerful probe of the origin of cosmic structure. Stage-IV surveys like \Euclid will measure galaxy $2$- and $3$-point clustering at high signal-to-noise, but exploiting this requires a robust joint analysis. We prepare for analysis of \Euclid\/'s spectroscopic sample by validating a redshift-space power-spectrum and bispectrum pipeline (one-loop $P_\ell$, tree-level $B_\ell$) on \Euclid-like mocks from Abacus-PNG $N$-body simulations with Gaussian and local-PNG initial conditions, using a halo occupation distribution (HOD) tuned to Euclid Flagship 2. We stress-test analysis choices -- PNG-bias parametrisation, priors, and scale cuts -- and perform null tests without PNG. In a `prior-agnostic setup', high-significance detection of the dominant PNG term $\propto f_{\rm NL} \, b_\phi$ in single redshift bins is difficult; nevertheless, the bispectrum provides constraints on other PNG combinations that partially lift degeneracies. We propose a physically motivated prior on $b_\phi$ that yields unbiased $f_{\rm NL}$ while accounting for theory uncertainty, and determine scale cuts that give unbiased $\Lambda$CDM and $f_{\rm NL}$. With $V_{\rm eff}=16\,h^{-3}\,{\rm Gpc}^3$ across four snapshots ($0.8\le z\le1.7$), our likelihood analyses recover $<1\sigma$ bias in $f_{\rm NL}$ and $\Lambda$CDM. At fixed cuts, $B_\ell$ alone reduces $\sigma({f_{\rm NL}})$ by $\sim29$--$46\%$ relative to $P_\ell$, and joint
power spectrum-bispectrum analysis tightens a further $\sim8$--$13\%$; the cumulative gain from $z=0.8$ to $1.7$ is $\sim2.3$ for the joint case. The bispectrum quadrupole is a key contributor. Our strongest results are at $z=1.7$: $1.9\sigma$ for $\fnl \, b_\phi$ (prior-agnostic) and $2.35\sigma$ for $\fnl$ (prior-based). Joint analyses thus offer strong prospects for testing multi-field inflation, pending end-to-end validation in the full \Euclid geometry with observational systematics. \\ }

%
%
%

\titlerunning{Testing multi-field inflation with galaxy power spectrum and bispectrum} 
\authorrunning{Euclid Collaboration: D. Linde et al.}
   
\maketitle
%
%
%
%
\section{Introduction}

The ongoing and upcoming Stage-IV galaxy surveys -- including the Dark Energy Spectroscopic Instrument \citep[DESI;][]{DESI:2016fyo}, \Euclid~\citep{EUCLID:2011zbd,Euclid:2024yrr}, the Spectro-Photometer for the History of the Universe, Epoch of Reionization and Ices Explorer \citep[SPHEREx;][]{SPHEREx:2014bgr}, the \textit{Nancy Grace Roman} Space Telescope \citep{Akeson:2019biv,Wang:2021oec}, and the Vera C.\ Rubin Observatory Legacy Survey of Space and Time \citep[LSST;][]{LSST:2008ijt} -- will deliver high-precision maps of the large-scale structure (LSS) of the Universe by measuring galaxy clustering and weak-lensing shear across very large cosmological volumes. These data sets provide a compelling route to stress test the standard cosmological model ($\Lambda$CDM) and to search for departures indicative of new physics -- tightening bounds on neutrino masses and the dark energy equation of state, and detecting or tightly constraining deviations from Gaussianity in the primordial fluctuations. 

The simplest models of inflation \citep{Starobinsky:1980te,Guth:1980zm,Linde:1981mu}, characterised by a canonical single scalar field originating from the Bunch--Davies vacuum and slowly rolling down its potential, predict a nearly Gaussian distribution of primordial fluctuations. Deviations from any of these assumptions produce distinctive non-Gaussian signatures, commonly characterised in terms of the dependence of the primordial higher-order statistics on the shape of the configuration (triangle in the case of bispectrum) formed by the three wavenumbers forming a triangle. The so-called local shape, characterised by large bispectrum values for squeezed triangular configurations, is considered a smoking gun for multi-field models of inflation, since single-field models predict a nearly zero local-type primordial bispectrum \citep{Maldacena:2002vr,Creminelli:2004yq,Cabass:2016cgp}. The tightest current bounds on local primordial non-Gaussianity (PNG) come from the cosmic microwave background (CMB) by the {\it Planck} satellite \citep[][$\fnl = -0.9\pm 5.1$ at $68\%$ confidence]{Planck:2019kim}, and are nearly cosmic-variance limited. Three-dimensional maps of LSS from Stage-IV spectroscopic surveys promise to push beyond the CMB by exploiting vastly more modes over a broad redshift range \citep[see, e.g.,][for recent CMB and LSS forecasts]{Achucarro:2022qrl}.

PNG impacts late-time structure in two ways. First, linear growth transports the initial higher-order matter correlators forward in time, yielding a primordial contribution to the late-time bispectrum (and beyond). Second, PNG modifies halo abundances and, therefore the bias relation between matter and tracers. For local PNG, the leading effect in the galaxy power spectrum is the large-scale, scale-dependent bias $\Delta b_1 \propto \fnl \, k^{-2}$ \citep{Dalal:2007cu, Matarrese:2008nc, Slosar:2008hx, Afshordi:2008ru,McDonald:2008sc}, which has underpinned most current constraints from galaxy and quasar power spectra \citep{Slosar:2008hx,eBOSS:2021jbt,Cagliari:2023mkq,Chaussidon:2024qni,Cagliari:2025rqe,Fabbian:2025fdk}, as well as cross-correlations of galaxies/quasars with CMB lensing data \citep{DESI:2023duv,Bermejo-Climent:2024bcb,Fabbian:2025fdk}. The galaxy bispectrum, however, carries both the primordial matter contributions and PNG-induced bias terms, offering shape information that is complementary to the power spectrum. A joint analysis of redshift-space power spectrum and bispectrum, therefore, leverages distinct dependencies on $\fnl$ and bias parameters, improving robustness and tightening constraints relative to either observable alone. 

While the study of the bispectrum for constraining PNG \citep{Scoccimarro:2000qg, Verde:1999ij, Scoccimarro:2003wn, Sefusatti:2007ih} and gravitational evolution \citep{Fry:1992vr,Matarrese:1997sk,Scoccimarro:1999ed} has a long history, and galaxy bispectrum measurements date back more than thirty years \citep{Baumgart:1991rz,Matarrese:1997sk,Feldman:2000vk,Scoccimarro:2000sp,Verde:2001sf}, its use in modern survey analyses has long lagged behind that of the power spectrum. This reflects both modelling difficulties and the high dimensionality of the bispectrum, which complicates measurement, covariance estimation (especially from mocks), and parameter inference.

The perturbative modelling of LSS, particularly within the effective field theory framework \citep[EFTofLSS;][]{McDonald:2009dh,Carrasco:2012cv,Vlah:2015sea,Perko:2016puo,Chen:2020zjt},\footnote{See also \citet{McDonald:2008sc},\citet{Chan:2012jj}, \cite{McDonald:2014dxa}, \citet{Assassi:2014fva}, \citet{Assassi:2015fma}, \citet{Vlah:2015zda}, \citet{Desjacques:2016bnm}, \citet{Senatore:2017pbn}, and \citet{Vlah:2018ygt}.} has advanced substantially in recent years. By consistently modelling nonlinear gravitational evolution, galaxy bias, stochasticity, and redshift-space distortions (RSD), together with baryon acoustic oscillations through infrared resummation, both the power spectrum and bispectrum can now be described within the perturbative regime (see, e.g., \citealp{Ivanov:2022mrd} and \citealp{Cabass:2022avo} for recent reviews). Furthermore, an alternative RSD modelling that partially preserves the nonlinear redshift-space mapping has been shown to notably improve full-shape analyses of both the power spectrum and bispectrum \citep{Eggemeier2025}, reflecting that RSD imprint nonlinearity-sensitive features across all scales, unlike dynamical evolution or galaxy-bias terms.

One-loop power-spectrum calculations are now standard in spectroscopic survey analyses, enabled by fast loop-evaluation algorithms \citep{Hamilton:1999uv,McEwen:2016fjn,Schmittfull:2016yqx,Simonovic:2017mhp,Bakx:2024zgu}, now implemented in several publicly available codes (e.g., \texttt{CLASS-PT}, \citealp{Chudaykin:2020aoj}; \texttt{PyBird}, \citealp{DAmico:2020kxu}; \texttt{Velocileptors}, \citealp{Chen:2020zjt}; \texttt{CLASS-OneLoop},\footnote{To be released soon.} \citealp{Linde:2024uzr}). In contrast, bispectrum modelling has until recently been mostly limited to tree-level implementations. Although one-loop predictions have been explored theoretically \citep{Scoccimarro:1995if,Sefusatti:2009qh,Eggemeier:2021cam,Philcox:2022frc,DAmicoEtal2022C},\footnote{See also \citet{Angulo:2014tfa}, \citet{Baldauf:2014qfa}, \citet{Hashimoto:2017klo}, \citet{Eggemeier:2018qae}, \citet{Alkhanishvili:2021pvy}, and \citet{DAmico:2022osl}.} their application to data has been hindered by the lack of efficient loop algorithms and by the complexity of expanding matter density, bias, counterterm, stochastic, and velocity contributions to fourth order. Recent computational advances \citep{Bakx:2024zgu,Bakx:2025pop,Bakx:2025jwa,Anastasiou:2025jsy} now promise to enable consistent full-shape analyses of the power spectrum and bispectrum at two- and one-loop order, respectively.

The recent joint analyses of power spectrum and bispectrum of BOSS data as well as the first data release of DESI have shown that adding the bispectrum can significantly improve cosmological constraints, in particular on selected $\Lambda$CDM parameters, dark energy, and the sum of neutrino masses \citep{PhilcoxIvanov2021,Ivanov:2023qzb,DAmico:2022osl,Chudaykin:2025lww,Ivanov:2026dvl,Chudaykin:2026nls,Chudaykin:2025aux}, and improve bounds on primordial bispectra from single- and multi-field inflation \citep{CabassEtal2022A,Cabass:2022ymb,DAmicoEtal2022,Cagliari:2025rqe,Chudaykin:2025vdh}. Looking ahead to the full datasets from Stage-IV surveys, with far higher signal-to-noise in both $2$- and $3$-point statistics, robust analyses require scaling the methodology in both accuracy and speed, while controlling theory systematic effects and carefully assessing observational systematic effects.\footnote{See also \cite{Euclid:2025hlc} for recent \Euclid-specific forecasts on constraints on initial conditions, including analyses based on the bispectrum, and \cite{Euclid:2024ris} for the application of field-level inference to local PNG.}

In this paper, we validate the theoretical modelling and inference pipeline, and assess the impact of key analysis choices on constraining local-type PNG through a joint analysis of redshift-space multipoles of the galaxy power spectrum and bispectrum, focusing on the spectroscopic sample of \Euclid. This work extends the analysis of \citet{MoradinezhadDizgah:2020whw}, which compared the perturbative predictions of the real-space power spectrum and bispectrum with the measurements on halo catalogues from \textsc{Eos Dataset}, suite of $N$-body simulations with Gaussian and non-Gaussian initial conditions for a fixed cosmology, to include RSD and account for the simultaneous variation of the amplitude of the primordial bispectrum, $\Lambda$CDM, and nuisance galaxy parameters. Our goal here is to provide a thorough assessment of the key theoretical systematics affecting PNG inference, rather than to present forecasts for the full Euclid Survey.

This paper is part of a series of Euclid Collaboration preparation papers dedicated to validating clustering analyses on simulated data, representative of the spectroscopic H$\alpha$ galaxy sample. These works investigate the impact of analysis choices and modelling systematic effects on baryon acoustic oscillations (BAO) and full-shape analysis of $2$-point and $3$-point statistics, assuming $\Lambda$CDM cosmology with Gaussian initial conditions (\citealp{Euclid:2023tog,Euclid:2025diy}; \citealp{Camacho2026Euclid}; \citealp{Euclid:2026yna}; \citealp{Pardede2026Euclid}; Euclid Collaboration: Pugno et al. in prep; Euclid Collaboration: Moresco et al. in prep). In contrast, the other works in the series analysed the Euclid Flagship 1 simulations \citep{Euclid:2024few}. In this paper, we construct new halo occupation distribution (HOD) galaxy mocks designed to reproduce the expected properties of \Euclid H$\alpha$ galaxies using Abacus $N$-body simulations with local-type PNG \citep{Hadzhiyska:2024kmt}, along with the counterparts using Gaussian initial conditions. We quantify how analysis choices (e.g., parameter priors, scale cuts) affect the inferred value of $\fnl$ and investigate the information content of the power spectrum and bispectrum, both separately and in combination. The purpose of this work is to provide a thorough assessment of the key theoretical systematics affecting PNG inference, rather than to deliver forecasts for the full Euclid Survey.

The rest of the paper is organised as follows. In Sect.~\ref{sec:th} we present the theoretical modelling of the redshift-space one-loop power spectrum and tree-level bispectrum in the presence of local PNG. In Sect.~\ref{sec:data_like} we describe the simulation suites and HOD construction for the \Euclid spectroscopic sample and the likelihood framework, and the employed performance metrics. In Sect.~\ref{sec:code} we introduce our new code, \texttt{CosmoSIS-gClust}, which enables joint analysis of the galaxy power spectrum and bispectrum in the presence of PNG. In Sect.~\ref{sec:results} we present our results, including validation tests that delineate the relevant model parameter space, assess the impact of prior choices and adopted scale cuts, and compare parameter constraints from the power spectrum and bispectrum separately and jointly at different redshifts. We conclude in Sect.~\ref{sec:conclusions}. Several appendices present further details on modelling (Appendix~\ref{app:theory}), measurements (Appendix~\ref{app:measurment-all}), likelihood analysis (Appendix~\ref{app:inference}), and some auxiliary results (Appendix~\ref{app:add_joint}).

\section{Galaxy clustering in redshift space with  PNG}\label{sec:th}

We adopt the perturbative model of the redshift-space galaxy power spectrum and bispectrum within the Eulerian-space formulation of the EFTofLSS. For the one-loop power-spectrum correction, we base our model on the velocity-moment expansion of \citet{Chen:2020zjt} and extend the computation to include contributions from PNG of local type. The bispectrum is modelled at tree level, while we include a counterterm to capture the effect of nonlinear RSD due to large small-scale velocities. The impact of large bulk flows, infrared (IR) resummation, as well as the mismatch between fiducial and true cosmologies, known as the Alcock--Paczynski (AP) effect \citep{Alcock:1979mp}, is accounted for in both the power spectrum and the bispectrum. The modelling components that have a dependence on PNG are reviewed below, while the IR resummation and the AP effect are described in Appendices~\ref{app:IR} and \ref{app:AP}.

\subsection{Renormalised bias expansion}\label{sec:bias_exp}
In the presence of PNG, the galaxy bias expansion must be extended beyond operators constructed from the late-time matter density and tidal fields \citep{McDonald:2009dh,Chan:2012jj,Baldauf:2012hs}. One must include operators built from the primordial gravitational potential and mixed operators that correlate it with the standard density/tidal operators \citep{McDonald:2008sc,Giannantonio:2009ak,Assassi:2015fma}. The precise form of the operators depends on the PNG shape.

Retaining both contributions, we write the galaxy overdensity field, $\delta_{\mathrm{g}}$, as 
\be\label{eq:supgen}
\delta_{\mathrm{g}}(\bx) = \delta_{\mathrm{g}}^{\mathrm{G}}(\bx) + \delta_{\mathrm{g}}^{\mathrm{PNG}}(\bx)\,,
\ee
where $\bx$ denotes the comoving Eulerian coordinate. The Gaussian part collects the contributions from nonlinear gravitational evolution, the galaxy biases, and stochastic terms. Truncated at third order, it reads as \citep{Desjacques:2016bnm}
\begin{align}\label{eq:G_bias}
\delta_{\mathrm{g}}^{\mathrm{G}}(\bx) =& \,b_1 \, \delta(\bx)+b_{\nabla^2\delta} \, \nabla^2\delta(\bx)+\epsilon(\bx) + \frac{b_2}{2} \, \delta^2(\bx)+ b_{\mathcal{G}_2} \, \mathcal{G}_2(\bx) \nn \\
& + \epsilon_\delta(\bx) \, \delta(\bx)+\frac{b_3}{6} \, \delta^3(\bx) +b_{\mathcal{G}_3} \, \mathcal{G}_3(\bx)+b_{\mathcal{G}_2\delta} \, \mathcal{G}_2(\bx) \, \delta(\bx) \nn \\ 
& +b_{\Gamma_3} \, \Gamma_3(\bx)+ \epsilon_{\delta^2}(\bx) \,  \delta^2(\bx)+ \epsilon_{\mathcal G_2}(\bx) \, {\mathcal G_2}(\bx) \, . 
\end{align}
Here, $\cG_2$ and $\cG_3$ are second- and third-order Galileon operators,
\begin{align}
\mathcal{G}_2(\Phi) \equiv&\,(\partial_i\partial_j\Phi)^2-(\partial^2\Phi)^2 \, , \\
\mathcal{G}_3(\Phi) \equiv&\,-\partial_i\partial_j\Phi \, \partial_j\partial_k\Phi \, \partial_k\partial_i\Phi-\frac{1}{2} \,(\partial^2\Phi)^3 \nn \\
&\,+\frac{3}{2} \, (\partial_i\partial_j\Phi)^2 \, \partial^2\Phi\,, 
\end{align}
with $\Phi$ the late-time gravitational potential, and $\Gamma_3$ the difference between the density and velocity tidal tensors \citep{Chan:2012jj}, 
\begin{align}
\Gamma_3 &\equiv\mathcal{G}_2(\Phi)-\mathcal{G}_2(\Phi_v)\, ,
\end{align}
where $\Phi_v$ is the velocity potential, and differs from $\Phi$ starting at second order. The fields $\epsilon$, $\epsilon_\delta$, $\epsilon_{\delta^2}$, and $\epsilon_{\mathcal G_2}$ denote stochastic contributions.

The PNG-induced part at the same order reads
\begin{align}\label{eq:PNG_bias} 
\delta_{\mathrm{g}}^{\mathrm{PNG}}(\bx) =& \, f_{\rm NL} \, \left[ b_\phi  \, \phi(\bp) + b_{\nabla^2\phi} \, \nabla_p^2 \phi({\bp}) + b_{\phi \delta} \, \phi(\bp) \, \delta(\bx) \right. \nn \\ 
& + \left. \epsilon_\phi(\bx) \, \phi(\bp) + b_{\phi \delta^2} \, \phi(\bp) \, \delta^2(\bx) + b_{\phi{\mathcal G}_2} \, \phi(\bp) \, {\mathcal G}_2(\bx) \right. \nn \\ 
& + \left. \epsilon_{\phi \delta}(\bx) \, \phi(\bp) \, \delta(\bx) \right] + \frac{1}{2} \, f_{\rm NL}^2 \, b_{\phi^2} \, \phi^2(\bp)\, .
\end{align}
Here $\phi$ is the primordial gravitational potential, evaluated at the Lagrangian coordinate $\bp$ corresponding to $\bx$, and $\nabla_{p}$ denotes derivatives with respect to $\bp$. At linear order, the late-time density is related to $\phi$ via 
\begin{equation}
    \delta_0(\bk,z) = \cM(k,z)\,\phi(\bk)     
\end{equation}
with
\begin{equation}
    \cM(k,z) = \frac{2}{3} \frac{c^2\,k^2\,T(k)\,D(z)}{\Omega_{\mathrm{m}} \, H_0^2} \, ,    
\end{equation}
where $T(k)$ is the transfer function that converges to unity at large scales, and $D(z)$ is the linear growth factor normalised to $(1+z)^{-1}$ in the matter-dominated era. We drop the explicit redshift dependence in $\cM$ from now on for brevity.

In Eq.~\eqref{eq:PNG_bias}, the primordial fluctuations $\phi$ and the corresponding PNG stochastic fields are evaluated at the Lagrangian position $\bp$, while the density, tidal fields, and their stochastic terms are evaluated at the Eulerian position $\bx$. These are related by the matter displacement field,
\begin{equation}
\bx = \bp + \boldsymbol\psi(\bp) \,.
\end{equation}
Consequently, writing $\phi$ in terms of $\bx$ gives
\begin{equation}
\phi(\bp) = \phi[\bx-\boldsymbol\psi(\bp)]
= \phi\{\bx-\boldsymbol\psi[\bx-\boldsymbol\psi(\bx)]\} \, ,
\end{equation}
and a perturbative expansion up to third order yields\footnote{We retain the cubic contributions (including those arising from the second-order displacement field in the second term), which were neglected in \citet{MoradinezhadDizgah:2020whw}; while small for the power spectrum, they are needed for a consistent IR resummation \citep{Cabass:2022ymb}.}
\begin{align}
\phi(\bp) =& \, \phi(\bx) - \psi^i(\bx) \, \partial_i\phi(\bx) + \psi^k(\bx) \, \partial_k\psi^i(\bx) \, \partial_i\phi(\bx) \notag \\ 
&+ \frac{1}{2} \, \psi^i(\bx) \, \psi^j(\bx) \, \partial_i\partial_j \phi(\bx) \, .
\end{align}
At this order, the term $-\psi^i \, \partial_i\phi$ receives a contribution from the second-order displacement field. The first- and second-order Lagrangian displacements are
\begin{align}
\boldsymbol\psi^{(1)}(\bk) &= \immag\,\bk\,k^{-2}\,\delta_0(\bk)\,, \\
\boldsymbol\psi^{(2)}(\bk) &= \frac{\immag^2}{2} \int_{\bq}  \mathbf L^{(2)}(\bq,\bk-\bq)\,\delta_0(\bq)\,\delta_0(\bk-\bq)\,,
\end{align}
with the second-order Lagrangian perturbation theory (LPT) kernel
\be
\mathbf L^{(2)}(\bq_1,\bq_2) = \frac{3}{7}\,
\frac{\bq_1+\bq_2}{|\bq_1+\bq_2|^2}\,\left[1-\left(\frac{\bq_1 \cdot \bq_2}{q_1 \, q_2}\right)^{\,2}\right]\,,
\ee
where we denoted $\int_\bq \equiv \int \diff^3 q \,(2\pi)^{-3} $.

At the one-loop level, only one third-order bias contributes $b_{\Gamma_3}$ to the power spectrum  since the contributions of other cubic terms amounts to a scale-independent correction to the linear bias $b_1$. As noted in \citet{MoradinezhadDizgah:2020whw}, none of the third-order PNG biases contribute to the one-loop PNG terms, since they are absorbed in the definition of the renormalised linear non-Gaussian biases, $b_1$ and $b_\phi$. Moreover, the contributions from $b_{\phi \delta}$, where either $\phi$ or $\delta$ is expanded to second order in perturbations, are also absorbed by the renormalisation of the two linear biases \citep[see also][]{Cabass:2022ymb}.

\subsection{Galaxy power spectrum}\label{sec:Galaxy_PowerSpectrum}

The line-of-sight (LoS) velocities of galaxies add a Doppler component to their observed redshifts, on top of the cosmological redshift from expansion. Consequently, an object’s redshift-space position $\bs$ is displaced relative to its real-space position $\bx$ by its peculiar velocity $\bv$, such that $\bs = \bx + \hat{\bn} \, (\bv \cdot \hat{\bn}) \, \cH^{-1}$, where $\hat{\bn}$ is the LoS direction and $\cH\equiv a \, H$ is the conformal Hubble parameter. This velocity-induced mapping from real to redshift space imprints an anisotropy in the clustering pattern, known as redshift-space distortions.

Starting from the relationship between real-space and redshift-space galaxy overdensities, 
\begin{align}
    1 +\delta_{\mathrm{g}}^s(\bs) &= \int \diff^3 x \,  \left[1+\delta_{\mathrm{g}}(\bx) \right] \,  \delta_{\mathrm{D}}(\bs-\bx-\bu) \, , \notag \\
    (2\pi)^3 \, \delta_{\mathrm{D}}(\bk) + \delta_{\mathrm{g}}^s(\bk) &= \int \diff^3 x \, \left[1+\delta_{\mathrm{g}}(\bx) \right]  \, \e^{\immag \, \bk \cdot [\bx+\bu(\bx)]} \, ,
\end{align}
with $\delta_{\mathrm{D}}$ being the Dirac delta, $\bu = \hat{\bn} \, (\bv \cdot \hat{\bn}) \, \cH^{-1}$, the redshift-space galaxy power spectrum,
\be
\left\langle \delta_{\mathrm{g}}^s(\bk)  \, \delta_{\mathrm{g}}^s(\bk')\right\rangle \equiv (2\pi)^3 \,   \delta_{\mathrm{D}}(\bk+\bk') \, P_{\mathrm{g}}^s(\bk) \, ,
\ee
can be computed from the velocity moment-generating function, written as \citep{Chen:2020zjt}
\be
P_{\mathrm{g}}^s(\bk) = \int \diff^3 r \, \left\langle \left[1+\delta_{\mathrm{g}}(\bx_1)\right] \, \left[1+\delta_{\mathrm{g}}(\bx_2)\right] \, \e^{\immag \, \bk \cdot \Delta \bu} \right\rangle_{\bx_2 - \bx_1 = \br} \, ,
\ee
where $\Delta \bu \equiv \bu(\bx_2) - \bu(\bx_1)$ is the pairwise relative velocity. At one-loop level, retaining only contributions that contain at most four velocity fields, we have
\begin{align}
    P_{\mathrm{g}}^s(k,\mu) \simeq& \, \Sigma_{\hat{\bn}}^{(0)} + \immag \, k \, \mu \, \Sigma_{\hat{\bn}}^{(1)} + \frac{\immag^2}{2} \, k^2 \, \mu^2  \, \Sigma_{\hat{\bn}}^{(2)} + \frac{\immag^3}{3!} \,  k^3 \, \mu^3 \, \Sigma_{\hat{\bn}}^{(3)} \nn \\
    &+ \frac{\immag^4}{4!} \, k^4 \, \mu^4 \, \Sigma_{\hat{\bn}}^{(4)} \, , 
\end{align}
where $\mu\equiv \hat{\bk} \cdot \hat{\bn}$ corresponds to the cosine of the angle between the LoS and the wavenumber. The velocity moments are given by $\Sigma_{\hat{\bn}}^{(n)} \equiv \left\langle \left(1+\delta_1\right) \,  \left(1+\delta_2\right) \, \Delta \bu_1 \dotsc \Delta\bu_n \right\rangle$, and include stochastic and EFTofLSS counter terms in addition to deterministic terms. In the presence of PNG, each of the velocity moments receives two contributions: in addition to contributions present for Gaussian initial conditions (explicit forms in \citealp{Chen:2020zjt} and \citealp{Linde:2024uzr}), there are several PNG loops (explicit form in the \texttt{CosmoSIS-gClust} code paper, Linde \& Moradinezhad in prep.). Except for two contributions to the zeroth-order velocity moment, we only retain terms linear in $\fnl$ in the galaxy power spectrum. By adapting the notation introduced in \citet{Linde:2024uzr}, the one-loop power spectrum in redshift space requires computing $32$ loops for Gaussian initial conditions and an additional $45$ loops when considering local PNG.

In what follows, we denote the three contributions from clustering (deterministic), EFTofLSS counterterms, and shot noise (stochastic) as
\be
P_{\mathrm{g}}^s(k,\mu) = P_{\mathrm{clust}}^s(k,\mu) + P_{\mathrm{ctr}}^s(k,\mu) + P_{\mathrm{shot}}^s(k,\mu) \, .
\ee
In the presence of local PNG, each of these terms receives additional contributions to those for Gaussian initial conditions. We neglect the PNG stochastic contributions; however, as discussed below, we include new PNG counterterms, which prove important in obtaining unbiased constraints on simulated data.

Since several counterterms in the moment expansion exhibit identical $k$ and $\mu$ dependence, we collect them into a single term characterised by free amplitude, $c_n$,
\begin{align}
P_{\rm ctr, \, G}^s(k,\mu) =& \, - 2 \, k^2 \, P_{\rm lin}(k) \, \Biggl[ c_0 + c_2 \, f \, \mu^2 + c_4 \, f^2 \, \mu^4 \nn \\
& \,  + c_6 \, f^4 \, \mu^4 \, \left(b_1 + f \, \mu^2\right)^2\,k^2 \Biggr] \, .
\end{align}
For the shot noise, we absorb the growth rate, $f$, dependence into the amplitudes, $s_n$, and consider
\be
P_{\rm shot}^s(k,\mu) = \frac{1}{\bar n} \, \left(1 + s_0 + s_{02} \, k^2 + s_2 \, \mu^2 \, k^2 + s_4 \, \mu^4 \, k^4\right) \, ,
\ee
where $\bar n$ is the mean number density of the tracer.\footnote{We note that we use a slightly different parametrisation for the shot noise and counterterms compared to \citet{Chen:2020zjt} and \citet{Linde:2024uzr}. In particular, we neglect the $k$-independent $\mu^6$ counter term and model the next-to-leading order counterterm accounting for higher-derivative terms from the nonlinear RSD mapping, as done in \citet{Pardede2026Euclid}. This form is motivated by the terms arising from the expansion of the velocity-difference generator \citep{Eggemeier2025}.} 

Accounting for the additional one-loop terms that couple the primordial potential to the matter density requires introducing extra EFTofLSS counterterms to absorb the associated ultraviolet (UV) sensitivities. Each PNG loop is renormalised by adding counterterms that cancel its short-distance dependence; in the process, previously introduced Gaussian counterterms acquire finite renormalisations, and two PNG-specific pieces are needed: (i) a classical response that cancels the contact-type PNG contributions to the velocity moments, and (ii) a semi-stochastic piece-independent of the linear $P(k)$ that approaches a constant as $k\to0$, thereby renormalising the shot noise. We write
\be
P_{\rm crt, \, PNG}^s(k,\mu) = P_{\rm ct, \, \phi}^s(k,\mu) + P_{\rm ct, 12}^s(k,\mu)\,,
\ee
where
\begin{align}
P_{\rm crt, \, \phi}^s(k,\mu) &= -\,k^{2}\,\cM^{-1}(k)\,P(k)\,\left(c_{0}^{\phi}+c_{2}^{\phi}\,\mu^{2}+c_{4}^{\phi}\,\mu^{4}\right)\,, \\
P_{\rm crt,12}^s(k,\mu) &= -\,k^{2}\,\cM(k)\,\left(c_{0}^{12}+c_{2}^{12}\,\mu^{2}+c_{4}^{12}\,\mu^{4}\right)\,.
\end{align}
The factors $\cM^{\pm 1}(k)$ track whether the counterterm dresses an insertion of the primordial potential $\phi$ (yielding $\cM^{-1}$) or of the density $\delta$ (yielding $\cM$). Putting all together, our counterterms are given by
\be
P_{\rm crt}^s(k,\mu) =  P_{\rm crt, \, G}^s(k,\mu) + P_{\rm crt, \, PNG}^s(k,\mu) \,. 
\ee
The power spectrum multipoles are then defined as
\be
P_{\ell}(k) = \frac{2\ell + 1}{2} \, \int_{-1}^1 \diff \mu \, \cL_{\ell}(\mu) \,  P_{\mathrm{g}}^s(k,\mu) \, , 
\ee
where $\cL_\ell$ are the Legendre polynomials, and $\ell$ identifies the multipole order.

\subsection{Galaxy bispectrum}\label{sec:Galaxy_Bispectrum}

We model the galaxy bispectrum at tree level in perturbation theory, including contributions from the matter bispectrum, $B_0$, and bias terms induced by the PNG. We also account for stochastic contributions beyond Poisson shot noise and include a counterterm to capture the effect of small-scale velocities in redshift space. Defining the bispectrum as
\be
\left\langle \delta_{\mathrm{g}}^s(\bk_1) \, \delta_{\mathrm{g}}^s(\bk_2) \, \delta_{\mathrm{g}}^s(\bk_3) \right\rangle \equiv (2\pi)^3 \, \delta_{\mathrm{D}}\left(\bk_{123}\right) \, B_{\mathrm{g}}^s\left(\bk_1,\bk_2,\bk_3\right) \, ,
\ee
where $\bk_{123} = \bk_1+\bk_2+\bk_3$ and the three components are denoted as 
\begin{align}
    B_{\mathrm{g}}^s(\bk_1,\bk_2,\bk_3) =& \, B_{\mathrm{clust}}^s(\bk_1,\bk_2,\bk_3) + B_{\mathrm{ctr}}^s(\bk_1,\bk_2,\bk_3) \nn \\
    &+ B_{\mathrm{shot}}^s(\bk_1,\bk_2,\bk_3) \, . 
\end{align}

To express the (deterministic) clustering component of the bispectrum compactly, we use the redshift-space perturbation-theory kernels \citep{Scoccimarro:1999ed,Bernardeau2002},
\begin{align} 
\delta_{\mathrm{g, \, clust}}^s(\bk) =&\, \sum_{n=1} \int_{\bq_1} \dotsi \int_{\bq_n} \delta_{\mathrm{D}}(\bk-\bq_{1\dotso n}) \, Z_n(\bq_1,\dotsc,\bq_n) \nn \\
&\times \delta_0(\bq_1) \dotsm \delta_0(\bq_n)\, ,
\end{align}
where $\bq_{1\dotso n} = \bq_1+\bq_2+\dotsb +\bq_n$. In the presence of PNG, the kernels receive additional terms from PNG biases. At leading order in $f_{\rm NL}$, we have
\begin{align}
Z_1(\kv_1) =& \, b_1 + f \, \mu_1^2 + \fnl \, b_\phi \, {\mathcal M}^{-1}(k_1) \, ,\\
Z_2(\kv_1,\kv_2) =& \, K_2(\kv_1,\kv_2) + f \, \mu^2 \, G_2(\kv_1,\kv_2) \nn\\
&- \frac{1}{2} \, f \, \mu_3 \, k \, \left[\frac{\mu_1}{k_1} \, \left(b_1+f \, \mu_2^2\right)+\frac{\mu_2}{k_2} \, \left(b_1+f \, \mu_1^2\right)
\right], \nn \\
&+ \frac{1}{2} \, b_\phi \, \fnl \, \mu_{12} \, \left[\frac{k_1}{k_2} \, {\mathcal M}^{-1}(k_1) + \frac{k_2}{k_1} \, {\mathcal M}^{-1}(k_2) \right] \nn \\
&- \frac{1}{2} \, b_\phi \, \fnl \, f \, \mu_3 \, k \,  \left[\frac{\mu_1}{k_1} \,{\mathcal M}^{-1}(k_2) + \frac{\mu_2}{k_2} \, {\mathcal M}^{-1}(k_1) \right]\nn \\
&+ \frac{1}{2} \, b_{\phi \delta} \, \fnl \, \left[{\mathcal M}^{-1}(k_1) + {\mathcal M}^{-1}(k_2)\right],
\end{align}
where $\mu_i = \hat \bk_i \cdot \hat \bn$ is the cosine of the angle of each wave-vector with respect to the LoS direction, with $\mu_3$ satisfying the relation $\mu_3 = - k_1 \, k_3^{-1} \, \mu_1 - k_2 \, k_3^{-1} \, \mu_2$, and $\mu_{12} = \hat \bk_1 \cdot \hat \bk_2$. Therefore, the clustering contribution to the galaxy bispectrum at tree-level is given by 
\begin{align}\label{eq:Bgrav}
	B_{\mathrm{clust}}^s (\kv_1,\kv_2,\kv_3) 
    =&\; Z_1(\bk_1) \, Z_1(\bk_2) \, Z_1(\bk_3) \, B_0(k_1,k_2,k_3) \nn \\
    & + \left[2 \, Z_1(\kv_1) \, Z_1(\kv_2) \, Z_2(\kv_1,\kv_2) \, P(k_1) \, P(k_2) \right. \nn \\
    & \left. + 2 \, \text{perms.}\right]\,,
\end{align}
where $B_0$ is the linear matter bispectrum sourced by a given non-zero primordial gravitational potential bispectrum $B_\phi$,
\be
B_0(k_1,k_2,k_3) = \cM(k_1) \, \cM(k_2) \, \cM(k_3) \, B_\phi(k_1,k_2,k_3) \, .
\ee
For local PNG, we have 
\be
B_\phi^{\rm loc}(k_1,k_2,k_3) = 2 \, f_{\rm NL}^{\rm loc} \, \left[P_\phi(k_1) \, P_\phi(k_2) + 2 \, {\rm perms.}\right]\,.
\ee
Retaining all linear-in-$\fnl$ terms in $Z_1$ and $Z_2$, we have terms up to cubic order in $\fnl$ in the bispectrum. Nevertheless, for the fiducial value of $\fnl=30$ in our simulations, the higher-order $\fnl$ terms are not relevant. The (stochastic) shot noise contribution to the bispectrum is given by 
\begin{align}
B_{\mathrm{shot}}^s(\kv_1,\kv_2,\kv_3) =&\; \frac{1}{\bar{n}} \, \Biggl\{\left[(1+{\tilde s}_1) \, b_1 + (1+{\tilde s}_3) \, f \, \mu_1^2\right] \, Z_1(\kv_1) \, P(k_1) \nn \\ 
& + 2 \, \text{perms.}\Biggr\}\, + \frac{1+{\tilde s}_2}{\bar{n}^2} \, ,
\end{align}   
where we used ${\tilde s}_i$ to denote the free amplitudes to distinguish them from shot parameters of the power spectrum $s_i$.

While the tree-level bispectrum does not strictly require counterterms, phenomenological corrections are often introduced to extend the perturbative reach of the model to smaller scales by accounting for the impact of small-scale velocities, known as the Fingers of God (FoG). The work of \citet{IvanovEtal2022B} introduces a FoG counterterm directly in the $Z_1$ kernel,
\begin{equation}
\label{Z1_FoG}
Z_1(\bk)\ \rightarrow\ Z_1^\mathrm{FoG}(\bk) = Z_1(\bk) - c_1^\mathrm{FoG}\,\mu^2\, k^2\,.
\end{equation}
In contrast, \citet{Eggemeier2025} adopt a non-perturbative modelling of redshift-space distortions via the velocity-difference generator (VDG). A low-$k$ expansion of the VDG motivates a bispectrum-level counterterm, while leaving the standard kernel $Z_1$ unchanged,
\begin{align}
B_{\mathrm{ctr}}^s\!\left(\bk_{1},\bk_{2},\bk_{3}\right)
=\; c_{\text{VDG}}^{B}\,f^{2}\,[k\mu]_{123}^{2}\,
B_{\mathrm{clust}}^s\left(\bk_{1},\bk_{2},\bk_{3}\right)\,,
\label{eq:VDG_counter}
\end{align}
where $[k\mu]_{123}^{2}\equiv k_1^2 \, \mu_1^2+k_2^2 \, \mu_2^2+k_3^2 \, \mu_3^2$. This should be interpreted as an effective angle- and scale-dependent low-$k$ FoG/nonlinear-RSD correction motivated by the VDG framework, rather than as the expansion of a phenomenological Gaussian damping ansatz. \citet{Eggemeier2025} showed that this counterterm extends the reach of the perturbative model non-negligibly compared to the case where the FoG correction is introduced through a modification of $Z_1$.

To determine the preferred form of the bispectrum counterterm, we performed scale-cut tests comparing the two parametrisations and found -- consistent with \citet{Eggemeier2025} -- that the VDG-motivated form extends the model reach. For example, in the lowest-redshift snapshot of our H$\alpha$ HOD mocks, the maximum wavenumber used in the monopole and quadrupole fits increases from $k_{\max}^{B_0,B_2}=0.10\,h\,\mathrm{Mpc}^{-1}$ to $k_{\max}^{B_0,B_2}=0.12\,h\,\mathrm{Mpc}^{-1}$ with the VDG counterterm, without introducing cosmological bias. For the $k$-binning we chose in this work ($\Delta k = k_{\rm f}$, with $k_{\rm f} = 2 \pi \, V^{-3}$ being the box fundamental wavelength and $V$ its volume), this higher $k_{\max}$ adds $\sim 2500$ triangles to the data vector, substantially improving the constraints on cosmological parameters. We therefore adopt the VDG counterterm, Eq.~\eqref{eq:VDG_counter}, in the analyses presented in Sect.~\ref{sec:results}.
  
In this work, we adopt the redshift-space bispectrum–multipole estimator introduced by \citet{Scoccimarro:1999ed}. Other bispectrum estimators introduced previously include \citet{Hashimoto:2017klo} and \citet{Sugiyama:2018yzo}, while several data compression schemes to reduce the dimensionality of the bispectrum were also introduced \citep[e.g.,][]{Schmittfull:2014tca,Byun:2017fkz, Gualdi:2017iey,Schmittfull:2020hoi,Philcox:2020zyp}. 

For the orientation of the wavevector and the line of sight, we choose
\begin{align}
\bk_1 &=(0,0,k_1) \, , \\
\bk_2 &= (0,k_2\sin \theta_{12}, k_2\cos \theta_{12}) \, , \\ 
\bf{n} &=(\sin \theta \cos \varphi, \sin \theta  \sin \varphi, \cos \theta) \, .
\end{align}
The bispectrum multipoles are thus defined via averaging over the azimuthal angle $\varphi$,
\be
B_\ell(k_1,k_2,k_3) = \frac{2\ell+1}{4\pi}\int_0^{2\pi} \diff \varphi \int_{-1}^1 \diff \mu \, {\mathcal L}_\ell(\mu) \,  B_{\mathrm{g}}^s\,,
\ee
where $B_{\mathrm{g}}^s = B_{\mathrm{g}}^s(k_1,k_2,k_3,\mu,\varphi)$.

\begin{table}
    \centering
    \caption{Summary of simulations used in this work.}
    \begin{tabular}{cccc} \toprule 
         Name & $N_\mathrm{part}$ & $N_\mathrm{real}$ & $f_\mathrm{NL}$ \\ \midrule
         AbacusSummit  $\Lambda$CDM & $6912^3$ & 2 & 0 \\
         AbacusSummit $\Lambda$CDM$+2\%$ $\sigma_8$ & $6912^3$ & 1 & 0 \\
         AbacusSummit  $\Lambda$CDM$-2\%$ $\sigma_8$ & $6912^3$ & 1 & 0 \\
         AbacusPNG $\Lambda$CDM & $4096^3$ & 2 & 0 \\ 
         AbacusPNG $f_\mathrm{NL}$ & $4096^3$ & 2 & 30 \\ \bottomrule
    \end{tabular}
    \label{tab:abacus_sims}
    \vspace{0.1in}
\end{table}
\begin{table}
    \centering
    \caption{Fiducial cosmological parameters.}
    \begin{tabular}{cccccccc} \toprule 
         $h$ & $\omega_\mathrm{c}$ & $\omega_\mathrm{b}$ & $A_{\mathrm{s}} \times 10^{9}$\tablefootmark{\rm(*)} & $n_{\mathrm{s}}$ & $M_\nu \,[\mathrm{eV}]$ \\ \midrule
         0.6736 & 0.12 & 0.02237 & 2.0830 & 0.9649 & 0.06 \\ \bottomrule
    \end{tabular}
    \tablefoot{The reference flat $\Lambda$CDM model used by the Abacus Simulations. \\
    \tablefoottext{\rm *}{Except the AbacusSummit  $\Lambda$CDM$\pm 2\% \ \sigma_8$ simulations, for which $A_{\mathrm{s}} \times 10^9 = 2.1672$ and $A_{\mathrm{s}} \times 10^9 = 2.0021$, respectively.}}
    \label{tab:abacus_params}
\end{table}

\section{Dataset and likelihood inference}\label{sec:data_like}

In this section, we introduce the new \Euclid-like H$\alpha$ synthetic mocks with PNG and summarise the core elements of our likelihood pipeline -- free parameters, priors, and performance metrics. Additional components are detailed in the appendices: measurements (Appendix~\ref{app:measurment}), binning effects (Appendix~\ref{app:binnig}), and analytic marginalisation (Appendix~\ref{app:analytic_marg}).

\subsection{Euclid-H$\alpha$ HOD mocks from Abacus PNG simulations}

We construct a set of \Euclid-like galaxy mocks from the Abacus simulations with local-type PNG \citep{Hadzhiyska:2024kmt}. The PNG suite comprises ten simulations spanning five $f_{\rm NL}$ values, $\{-100,-30,0,30,100\}$, with two realisations each.  Extending the AbacusSummit suite \citep{Garrison:2021lfa,Maksimova:2021ynf}, these runs evolve $4096^3$ particles in boxes of volume $V=2^3\,h^{-3}\,{\rm Gpc}^3$ at slightly reduced mass resolution, yielding a particle mass of $M_{\mathrm{p}}\sim 10^{10}\,h^{-1}\,M_\odot$.
In this work, we focus on $f_{\rm NL}=0$ (for the null tests) and $f_{\rm NL}=30$ (for the main analysis), using snapshots at $z\in\{0.8,1.1,1.4,1.7\}$.
To test for resolution effects, we also create similar mocks from the first two realisations of the $\Lambda$CDM AbacusSummit suite (see Table \ref{tab:abacus_sims}).
Finally, we use two more AbacusSummit simulations that share the same cosmology of $\Lambda$CDM AbacusSummit one, except for the value of $\sigma_8$, which has been increased by $2\%$ in one and decreased by the same amount in the other. These two realisations will be used as separate universe (SU) simulations to estimate the parameter $b_\phi$. We summarise in Table~\ref{tab:abacus_sims} the parameters (number of particles, number of realisations, and $\fnl$ values) of the simulations we have used in this work.
The values of cosmological parameters for this suite are given in Table~\ref{tab:abacus_params}.

\begin{figure}
    \centering
    \includegraphics[width=0.5\textwidth]{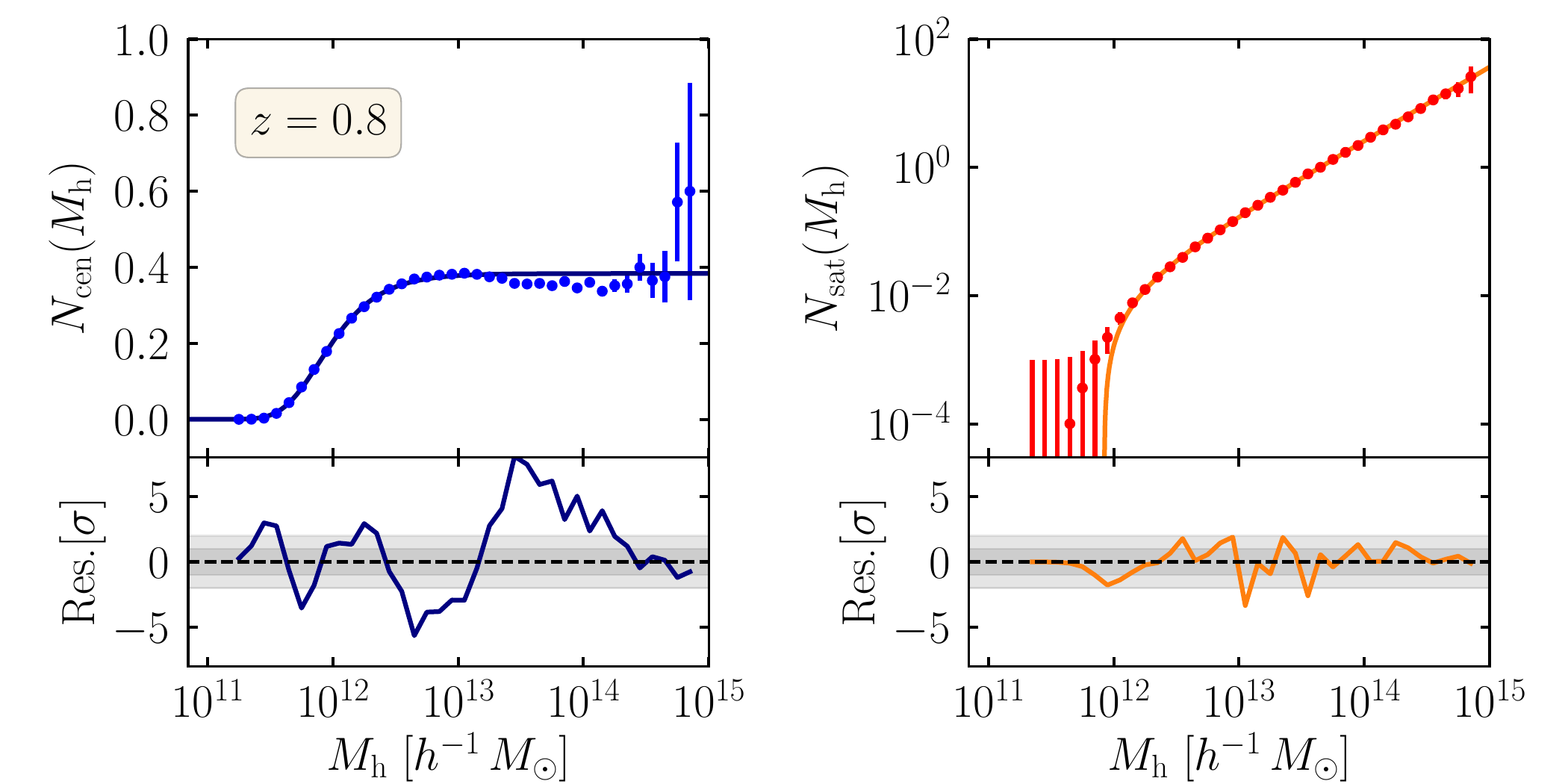}
    \caption{Measured HOD from the Euclid FS2 simulation and its fit at $z=0.8$. The upper panels show the measured central and satellite occupations, $N_{\rm cen}(M_{\rm h})$, and $N_{\rm sat}(M_{\rm h})$, respectively, with solid curves denoting the best-fitting HOD models. The lower panels show the corresponding residuals in units of the statistical uncertainty, $\sigma$.}
    \label{fig:HOD_z0p8}
\end{figure}

The mocks are built starting from H$\alpha$ galaxies selected from the Euclid Flagship 2 simulation \citep[FS2;][]{Euclid:2024few}, which will represent our bona fide model and of which we want to reproduce the clustering amplitude, in the $\Lambda$CDM case.\footnote{We use CosmoHub to retrieve data from the Euclid FS2 mocks.} 
To do so, we first select all FS2 galaxies with an H$\alpha$ flux larger than $f_{\mathrm{H}\alpha, \, \rm min}=2\times10^{-16} \, \rm{erg \, s^{-1} \, cm^{-2}}$.
Since the FS2 simulation has a lightcone geometry, we then cut small redshift slices with size $\Delta z=0.2$, centred on the redshifts of the Abacus-PNG snapshots we are going to use. Thanks to the information available in FS2, we can measure the HOD of these selected galaxies in these redshift slices by directly counting the number of central and satellite galaxies per halo of a given mass bin.
We use the viral mass, $M_\mathrm{vir}$, definition of halo mass in FS2 as the mass variable. We separately fit the obtained curves as a function of halo mass, $M_\mathrm{h}$, with an asymmetric sigmoid function and a power law for central and satellite galaxies,
\begin{eqnarray}
    N_\mathrm{cen}(M_\mathrm{h}) &=& \frac{f_\mathrm{max}}{\left\{1+\left[\mathcal C(a,b) \, \frac{\logten \!\left(M_\mathrm{min}/(h^{-1}M_\odot)\right)}{\logten \!\left(M_\mathrm{h}/(h^{-1}M_\odot)\right)}  \right]^{1/a}\right\}^{1/b}} \, ,
    \\
    N_\mathrm{sat}(M_\mathrm{h}) &=& \left(\frac{M_\mathrm{h}-M_\mathrm{cut}}{M_1}\right)^\alpha\, ,
    \label{eq:asymsig_hod}
\end{eqnarray}
where $f_\mathrm{max}$, $\logten M_\mathrm{min}$, $a$, $b$, $M_\mathrm{cut}$, $M_1$, $\alpha$ are the parameters of the model. The function $\mathcal C(a,b) = \left[ (ab+b) \, (1-ab)^{-1}\right]^{\, a}$ is a numerical factor that ensures that $\logten \!\left(M_\mathrm{min}/(h^{-1}M_\odot)\right)$ is the mass where the mean occupation number of central galaxies is equal to $f_\mathrm{max}/2$.
In Fig.~\ref{fig:HOD_z0p8} we show an example of the fit performed on the FS2 HOD at $z=0.8$.
The left panel shows the fit for central galaxies and the right panel for satellite galaxies.

\begin{figure}[t!]
    \centering
    \includegraphics[width=0.9\columnwidth]{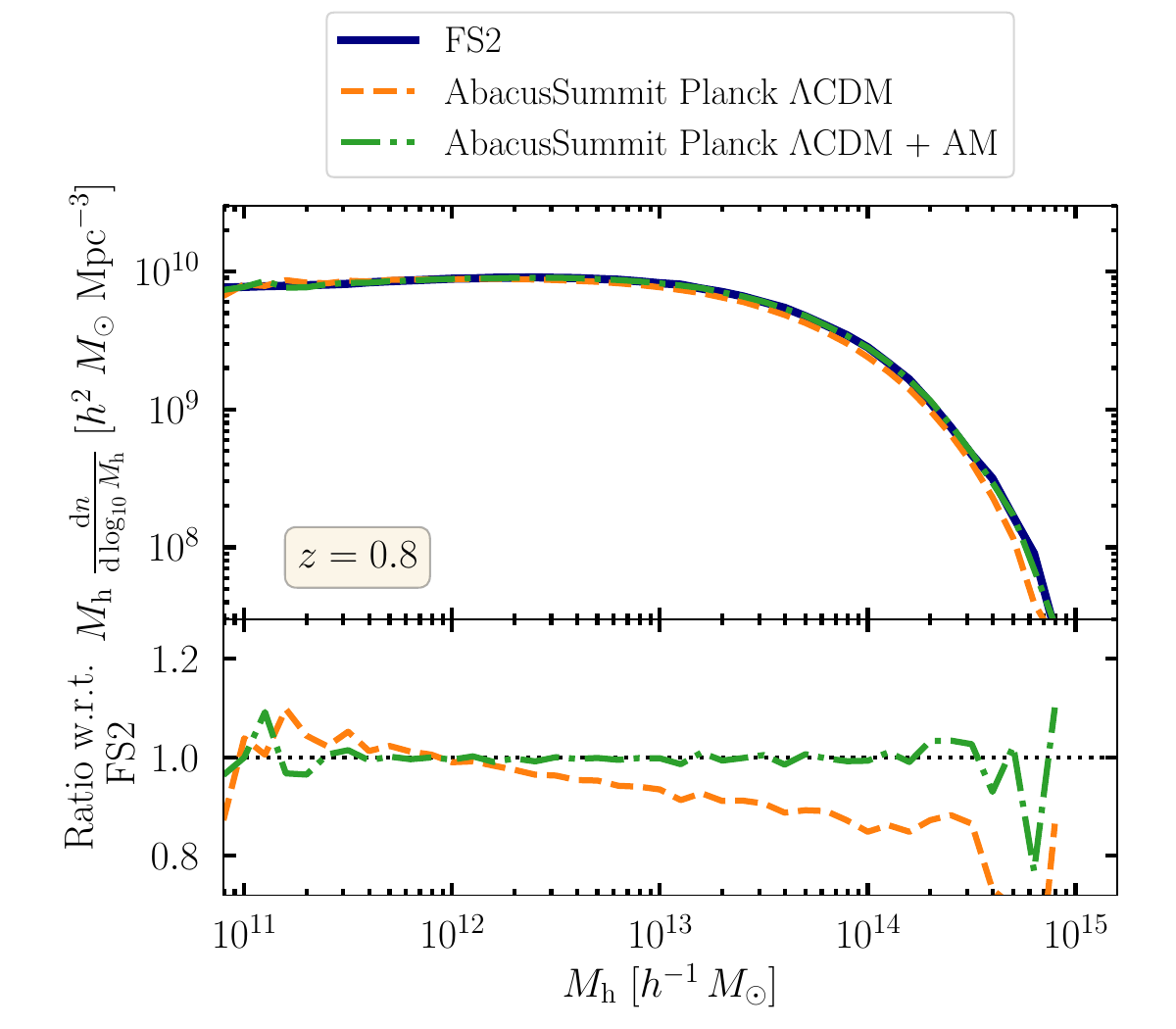}
    \caption{Abundance matching applied to the AbacusSummit halo mass function at $z=0.8$. The top panel shows the FS2 halo mass function (solid dark blue) and the AbacusSummit $\Lambda$CDM mass function (dashed orange). The bottom panel presents their ratio relative to FS2, illustrating differences of up to $\sim 20 \%$. Applying AM (dot-dashed green) brings the AbacusSummit mass function into excellent agreement with FS2.}
    \label{fig:AM_z0p8}
\end{figure}

Before using the obtained HOD to populate Abacus halos, we need to perform an intermediate step. For the two simulation suites, different halo finders were employed to identify density peaks. In fact, whereas the Rockstar halo finder was employed for the FS2 simulation \citep{Behroozi:2011ju}, a newly developed spherical overdensity, purpose-built halo finder was utilised for the Abacus suite \citep{Hadzhiyska:2021zbd}. The different nature of the two halo finders introduces relevant discrepancies in the estimate of the halo masses, especially at the high-mass end. The resulting halo mass functions can differ up to $\sim 20 \%$ for masses larger than $5\times 10^{14} \, h^{-1} \, M_\odot$. This, in turn, has a strong impact on the large-scale clustering and on the subsequent measurement of galaxy bias. To mitigate this inconsistency, we apply an abundance matching (AM) to the halo masses as measured in the $\Lambda$CDM AbacusSummit simulations. This map ensures that the FS2 and the $\Lambda$CDM AbacusSummit halo mass function match (within the statistical errors). We apply the same AM function to all the AbacusSummit and Abacus-PNG simulations in order to keep the relative differences between the $\Lambda$CDM and PNG models the same.
We show the effect of applying AM to the $\Lambda$CDM AbacusSummit simulation in Fig.~\ref{fig:AM_z0p8}.
In the top panel, we show the mass functions measured in FS2 and $\Lambda$CDM AbacusSummit with solid dark blue and dashed orange lines, respectively. Their ratio is plotted in the bottom panel.
Once we apply the AM to AbacusSummit, the two resulting mass functions match on the whole mass range.

The implementation of the AM, as outlined in this work, plays a critical role in PNG analyses, as the measurement of the $b_\phi$ parameter is inherently determined by the halo number density once a minimum mass is specified.
Different halo mass definitions will give rise to different number densities and therefore inconsistent estimates of $b_\phi$.

Once the AM operation is concluded, we make use of the \texttt{SciPIC-lite} code (Parimbelli et al. in prep.) to populate Abacus halos. This is a lighter, more galaxy clustering-oriented version of the \texttt{SciPIC} code \citep{Scipic} that was originally used to create the FS2 galaxy mock. We take the Abacus halo catalogues of the simulations described above and assign galaxies to halos according to the best-fit parameters of Eq.~\eqref{eq:asymsig_hod}, assuming Bernoulli and Poisson distributions for central and satellite galaxies, respectively. We use the same HOD parameters for all the simulations we use in this work. Central galaxies are assumed to sit in halo centres and to follow their dynamics; satellite galaxies are assumed to follow a Navarro--Frenk--White profile \citep{Navarro:1996gj} around halo centres. We do not consider assembly bias (i.e., galaxy statistics only depend on halo mass) and velocity bias. While previous work has shown that $b_\phi$ can depend sensitively on assembly bias and on tracer selection, this does not compromise the internal consistency of the present validation, since the same mass-only HOD is used to construct both the Gaussian and PNG mock suites. In applications to real data, however, assembly bias could affect the relation between $b_\phi$, $b_1$, and tracer selection, and hence the interpretation of $f_{\rm NL}$ constraints or the choice of priors on $b_\phi$.

\subsection{Likelihood analysis}

Considering the data vector consisting of the three lowest multipoles of the power spectrum and bispectrum, $\boldsymbol{\mathcal{P}} = \{\P_0,\P_2,\P_4,\B_0,\B_2\}$, and assuming the summary statistics to be Gaussian distributed, the likelihood function is given by
\be
\ln{\mathcal L} = -\frac{1}{2} \left( \hat{\boldsymbol{\mathcal{P}}} - \boldsymbol{\mathcal{P}}^{\rm th}\right)^{\rm T} \, \tens{C}^{-1} \, \left(\hat{\boldsymbol{\mathcal{P}}} - \boldsymbol{\mathcal{P}}^{\rm th}\right) -  \frac{n_{\mathcal{P}}}{2} \, \ln\left[ 2 \, \pi \, \det \left(\tens{C}\right) \right]\, ,
\ee
where $\hat{\boldsymbol{\mathcal{P}}}$ is the measured data vector ($\boldsymbol{\mathcal{P}}$ averaged over $N_{\mathrm{R}}=2$ realisations), $\boldsymbol{\mathcal{P}}^{\rm th}$ is the theoretical prediction, and $n_{\mathcal{P}}$ is their dimension. In our implementation, we neglect the second term on the right-hand side of the equation because, being a constant term, it does not affect the location of the likelihood stationary points. The covariance $\tens{C}$ is block-structured and includes auto- and cross-covariances of the multipoles. We include only the Gaussian (disconnected) contributions; hence, we neglect mode-coupling between different $k$-shells and between distinct triangle bins \citep{Sefusatti:2006pa}, and we also neglect the cross-covariance between power-spectrum and bispectrum multipoles. We will also neglect $B_4$ as it is noise-dominated and consistent with zero in our simulations.

In the weakly nonlinear regime, non-Gaussian contributions to the power-spectrum variance are subdominant to the Gaussian term; for realistic surveys, the non-Gaussian clustering contributions are further outweighed by shot noise and super-sample covariance \citep{Wadekar:2019rdu}. For the bispectrum, however, this is not the case: adopting a Gaussian covariance typically overestimates the information content \citep[][see also \citealp{Pardede2026Euclid}]{Chan:2016ehg,Byun:2017fkz,Yankelevich:2018uaz,Floss:2022wkq,Novell-Masot:2023gmj}. In the squeezed configurations, most relevant for local PNG, the non-Gaussian variance exceeds the Gaussian contribution, and the cross-covariance with the power spectrum is substantial \citep{Barreira:2020ekm,Biagetti:2021tua,Floss:2022wkq,Salvalaggio:2024vmx}. Given the high computational cost of estimating the bispectrum covariance from large ensembles of mocks for surveys like \Euclid and DESI, accurate approximate modelling of the covariance -- especially the squeezed-limit clustering piece, alongside shot-noise and super-sample terms -- is crucial for robust $\fnl$ constraints. In this work, we adopt the commonly used Gaussian covariance and leave a full redshift-space treatment of the non-Gaussian covariance to future work. The explicit forms of Gaussian covariance matrices of power spectrum and bispectrum multipoles can be found in Appendix~\ref{app:cov}.

\subsection{Free parameters and priors}\label{sec:prior}

We list our prior choices in Table~\ref{tab:Priors}. For the cosmological parameters $\{A_{\mathrm{s}},h,\omega_{\mathrm{c}},f_{\rm NL}\}$ and non-PNG bias parameters $\{b_1,b_2,b_{\mathcal G_2},b_{\Gamma_3}\}$ we adopt broad, uniform priors. For other nuisance parameters (counterterms and shot-noise corrections), we use wide Gaussian priors and marginalise over them analytically. For the PNG biases, whose amplitudes are strongly degenerate with $f_{\rm NL}$, the choice of parametrisation and priors is critical to achieving unbiased, high-significance constraints on $\fnl$ \citep{Barreira:2020ekm, Barreira:2022sey}. We therefore perform targeted validation tests (Sects.~\ref{sec:fnlbias} and \ref{sec:prior_choice}) to select a robust parametrisation and prior widths. To place our choices in context, we review below the theoretical predictions, tests on simulations, and the commonly adopted parametrisation of PNG biases in the analysis of the data. 

Under the universal halo mass function (UHMF) assumption, peak-background split (PBS) arguments imply that, for local-type PNG, the leading PNG bias is \citep{Matarrese:2008nc,Dalal:2007cu,Slosar:2008hx}
\begin{align}
    b_\phi^{\rm UHMF} = 2\,\delta_{\mathrm{c}}\,(b_1-1)\, ,
    \label{eq:UHMF_bphi}
\end{align}
with $\delta_{\mathrm{c}}=1.686$, and the next-order PNG bias in Eulerian variables is \citep{Giannantonio:2009ak,Baldauf:2010vn}
\begin{align}
    b_{\phi\delta}^{\rm UHMF} = b_\phi - b_1 + 1 + \delta_{\mathrm{c}}\,\left[b_2 - \frac{8}{21}\,(b_1-1)\right].
    \label{eq:UHMF_bphidelta}
\end{align}
These relations are only approximate, even for halos. Subsequent theoretical work generalised and refined them in several directions: for example, to include local-type PNG corrections at arbitrary orders and related consistency relations \citep{Nishimichi:2012da}; a missing, mass-dependent term in the PNG bias amplitude -- particularly relevant for non-local shapes \citep{Desjacques:2011mq}; treatments that incorporate non-Markovian effects \citep{Maggiore:2009rx} and departures from mass-function universality, and peak/excursion-set models with moving or stochastic collapse barriers \citep{Biagetti:2015exa}.

The $b_\phi(b_1)$ relation from UHMF has been extensively tested in $N$-body simulations, which collectively favour a rescaled mapping $b_\phi \simeq 2\,r\,\delta_{\mathrm{c}}(b_1-1)$, with $r$ varying with mass, redshift, and halo finder \citep{Grossi:2009an,Desjacques:2008vf,Pillepich:2008ka,Biagetti:2016ywx}. Moreover, $N$-body studies show a clear dependence of $b_\phi$ on properties beyond halo mass, which are the assemble bias: \citet{Reid:2010vc} found that recently formed halos have smaller $b_\phi$; \citet{Lazeyras:2022koc} reported based on SU simulations a strong concentration trend (high-concentration halos having larger $b_\phi$) and that concentration splits can drive $b_1$ and $b_\phi$ in opposite directions, with weaker trends for spin and sphericity; \citet{Hadzhiyska:2024kmt} confirmed the concentration trend, showed $b_\phi$ increases toward lower recent mass-accretion rate, and found a weaker dependence on tidal shear. A proposed explanation for (non-)universality is that high-mass halos, which dominate their local gravity, follow UHMF, whereas low-mass halos deviate due to growth influenced by nearby massive neighbours \citep{Shiveshwarkar:2025nac}.

Moving from halos to galaxies and quasars, a common parametrisation keeps the PBS form but replaces the universal intercept with a tracer-dependent one,
$b_\phi = 2\,\delta_{\mathrm{c}}\,(b_1 - p_{b_\phi})$. This was motivated by \citet{Slosar:2008hx} using extended Press--Schechter arguments (calibrated to merger-rate scalings) for tracers dominated by recent mergers (quasar hosts), implying a suppression of $b_\phi$ relative to UHMF and suggesting $p_{b_\phi}\sim1$--$1.6$ (with $p=1.6$ adopted for quasars). Recent quasar analyses (i.e., eBOSS, DESI) therefore quote $f_{\rm NL}$ constraints for $p_{b_\phi}\in\{1,1.6\}$ \citep{eBOSS:2021jbt,Cagliari:2023mkq,Cagliari:2025rqe,Chaussidon:2024qni}. For galaxies, hydrodynamical SU runs \citep[as the IllustrisTNG,][]{Nelson:2018uso} favour $p_{b_\phi}\sim0.55$ for stellar-mass selected galaxies \citep{Barreira:2020kvh}. Semi-analytic modelling with \textsc{GALACTICUS} \citep{Benson:2010kx} further shows strong selection dependence: colour-selected (older/redder) galaxies exhibit large non-Gaussian assembly bias consistent with the findings of \citet{Barreira:2020kvh}, whereas H$\alpha$-selected emission-line galaxies (relevant for \Euclid) show little to no detectable $b_\phi$ assembly bias over $10^{10} \,h^{-1}\,M_\odot \lesssim M \lesssim 10^{12} \,h^{-1}\,M_\odot$, consistent with their sensitivity to instantaneous stellar formation rate rather than formation history \citep{Marinucci:2023jag}. We will refer to the value of $p_{b_\phi} = 0.55$ in the subsequent sections as the galaxy formation model (GFM). Several recent works pursue complementary ways to obtain physically motivated constraints on $b_\phi$ -- and thereby tighten $f_{\rm NL}$ from the power spectrum -- including joint inference of non-Gaussian assembly bias \citep{Fondi:2023egm}, machine learning-based prediction of $b_\phi$ for survey selections \citep{Lucie-Smith:2023hil,Sullivan:2023qjr}, multi-tracer/self-calibration strategies \citep{Barreira:2023rxn,Sullivan:2023qjr}, field-level measurements of $b_\phi$ and $b_{\phi\delta}$ \citep{Sullivan:2024jxe}, and time-evolution (abundance-response) calibrations \citep{Sullivan:2025fie,Dalal:2025eve}.

As noted above, despite substantial progress, a broadly applicable model for $b_\phi$ across galaxy selections remains unsettled. To retain the theoretically motivated scaling of $b_\phi$ with $b_1$ while allowing tracer-dependent departures, we adopt $b_\phi \equiv 2\,\delta_{\mathrm{c}}\big(b_1 - p_{b_\phi}\big)$. As will be discussed in Sect.~\ref{sec:prior_choice}, we first test how the $f_{\rm NL}$ constraints (from the power spectrum, the bispectrum, and their combination) depend on the prior for $b_\phi$. We compare posteriors obtained with a wide flat prior (allowing both positive and negative $b_\phi$) to those from informative Gaussian priors whose means and widths we vary across the four \Euclid\/'s redshift bins. In agreement with \citet{Barreira:2022sey}, we find that a wide flat prior drives biased $f_{\rm NL}$ constraints toward zero.\footnote{In our earlier real-space analysis on \textsc{Eos Dataset} \citep{MoradinezhadDizgah:2020whw}, unbiased constraints for $f_{\rm NL} \in \{250,10\}$ were obtained using a wide flat prior with $b_\phi>0$. In redshift space, however, even enforcing $b_\phi>0$ leads to biased $f_{\rm NL}$ posteriors.} For our fiducial analysis, we therefore place a Gaussian prior on $p_{b_\phi}$ whose mean and width encompass $p_{b_\phi}=0.5$ and $1$ at the $1\sigma$ level. This parametrisation captures the leading redshift dependence of $b_\phi$ largely through that of $b_1$, allowing us to use the same prior across redshift bins and yielding an approximately redshift-independent best-fit value of $p_{b_\phi}$. 

For $b_{\phi\delta}$, our sensitivity tests show that contributions proportional to $f_{\rm NL}\,b_{\phi\delta}$ are loop-suppressed in the power spectrum and negligible for the bispectrum within our scale cuts. We therefore fix $b_{\phi\delta}$ to the UHMF expression in Eq.~\eqref{eq:UHMF_bphidelta} from now on.

\subsection{Performance metrics}

As in \citet{Pardede2026Euclid}, we use two measures to assess the validity of the theoretical models of the power spectrum and bispectrum, as well as the information content of the two statistics as a function of scale cuts. The systematic errors in the inferred parameters arising from inaccurate modelling in terms of the figure of bias (FoB),
\begin{equation}
    \mathrm{FoB}(\boldsymbol{\theta}) \equiv \left[\Big(\langle \boldsymbol{\theta} \rangle - \boldsymbol{\theta}_\mathrm{fid} \Big)^{\rm T} \, \tens{S}^{-1} (\boldsymbol{\theta}) \, \Big(\langle \boldsymbol{\theta} \rangle - \boldsymbol{\theta}_\mathrm{fid} \Big) \right]^{1/2}\,,
    \label{eq:FoB}
\end{equation}
where $\langle \boldsymbol{\theta} \rangle$ represents the posterior average of the parameters, $\boldsymbol{\theta}_\mathrm{fid}$ refers to their fiducial values, and $\tens{S}(\boldsymbol{\theta})$ denotes their covariance matrix. The information content of the two observables is quantified in terms of the figure of merit (FoM) derived from the parameter covariance matrix \citep{Wang:2008PhRvD},
\begin{equation}
  \mathrm{FoM}(\boldsymbol{\theta}) \equiv \left[ \mathrm{det} \, \Big(\tens{S}(\boldsymbol{\theta})\Big)\right]^{-1/2}\,.
  \label{eq:FoM}
\end{equation}

We evaluate in our study both the FoB and FoM for the combination of the three $\Lambda$CDM cosmological parameters and $\fnl$ by defining $\boldsymbol{\theta} = \{h,\,\omega_{\mathrm{c}},\,A_{\mathrm{s}}, \fnl\}$ in the previously mentioned equations. In addition to these two cumulative measures, we select the scale cuts such that we retrieve unbiased constraints on individual parameters within the 1$\sigma$ level. 

\section{Numerical implementation: \texttt{CosmoSIS-gClust}}\label{sec:code}

The results presented in this work are obtained using our new software package, \texttt{CosmoSIS-gClust}. A companion paper (Linde \& Moradinezhad in prep.) will describe the code in detail; here we summarise its main features.  

\texttt{CosmoSIS-gClust} extends the \texttt{CosmoSIS} package \citep{Zuntz_2015} to enable cosmological inference from $3$-dimensional power spectrum and bispectrum of biased tracers, in real and redshift space, for both Gaussian and non-Gaussian initial conditions. The current version models local-type PNG (non-zero primordial bispectrum); the extension to other PNG shapes is straightforward and in progress. At present, the code analyses simulated halo/galaxy catalogues in periodic boxes; support for observational (survey) data is under development. An earlier, real-space-only version without fast-theory computation was used in \citet{MoradinezhadDizgah:2020whw}. The code will be publicly available with the forthcoming manuscript (Linde \& Moradinezhad in prep.) and is planned for integration into a future public \texttt{CosmoSIS} release. For Gaussian initial conditions, \texttt{CosmoSIS-gClust} has been extensively validated against other public tools for Fourier-space modelling and parameter inference on $N$-body simulations (see \citealt{Euclid:2023tog}; \citealp{Pardede2026Euclid}, for comparisons against Euclid Flagship 1 simulations). \texttt{CosmoSIS-gClust} also underpins the new \texttt{CLASS-OneLoop} \citep{Linde:2024uzr} extension to \texttt{CLASS} \citep{Lesgourgues:2011re}. For local PNG, further validation of the Abacus halo catalogues \citep{Hadzhiyska:2024kmt} is presented in the companion paper (Linde \& Moradinezhad in prep.).

The package consists of two libraries: (i) \texttt{gclust}, which provides the theoretical predictions; and (ii) \texttt{gstats}, which computes the corresponding likelihoods.  Within \texttt{gclust}, the galaxy power spectrum and bispectrum are computed at one-loop and tree level, respectively, with linear matter power spectrum supplied either from Boltzmann solvers (\texttt{CLASS}, \citealp{Lesgourgues:2011re}; \texttt{CAMB}, \citealp{Lewis:1999bs}) or from emulators \citep[e.g., \texttt{BACCO},][]{Angulo:2020vky}. In this analysis, we employ the \texttt{BACCO} linear emulator and have verified the robustness of our results against \texttt{CLASS}. The accuracy of the FFTLog implementation for individual loop integrals is validated against direct integration, which is available as an option for the user. Covariance matrices are computed on the fly for the chosen binning of the data vector, taking into account the specific scale cuts and $k$-binning selected by the user. Parameter inference is run within \texttt{CosmoSIS} using any of its samplers. The results shown here use \texttt{pocoMC} \citep{Karamanis:2022pocoMC}, which we found to converge the fastest and most robustly for the extended and highly degenerate parameter space in our analysis. The code enables the user to analytically marginalise individual parameters entering linearly in the model and reconstruct their posteriors.

\section{Results}\label{sec:results}

In this section, we present validation tests and the main parameter constraints from our redshift-space power-spectrum and bispectrum analyses. We note that we have explicitly verified, in representative cases, that analytical marginalisation of parameters appearing linearly in the power-spectrum and bispectrum models yields contours consistent with those obtained from direct sampling of all parameters, by comparing the sampled posteriors with the reconstructed posteriors of the analytically marginalised parameters.

We begin with the lowest and highest redshift snapshots and perform targeted analyses to map the parameter space and its degeneracies -- especially among PNG parameters. We only show the results at $z=1.7$, which is the most constraining redshift. Similar conclusions are made at the lower redshift bin. Our goals are: (i) to test whether a non-zero PNG contribution can be detected without informative priors on PNG biases; (ii) to identify prior choices for the PNG biases that yield robust constraints on $\fnl$; and (iii) to validate these choices via a null test on simulations with Gaussian initial conditions and confirm that the analysis pipeline and assumptions do not lead to spurious detection of $\fnl$. Throughout, we analyse redshift-space multipoles of the power spectrum and bispectrum, both separately and jointly, to assess their complementarity. We discard the bispectrum hexadecapole since the measurement is noise-dominated and consistent with zero in our mocks.

Guided by the tests presented in the first three sections, below, we analyse all four snapshots ($z\in\{0.8,1.1,1.4,1.7\}$) of the Abacus-H$\alpha$ galaxy mocks with $\fnl=30$ local-type PNG. We first perform a scale-cut study for the power spectrum and bispectrum separately to determine $k_{\rm max}$ values that yield unbiased constraints on $\{A_{\mathrm{s}},h,\omega_{\mathrm{c}},\fnl\}$. Given the strong large-scale imprint of local $\fnl$, we also study the dependence of the power spectrum and bispectrum constraints on the large-scale cut-off $k_{\rm min}$. With these scale cuts fixed, we carry out combined analyses of the power spectrum and bispectrum multipoles to quantify the gain of the joint analysis with respect to the power spectrum and the bispectrum alone. Finally, by progressively adding multipoles, we quantify the information content of the components of the joint power spectrum-bispectrum data vector.

\subsection{Prior-agnostic detection of local PNG }\label{sec:fnlbias}

As discussed in Sect.~\ref{sec:prior}, on large scales the power spectrum of biased tracers encodes the scale-dependent signal $\propto \fnl\,b_\phi$, so $P$ alone cannot determine $\fnl$ without prior information on $b_\phi$; loop terms that depend on $\fnl$ without $b_\phi$ are too small for $\fnl=\mathcal{O}(1)$ to break this degeneracy in practice \citep{Barreira:2020ekm,MoradinezhadDizgah:2020whw,Barreira:2022sey}. At tree level and linear order in $\fnl$, the bispectrum contains three types of PNG contributions proportional to $\{\fnl, \fnl \, b_\phi, \fnl \, b_{\phi\delta}\}$, the pure-$\fnl$ piece matter/velocity bispectra and the two PNG-bias pieces. In the squeezed configurations that carry most of the local-PNG signal, the $\fnl \, b_\phi$ term typically dominates, so a prior-free $\fnl$ constraint from $B$ is expected to be challenging. 

In real space, \citet{MoradinezhadDizgah:2020whw} showed that unbiased $\fnl$ constraints from $B$ and joint power spectrum and bispectrum are possible with relatively broad priors on $b_\phi$ (positivity enforced) and $b_{\phi\delta}$, though for $\fnl=10$ wider priors degrade the constraint considerably; here, we revisit the question of detectability of PNG in redshift-space galaxy multipoles, where velocities modify both amplitudes and the degeneracy structure. We analyse the $\fnl=30$ simulations at $z=1.7$, sampling the three combinations $\{\fnl, \fnl \, b_\phi, \fnl \, b_{\phi\delta}\}$ directly, alongside cosmology and all non-PNG nuisances, using very broad flat priors ($\fnl: {\mathcal U}[-1000, 1000]$, $\fnl \, b_\phi:{\mathcal U}[-500, 500]$, and $\fnl \, b_{\phi\delta}: {\mathcal U}[-5000, 5000]$). We adopt the scale cuts of $k_{\rm max}^{P_0,P_2,P_4}=\{0.3,0.3,0.25\}\,h\,\mathrm{Mpc}^{-1}$ and $k_{\rm max}^{B_0,B_2}=\{0.1,0.1\}\,h\,\mathrm{Mpc}^{-1}$, which passed standard goodness-of-fit checks and recover inputs in our initial tests on the mocks. 

\begin{figure}[t!]
    \centering
    \includegraphics[width=\linewidth]{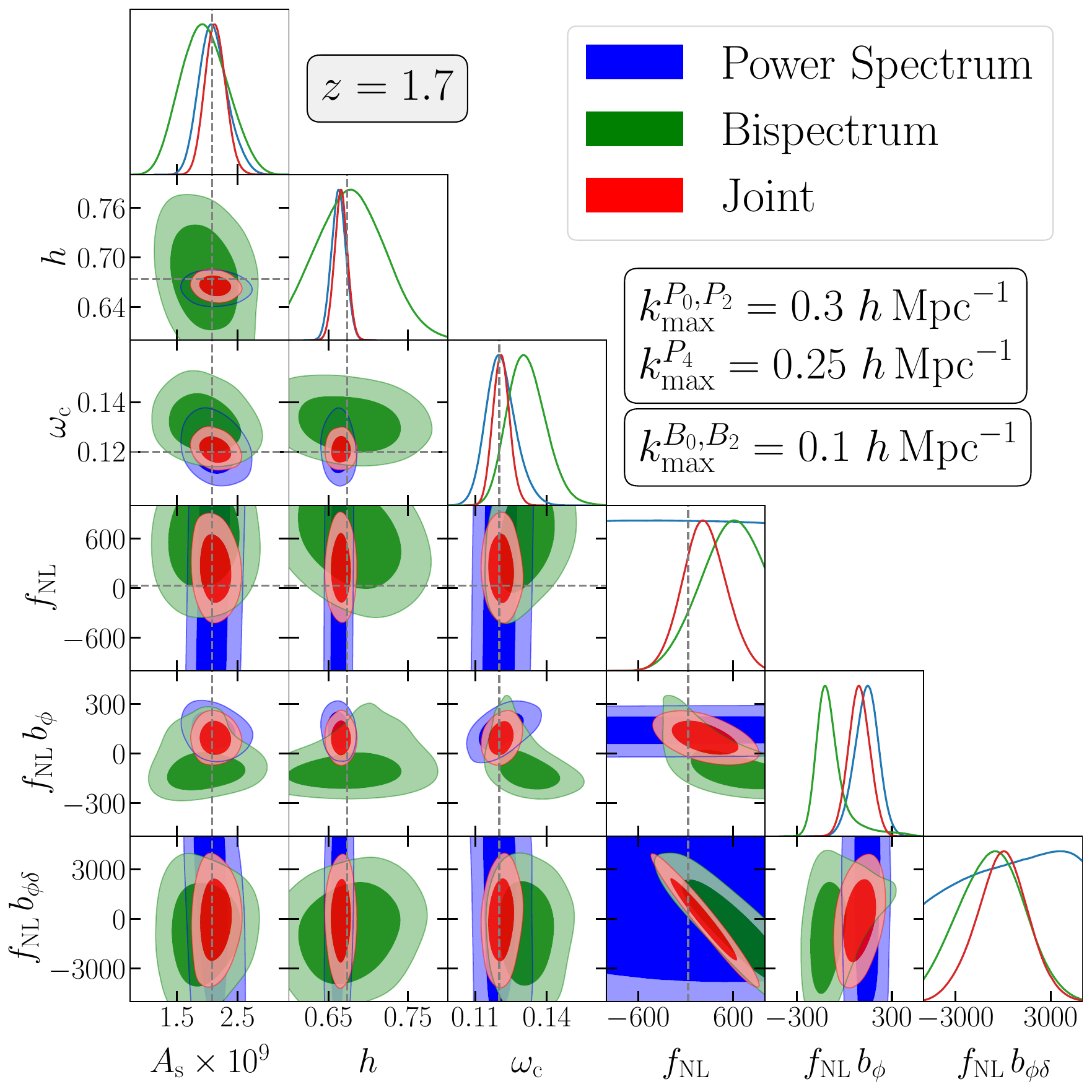}
    \caption{Posterior of the prior-agnostic analysis. Marginalised posterior distributions from the power spectrum (blue), bispectrum (green), and their combination (red) at the stated redshift and scale cuts. Shaded regions denote 68\% and 95\% credible intervals. Nuisance parameters are marginalised. Dashed lines mark the simulation fiducial values.}
    \label{fig:prior_agnostic_z1p7}
\end{figure}

Figure~\ref{fig:prior_agnostic_z1p7} shows the marginalised posteriors for $\Lambda$CDM and PNG parameters from the power spectrum (blue), the bispectrum (green), and their combination (red). For $\Lambda$CDM parameters, $B$-only yields weaker constraints overall: for $h$, the joint power spectrum-bispectrum result is effectively set by $P$; for $A_{\mathrm{s}}$ and particularly $\omega_{\mathrm{c}}$, the joint analysis delivers a noticeable gain. We also find a clear correlation between $\omega_{\mathrm{c}}$ and $f_{\rm NL} \, b_\phi$; because the degeneracy tilt differs in $P$ versus $B$, combining them reduces this coupling and tightens $\omega_{\mathrm{c}}$. In the PNG sector, as expected, $P$ effectively constrains only the product $f_{\rm NL}\,b_\phi$, leaving $f_{\rm NL}$ and $f_{\rm NL}\,b_{\phi\delta}$ essentially unconstrained. The $B$-only analysis weakly constrains all three combinations and reveal clear anti-correlations in the $(f_{\rm NL},f_{\rm NL}\,b_\phi)$ and $(f_{\rm NL},f_{\rm NL}\,b_{\phi\delta})$ planes, with the former tilting the posterior toward negative $f_{\rm NL}\,b_\phi$ and correspondingly positive $f_{\rm NL}$. The joint power spectrum-bispectrum analysis contracts the allowed volume in $\{f_{\rm NL},f_{\rm NL}\,b_\phi,f_{\rm NL}\,b_{\phi\delta}\}$ without shifting best fits by more than $1\sigma$, but it does not significantly tighten $f_{\rm NL}\,b_\phi$ beyond $P$-only at these cuts. The $f_{\rm NL}\,b_{\phi\delta}$ direction remains largely prior-dominated, supporting our choice to fix $b_{\phi\delta}$ in the main analysis.

The qualitative behaviour is similar at other redshifts, though the relative sizes of the PNG contributions -- and their ratio to the Gaussian contribution -- vary with $z$. In particular, the detection significance of $f_{\rm NL}\,b_\phi$ decreases toward lower $z$, consistent with smaller $b_\phi$ and a larger relative impact of the non-primordial (gravitational) bispectrum; see Appendix~\ref{app:add_joint} for a comparison of relative sizes of bispectrum components. At $z=1.7$, with an effective volume of $V=16\,h^{-3} \, \mathrm{Gpc}^3$, and using the largest-scale mode of half of this volume, we obtain an about $0.8\sigma$ indication of non-zero $f_{\rm NL}$ in $B$-only and a $1.9\sigma$ detection of a non-zero $f_{\rm NL}\,b_\phi$ in $P$-only. Lower-$z$ bins yield weaker constraints, where $f_{\rm NL}\,b_\phi$, extracted from $P$-only chains, can be differentiated from zero at significance levels of $\{1.3\sigma, 1\sigma, 0.6\sigma\}$ for redshifts of $z\in\{1.4,1.1,0.8\}$, respectively. 

From the above analysis, we conclude that for $\fnl=\mathcal{O}(1)$ and an effective volume $V_{\rm eff}=16\,h^{-3}\,{\rm Gpc}^{3}$ which is the average of two independent $8\,h^{-3}\,{\rm Gpc}^{3}$ realisations, comparable to the volume of the corresponding \Euclid redshift bin \citep[see][]{blanchard2020}, a prior-agnostic, high-significance detection of local-type PNG from $P$, $B$, or joint power spectrum-bispectrum is challenging. We note that while the statistical errors reflect $V_{\rm eff}$, the largest-scale mode used in our analysis is set by the fundamental mode of a single box; including larger-scale modes in the actual survey geometry could, in principle, increase the detection significance of $\fnl \, b_\phi$. Moreover, in survey analyses combining multiple redshift bins, $\fnl$ itself can be jointly constrained across bins (even though $\fnl \, b_\phi$ and $\fnl \, b_{\phi\delta}$ remain bin-specific due to redshift-dependent biases), which can further improve sensitivity to the pure-$\fnl$ component. We leave further investigation of detectability in a realistic survey setting to future work. 

Our finding that adding the bispectrum yields little additional detection significance for $f_{\rm NL}\,b_\phi$ beyond the power spectrum broadly agrees with the real-space forecasts of \citet{Barreira:2020ekm}. In redshift space, however, the bispectrum still carries useful information: the joint power spectrum-bispectrum analysis delivers weak but non-negligible constraints on the pure-$f_{\rm NL}$ and $f_{\rm NL}\,b_{\phi\delta}$ combinations, and the bispectrum quadrupole can be particularly impactful -- at some redshifts the PNG contribution to $B_2$ exceeds the Gaussian piece on the largest scales even when the monopole does not (see Sect.~\ref{subsec:Joint}). Hence, conclusions drawn in real space should be treated with caution when extrapolated to redshift space.

We also note that \citet{Barreira:2020ekm} assumed a volume of $V_s=100\,h^{-3}\,{\rm Gpc}^3$ at $z=1$ and reported a $1.1\sigma$--$1.25\sigma$ detection from real-space $P$-only for two samples with different biases. By contrast, our redshift-space $P$-only analysis achieves a higher significance in a substantially smaller volume, which likely owes to (i) the additional constraining power from velocities in redshift space and (ii) our inclusion of higher-redshift snapshots where $b_\phi$ is larger. Consistently, our lower-redshift results are closer to those of \citet{Barreira:2020ekm}. Differences in scale cuts, priors, and covariance modelling may also contribute to the quantitative offsets.

Since $\fnl \, b_\phi$ is the best-constrained PNG combination, a physically motivated prior on $b_\phi$ is required to obtain a robust constraint on $\fnl$. Accordingly, in the next subsection, we study prior choices for $b_\phi$. Given the negligible information on $\fnl \, b_{\phi\delta}$, we fix $b_{\phi\delta}$ to the UHMF expression (Eq.~\ref{eq:UHMF_bphidelta}); changing the $b_{\phi\delta}$ prescription (e.g., alternatives motivated by SU measurements, \citealp{Barreira:2021ueb}) does not affect $\fnl$ or the $\Lambda$CDM parameters and is largely absorbed by small shifts in $b_2$.

\begin{figure*}[t]
    \centering
    \includegraphics[width=0.7 \linewidth]{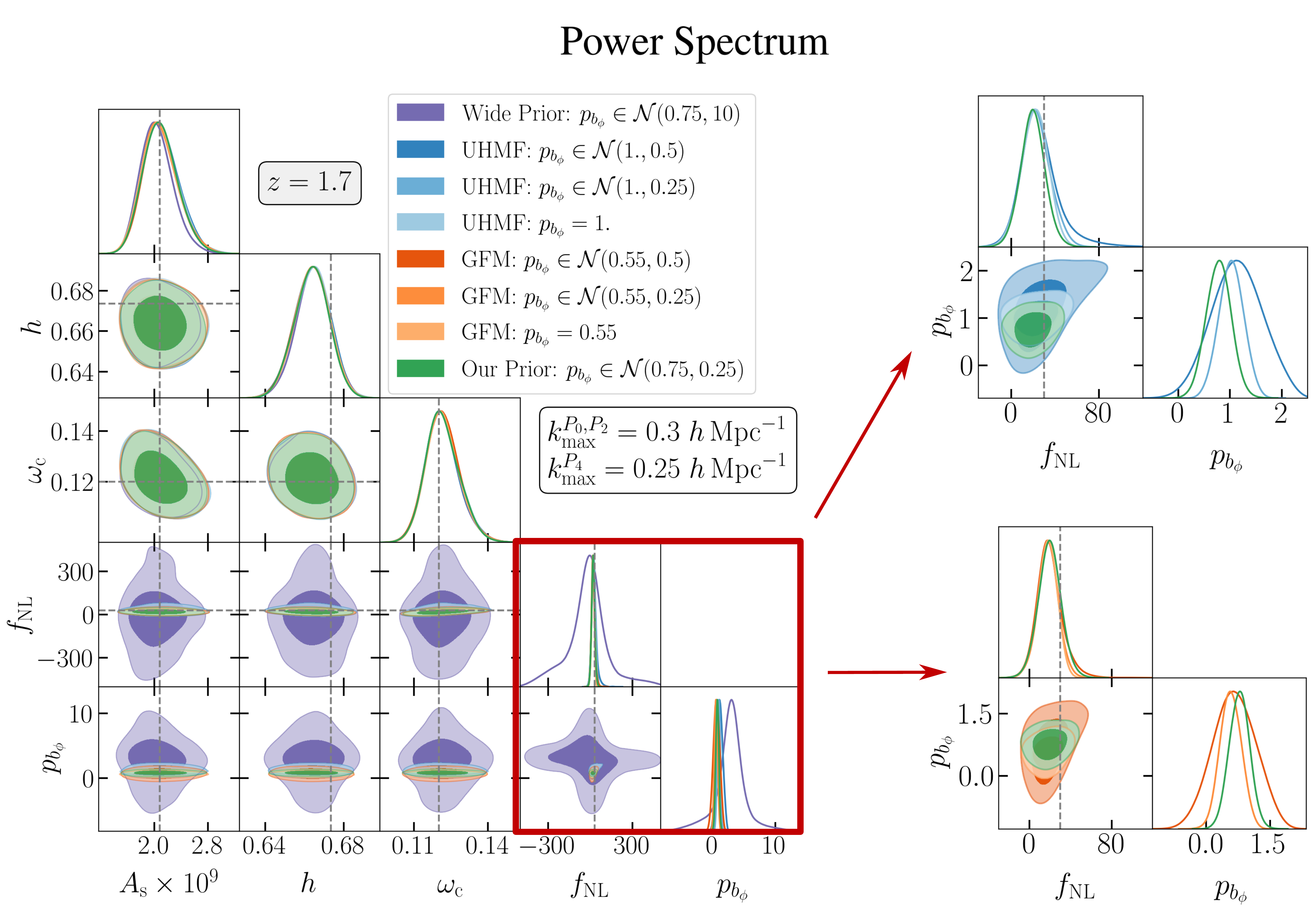}
    \caption{Impact of prior choice on power spectrum analysis. The plot shows the parameter posteriors from the power spectrum under different treatments of the PNG-bias intercept $p_{b_\phi}$: we either fix it to the UHMF or GFM value, or fit it with a Gaussian prior (mean and width shown in the legend). Blue contours use a Gaussian prior centred on the UHMF value; orange contours are centred on the GFM value; green contours are centred at $p_{b_\phi}=0.75$ with a width that covers both UHMF and GFM within $1\sigma$; magenta contours use a considerably broader prior (also centred at $p_{b_\phi}=0.75$).}
    \label{fig:Priors_PS}
\end{figure*}
\begin{figure*}[h!]
    \centering
    \includegraphics[width=0.7 \linewidth]{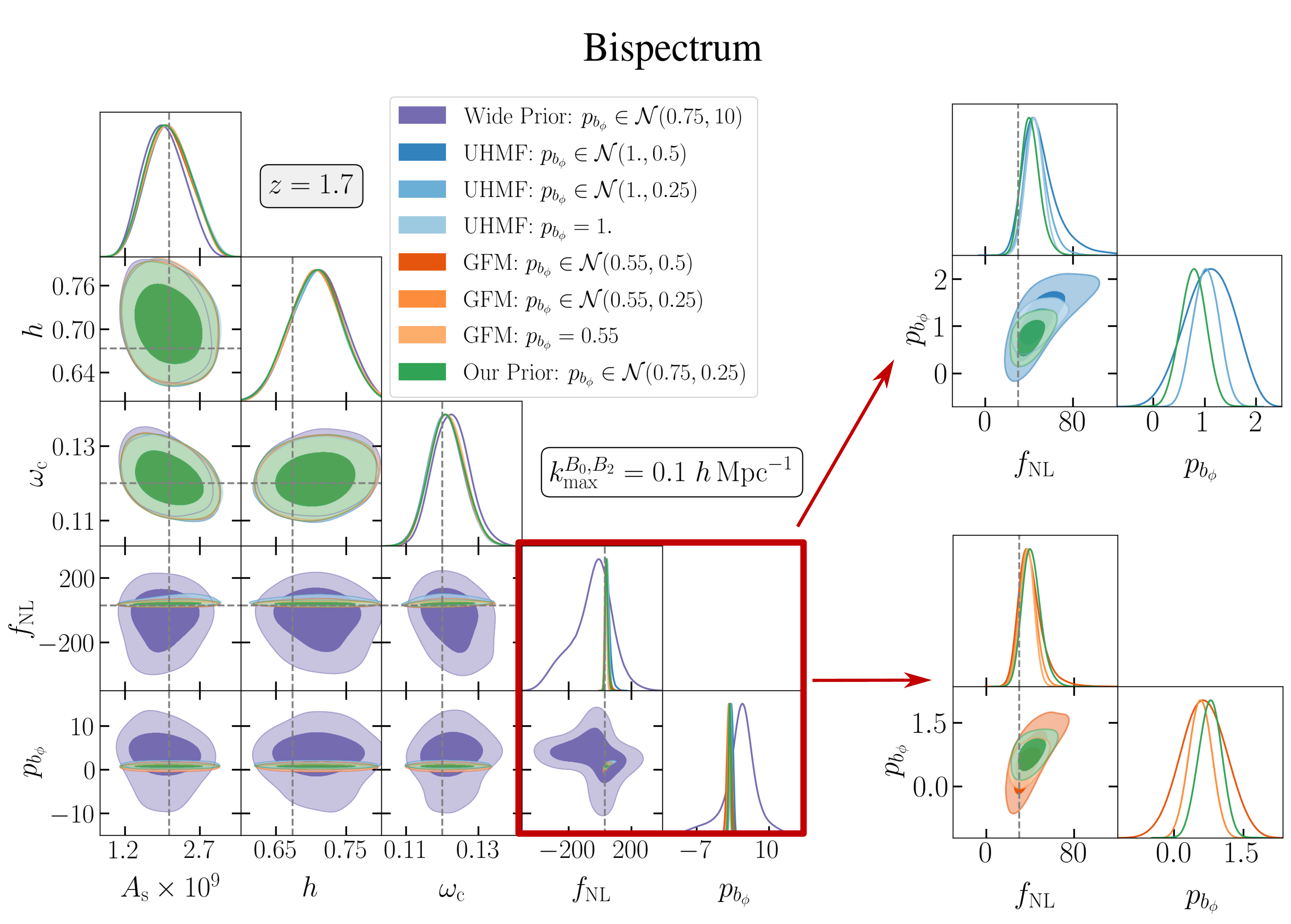}
    \caption{Same as Fig.~\ref{fig:Priors_PS} but for the bispectrum-only analysis.}
    \label{fig:Priors_BS}
\end{figure*}
\begin{figure*}[t]
    \centering
    \includegraphics[width=0.7 \linewidth]{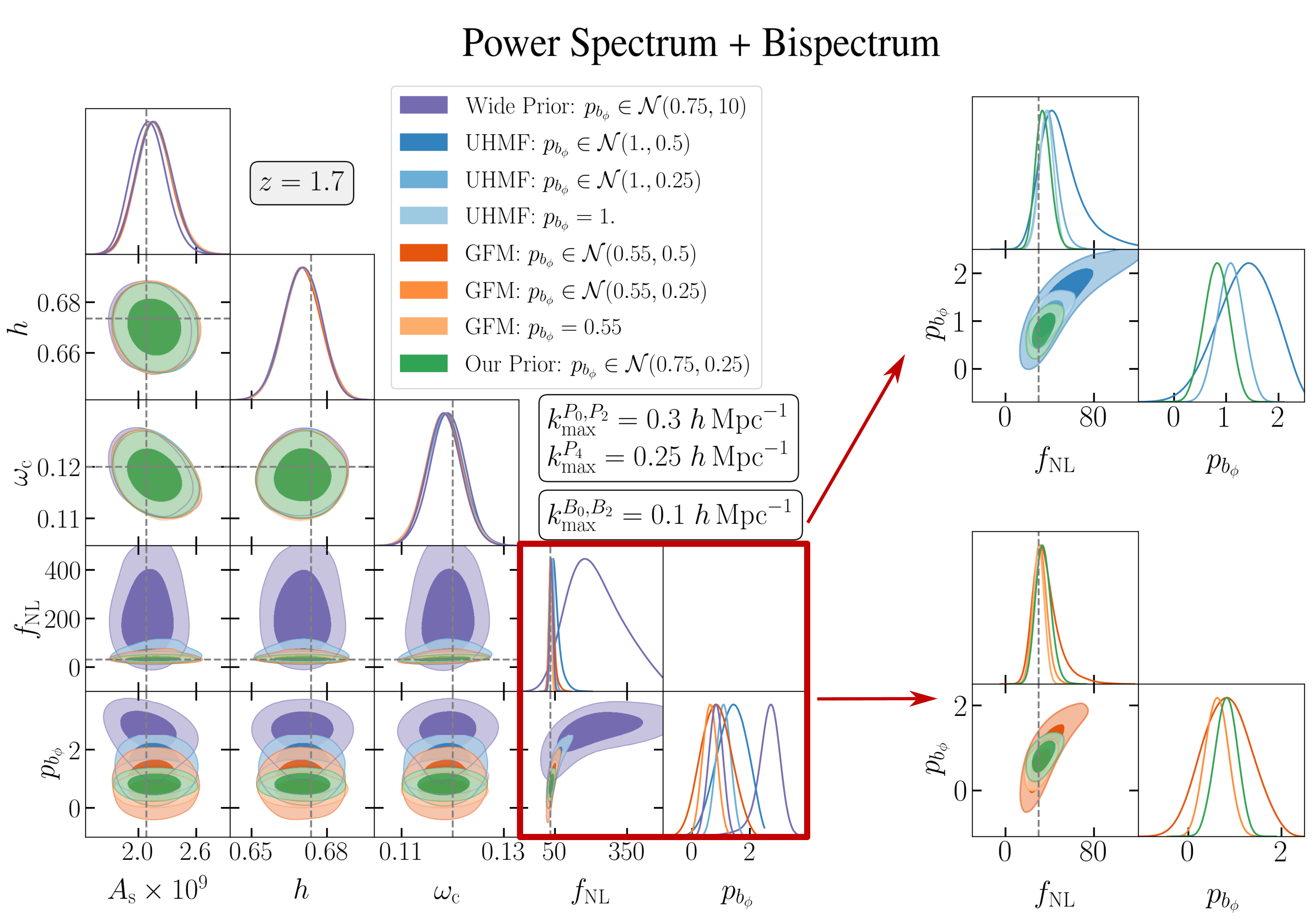}
    \caption{Same as Fig.~\ref{fig:Priors_PS} but for joint power spectrum plus bispectrum analysis.}
    \label{fig:Priors_Joint}
\end{figure*}

\subsection{Prior choice on PNG bias parameters for robustly constraining $\fnl$}\label{sec:prior_choice}

We test how the prior on the leading PNG bias affects constraints from the power spectrum, the bispectrum, and their combination. We adopt $b_\phi = 2\delta_{\mathrm{c}}\,(b_1-p_{b_\phi})$ and explore Gaussian priors on $p_{b_\phi}$ centred on UHMF ($p_{b_\phi}=1$), on the GFM/IllustrisTNG-motivated value ($p_{b_\phi}=0.55$), and on an intermediate value that contains both within $1\sigma$. We fix $b_{\phi\delta}$ to the UHMF prediction. Figures~\ref{fig:Priors_PS} to \ref{fig:Priors_Joint} display posteriors for several $p_{b_\phi}$ priors. All cosmological and nuisance parameters are varied (linear nuisances are analytically marginalised); for readability, figures show only $\Lambda$CDM parameters and $(\fnl,p_{b_\phi})$. 

With a very broad prior on $p_{b_\phi}$ (purple contours), the power spectrum is sensitive only to the product $f_{\rm NL} \, b_\phi$ leading to the familiar diamond-shaped degeneracy in the $f_{\rm NL}$-$p_{b_\phi}$ plane. Two practical consequences follow. First, the marginalised $f_{\rm NL}$ is pulled toward small values by projection along this extended direction (posterior weight accumulates near $f_{\rm NL} \sim 0$). Second, as $p_{b_\phi}$ drifts close to $b_1$, the mapping $b_\phi=2\delta_{\mathrm{c}} \, (b_1-p_{b_\phi})$ drives $b_\phi \to 0$, allowing $f_{\rm NL}$ to take large values with little likelihood penalty -- hence the long tails. This reproduces the known result: without information on $b_\phi$, $P$ alone cannot localise $f_{\rm NL}$. Imposing a tighter prior on $p_{b_\phi}$ collapses this direction and sharpens $f_{\rm NL}$. Centring on UHMF (blue) or GFM (orange) breaks the degeneracy, but the posterior mean of $f_{\rm NL}$ shifts mildly with the chosen centre because the implied $b_\phi$ differs between the two models. An intermediate Gaussian prior (green contours) that spans UHMF and GFM at $1\sigma$ strikes a useful balance: it stabilises $f_{\rm NL}$, acknowledges theoretical uncertainty on $b_\phi$, and avoids over-constraining with a single template. Cosmological parameters remain stable across these prior choices. 

For the bispectrum, the $(f_{\rm NL},p_{b_\phi})$ contours are already tilted even with a broad prior, reflecting sensitivity to both the pure-$f_{\rm NL}$ term and the $f_{\rm NL} \, b_\phi$ response (driven mainly by the redshift-space quadrupole). Very wide priors still inflate the marginalised $f_{\rm NL}$ error, while centring the prior on UHMF or GFM tightens constraints and shifts the mean along the same tilted direction; an intermediate prior spanning UHMF and GFM at $1\sigma$ stabilises the result without over-constraining $b_\phi$. Bispectrum constraints on $h$ are weaker than those from $P$, those on $A_{\mathrm{s}}$ are comparable but looser, and $\omega_{\mathrm{c}}$ are competitive yet mildly correlated with $f_{\rm NL}\,b_\phi$; all trends improve as the $p_{b_\phi}$ prior is tightened.

\begin{figure}
    \centering
    \hspace{-0.2in}\includegraphics[width=0.94\linewidth]{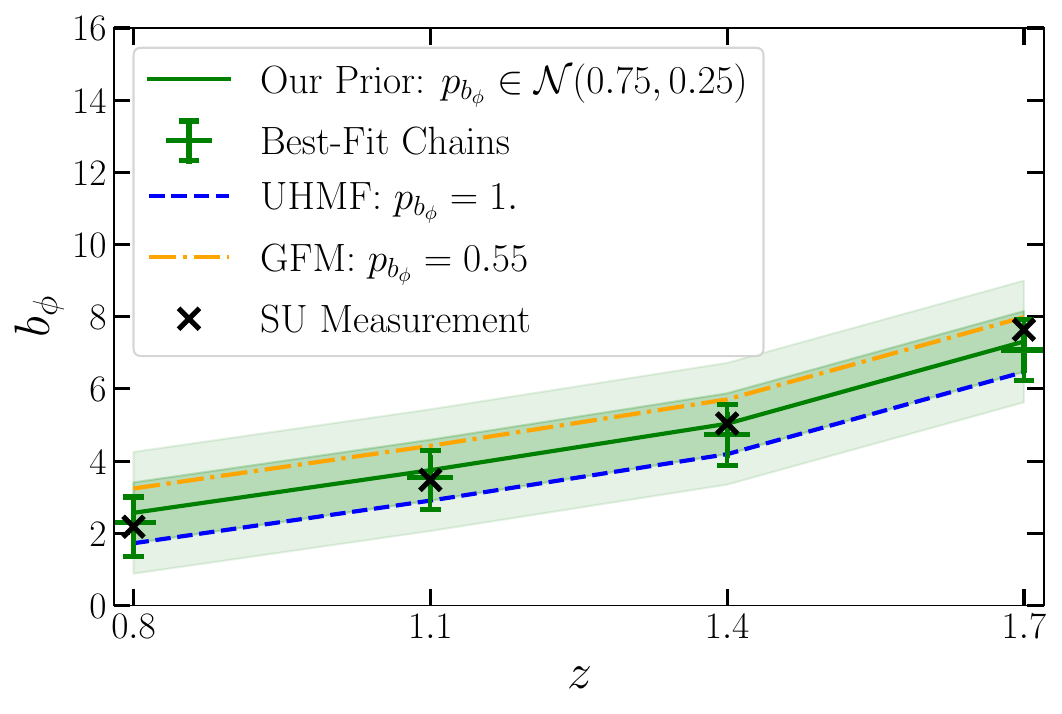}
    \caption{Measured versus assumed prior of $b_{\phi}$. We compare the directly measured values of $b_\phi$ using SU simulations at four redshift bins (crosses) with UHMF (dashed blue line) and GFM (dashed orange line) predictions. Values of $b_\phi$ corresponding to our prior centre (converted using inferred $b_1$ from joint power spectrum-bispectrum  chains) are shown in green line, with the shaded regions showing the $1$ and $2\sigma$ intervals. The best-fit values and the corresponding $1\sigma$ marginalised errors from the main analysis in Sect.~\ref{subsec:Joint} are shown in green crosses.}
    \label{fig:bphi_measurement}
\end{figure}

For the joint power spectrum-bispectrum analysis with a very wide prior on $p_{b_\phi}$, the star/diamond-shaped degeneracy seen in $P$ alone is replaced by a single elongated, banana-like ridge that extends toward positive $f_{\rm NL}$. Intuitively, $P$ fixes the sign of the product $f_{\rm NL}\,b_\phi$ through the large-scale response $\propto f_{\rm NL}\,b_\phi$, whereas $B$, through its pure-$f_{\rm NL}$ and $f_{\rm NL}\,b_{\phi\delta}$ pieces, breaks the $(f_{\rm NL},b_\phi)\to(-f_{\rm NL},-b_\phi)$ symmetry and, in our case, mildly prefers negative $f_{\rm NL}$. When combined, large negative $f_{\rm NL}$ would require $b_\phi<0$ (i.e., $p_{b_\phi}>b_1$) to satisfy the $P$ constraint, but such values are disfavoured by the $B$ likelihood (and carry little prior weight), so the overlap between $P$ and $B$ likelihoods resides primarily on the upper-right branch in the $(p_{b_\phi},f_{\rm NL})$ plane. Consequently, the joint posterior follows a single positive-$f_{\rm NL}$ ridge. This also explains why the $1$-dimensional marginalised constraint on $f_{\rm NL}$ from joint power spectrum-bispectrum fit can be broader than in $P$ or $B$ alone under a wide prior: the joint ridge is long and curved, so projecting it onto the $f_{\rm NL}$ axis spans a larger range (projection/prior-volume effect), even though the true joint credible region in $(f_{\rm NL},p_{b_\phi})$ is smaller. In the analysis with the informative priors, the posterior of the PNG parameters is very mildly sensitive to the prior centre, though the mean still shifts slightly along the residual $f_{\rm NL}$-$p_{b_\phi}$ tilt. 

Based on these tests, we adopt the intermediate Gaussian prior $p_{b_\phi} \sim \mathcal N(0.75,0.25)$ for our fiducial runs. It (i) encompasses both UHMF and GFM at $1\sigma$, (ii) mitigates projection along the $f_{\rm NL} \, b_\phi$ direction, and (iii) delivers stable, unbiased constraints across $P$, $B$, and $P+B$. 

To assess how well our assumed prior on $p_{b_\phi}$ reflects the actual PNG bias of our sample, we directly measured $b_\phi$ with SU Abacus-H$\alpha$ mocks by computing the difference of galaxy numbers in simulations with different values of $\sigma_8$. Figure~\ref{fig:bphi_measurement} compares the SU measurements (black crosses) to the UHMF prediction (dashed blue line) and the GFM prediction (dash-dotted orange line). The green line shows the prior mean mapped to $b_\phi(z)$ via Eq.~\eqref{eq:UHMF_bphi} using the inferred $b_1$ from the joint power spectrum-bispectrum chains; the shaded bands indicate the corresponding $1\sigma$ and $2\sigma$ ranges for our adopted Gaussian prior $p_{b_\phi}\sim\mathcal N(0.75,0.25)$. The SU points exhibit a mild redshift trend relative to this mapping -- slightly below (above) at lower (higher) $z$ -- indicating that the simple scaling with $b_1$ is not exact. Nevertheless, the SU measurements lie well within the prior envelope across all redshifts, supporting our use of this intermediate, theory-bracketing prior in the fiducial analysis. We emphasise that the choice of the prior centre was not informed by these SU measurements; the SU analysis was performed {\it a posteriori}, after completing our inference runs. As discussed in Sect.~\ref{sec:prior}, imposing selections on halos/galaxies can drive departures from the form $b_\phi = 2\delta_{\mathrm{c}}\, (b_1-p_{b_\phi})$. In our tests, halos are populated with galaxies using an HOD model that only depends on halo mass, imposing the expected flux cut for \Euclid H$\alpha$ galaxies. Under these conditions, one expects an approximately UHMF-like redshift evolution with a possible overall rescaling of the amplitude, consistent with previous $N$-body results (see Sect.~\ref{sec:prior}). This is precisely what we observe: the redshift dependence broadly follows the $b_1(z)$ scaling, with an overall offset captured by our prior.

\subsection{Null test on simulations with Gaussian initial conditions}

\begin{figure*}[htbp!]
    \centering
    \hspace{-0.1in}\includegraphics[width=0.9\columnwidth]{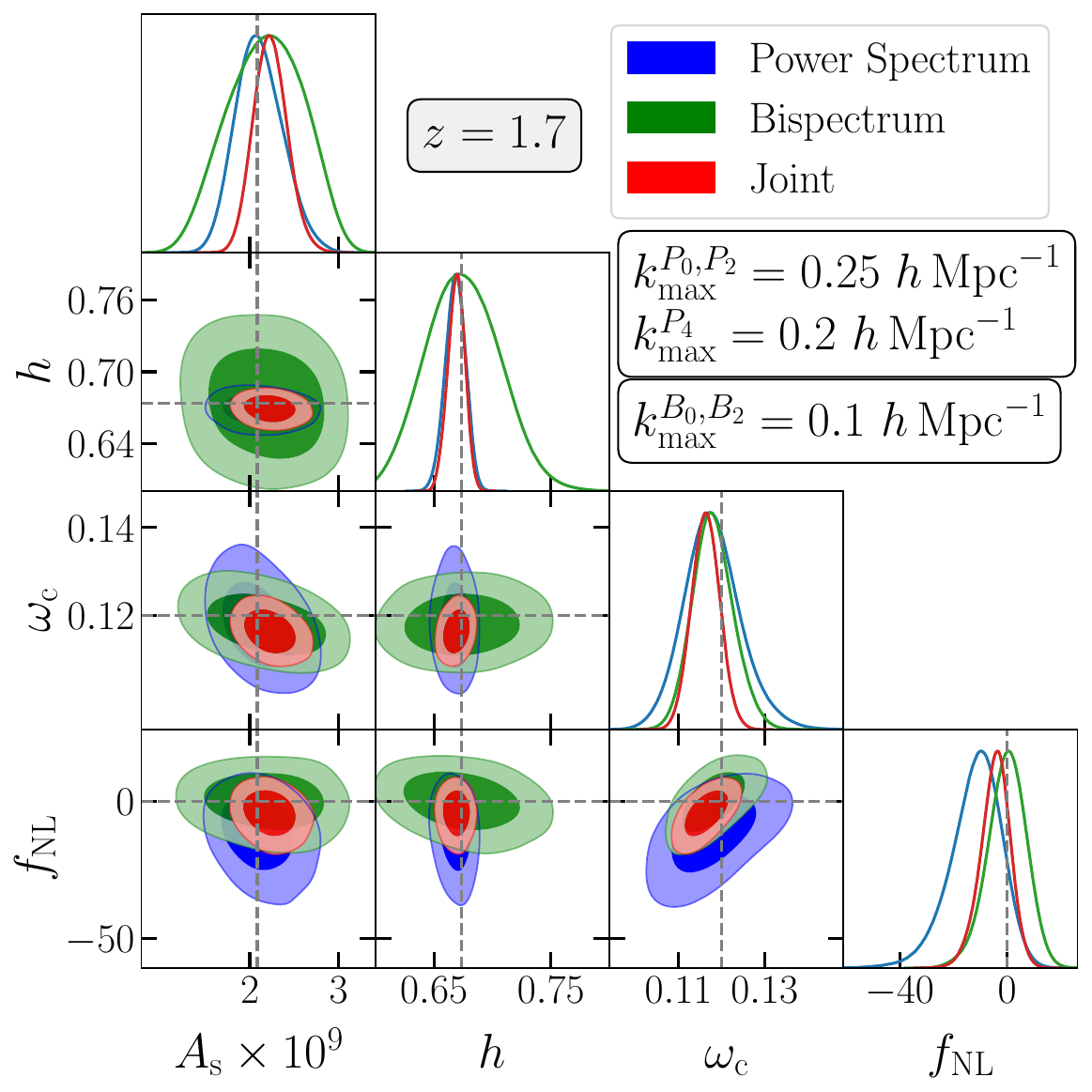}
    \hspace{0.2in}\includegraphics[width=0.84\columnwidth]{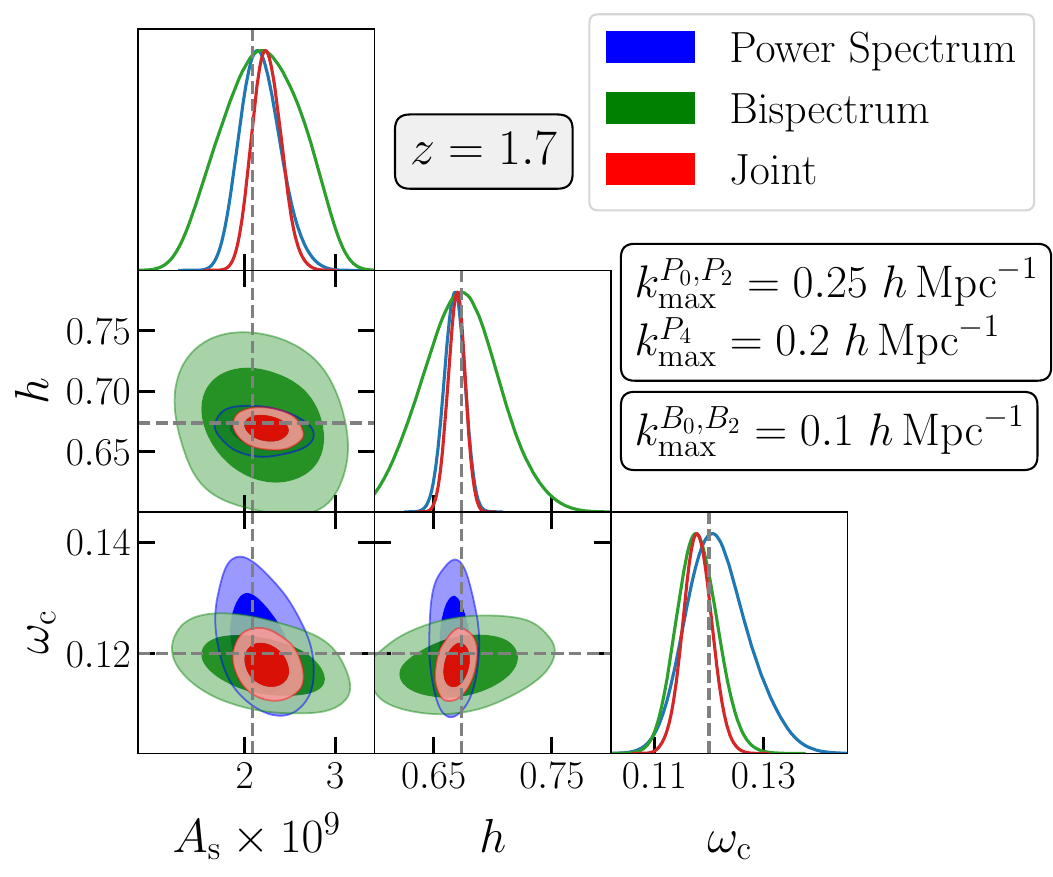}
    \caption{Null test on simulations with Gaussian initial conditions ($\fnl=0$). Parameter posterior distributions from power spectrum (blue), bispectrum (green), and their combination (red). \textit{Left}: constraints form the model with PNG. \textit{Right}: constraints from the model without PNG.}
    \label{fig:Gaussian_Posteriors}
\end{figure*}

To verify that our pipeline does not spuriously infer a PNG signal, we perform a null test on simulations with Gaussian initial conditions ($f_{\rm NL}=0$) at $z=1.7$ using the same scale cuts and the same $p_{b_\phi}$ prior as in Sect.~\ref{sec:prior_choice}, $p_{b_\phi} \sim \mathcal{N}(0.75,0.25)$. Figure~\ref{fig:Gaussian_Posteriors} shows posteriors from $P$ (blue contours), $B$ (green contours), and joint power spectrum and bispectrum (red contours). In all cases, the marginalised constraint on $f_{\rm NL}$ is centred on zero and consistent with $f_{\rm NL}=0$ within $1\sigma$, with no evidence for a spurious detection. The inclusion of the PNG sector (i.e., varying $f_{\rm NL}$ and the PNG-bias parameters as in the fiducial setup) leaves the $\Lambda$CDM parameters essentially unchanged: posterior means remain consistent with those from the Gaussian-only fits, and the credible intervals broaden only mildly, as expected when marginalising over partially degenerate directions. We also observe a modest, positive correlation between $\omega_{\mathrm{c}}$ and $f_{\rm NL}$ in the PNG-augmented fits (most visible for $P$, weaker for $B$), which the joint power spectrum-bispectrum analysis partially mitigates. Altogether, these null tests indicate that our redshift-space modelling and prior choices do not bias the analysis toward detecting PNG when it is absent, nor do they contaminate the inference of the standard cosmological parameters. 

We show the best-fit values of galaxy biases from joint analysis of power spectrum and bispectrum  on these simulations at the four considered redshift bins in Fig.~\ref{fig:Gbias}. The shaded regions are the predictions from co-evolution model assuming vanishing of Lagrangian tidal biases, and accounting for $1\sigma$ and $2\sigma$ uncertainties on $b_1$. We observe that deviation of of tidal biases $b_{\cG_2}$ and $b_{\Gamma_3}$ from co-evolution predictions is an indication for the presence of non-zero Lagrangian biases. Similar measurements on Flagship 1 simulations with additional comparisons with theoretical predictions can be found in \citet{Pardede2026Euclid}. 

\subsection{Analysis of simulations with non-Gaussian initial conditions}

We now proceed with the main analysis of the four snapshots of $\fnl=30$ simulations, setting $b_{\phi \delta}$ to the UHMF prediction and imposing the $p_{b_\phi}\sim\mathcal N(0.75,0.25)$ prior.  

\subsubsection{Scale-cut analysis for $P$ and $B$ individually}\label{subsec:scale_cuts}

\begin{figure}
    \centering
    \hspace{-0.12in}\includegraphics[width=0.92\columnwidth]{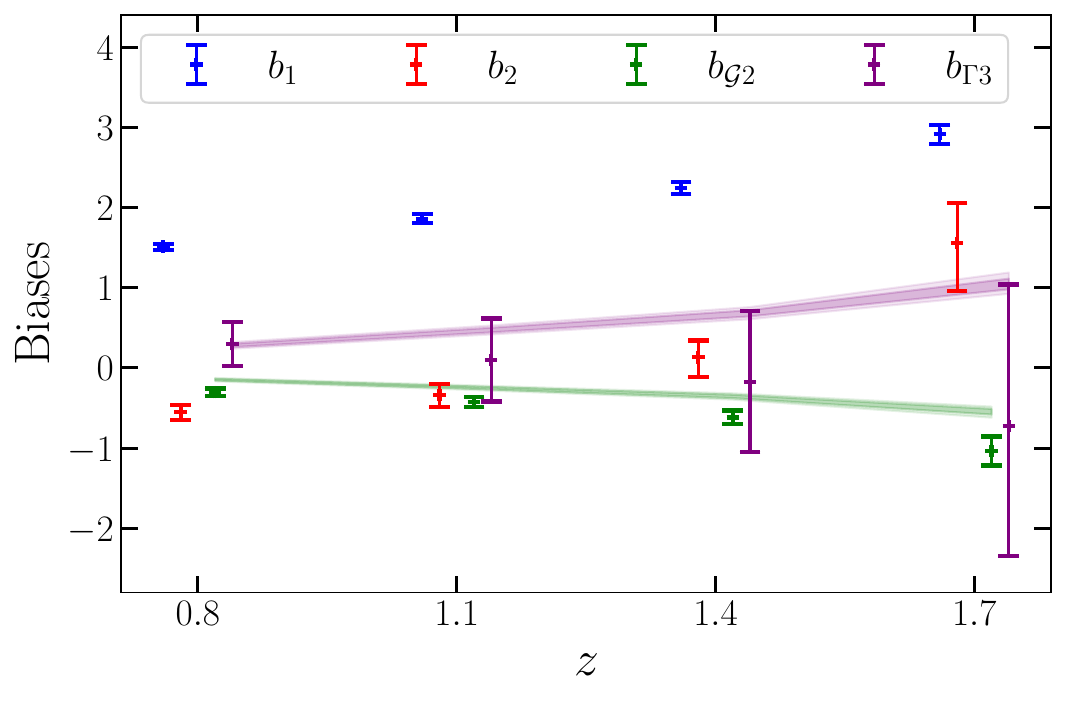}\vspace{-0.1in}
    \caption{Best-fit galaxy biases and the corresponding $1\sigma$ error bars from joint power spectrum-bispectrum fit to simulations with Gaussian initial conditions. To make the figure clearer, we have slightly shifted the data points horizontally within each redshift to avoid overlap. However, these measurements still correspond to the same redshift bin, respectively. The areas shaded in green and magenta show the co-evolution predictions (assuming vanishing Lagrangian tidal biases) for $b_{\cG_2}(b_1)$ and $b_{\Gamma_3}(b_1)$, incorporating the $1\sigma$ and $2\sigma$ ranges of $b_1$.}\label{fig:Gbias}
\end{figure}

\begin{figure*}[htbp!]
  \centering
    \includegraphics[width=0.97\linewidth]{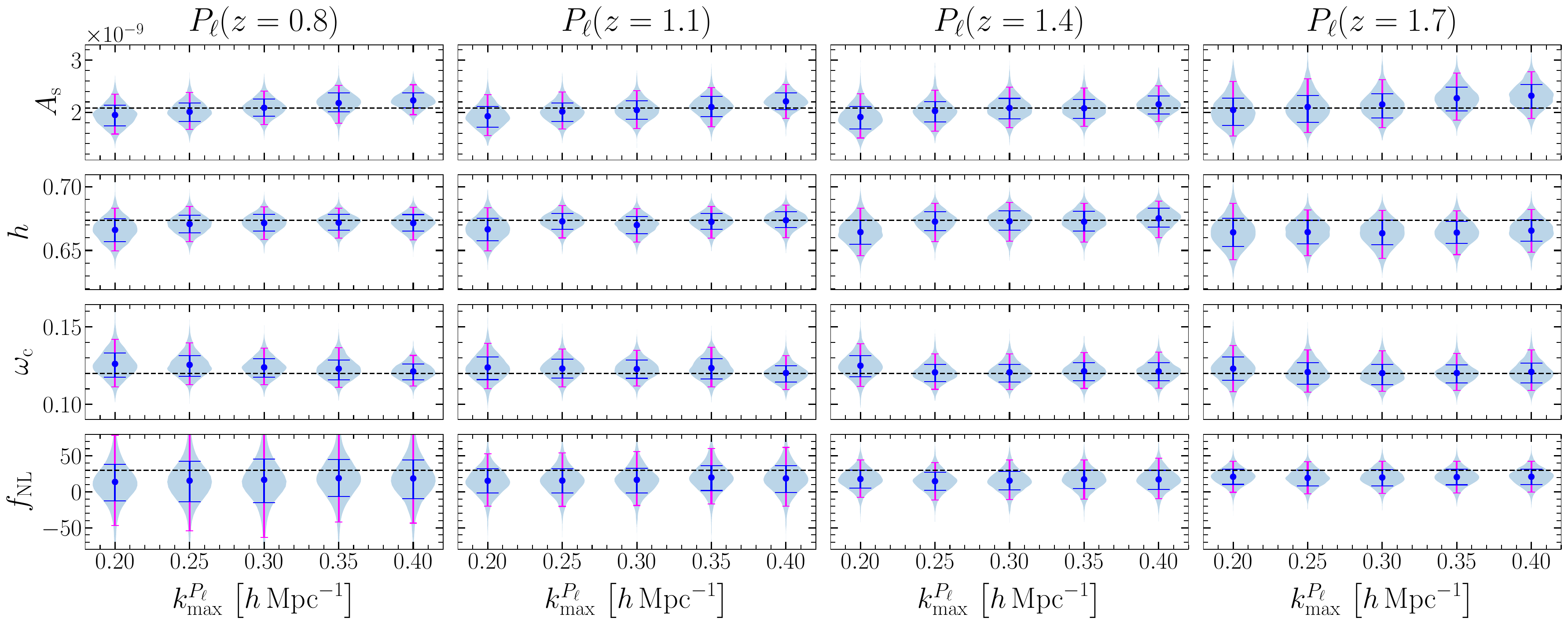}
    \caption{Marginalised $1$-dimensional posterior distributions in the four redshift bins and for five choices of scale cuts (for monopole and quadrupole). The violin errors represent the real posterior distribution for each parameter, with the blue and magenta error bars representing the $1\sigma$ and $2\sigma$ intervals of these distributions, respectively. The dashed lines are the fiducial values used in the simulations. The hexadecapole scale cut is always set to be $0.05\, h\ \mathrm{Mpc}^{-1}$ than the monopole and quadrupole one, $k^{P_4}_{\rm{max}} = k^{P_0, P_2}_{\rm{max}} - 0.05\, h\ \mathrm{Mpc}^{-1}$.\\}
    \label{fig:PS_scale_cuts_errors}

  \centering
    \includegraphics[width=0.97\linewidth]{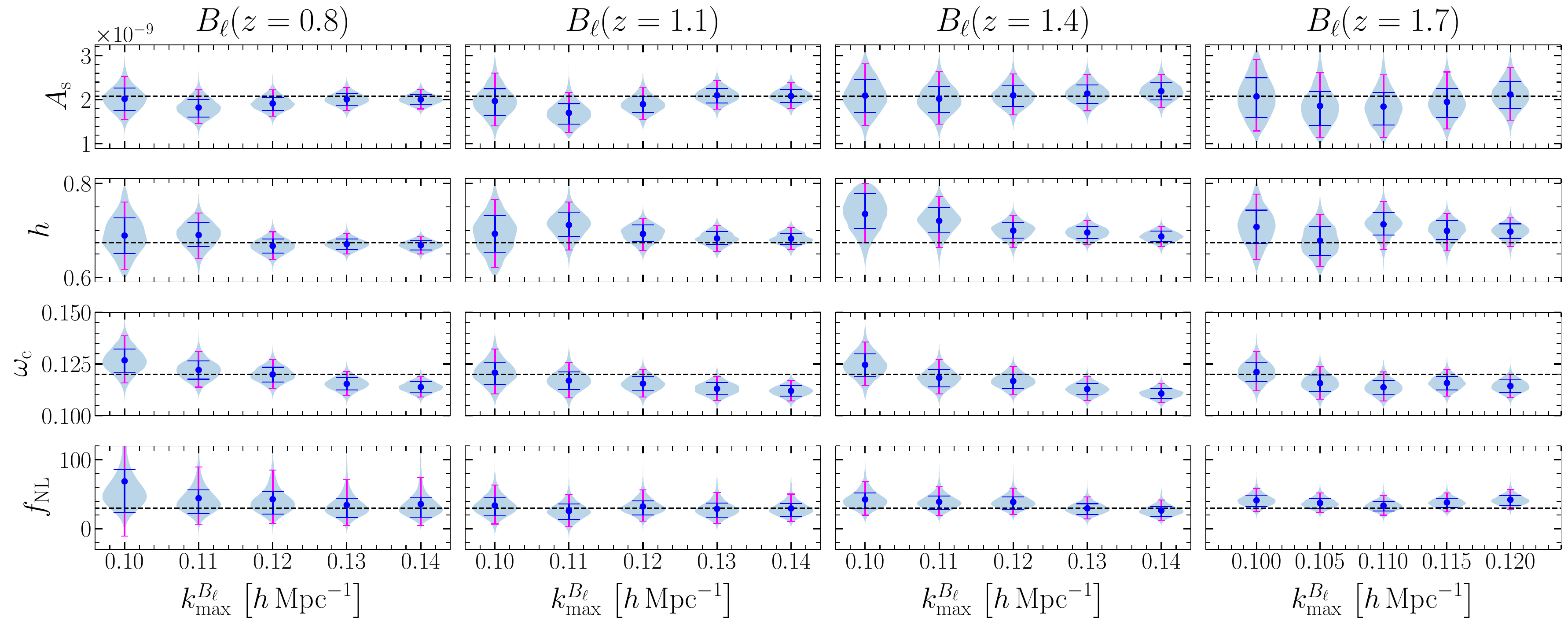}
    \caption{Marginalised $1$-dimensional posterior distributions in the four redshift bins and for five choices of scale cuts (for monopole and quadrupole). The hexadecapole is discarded due to noisy measurements. The rest of the plot styling matches Fig.~\ref{fig:PS_scale_cuts_errors}.}
    \label{fig:BS_scale_cuts_errors} 

    \centering
    \hspace{-0.2in}\includegraphics[width=0.9\columnwidth]{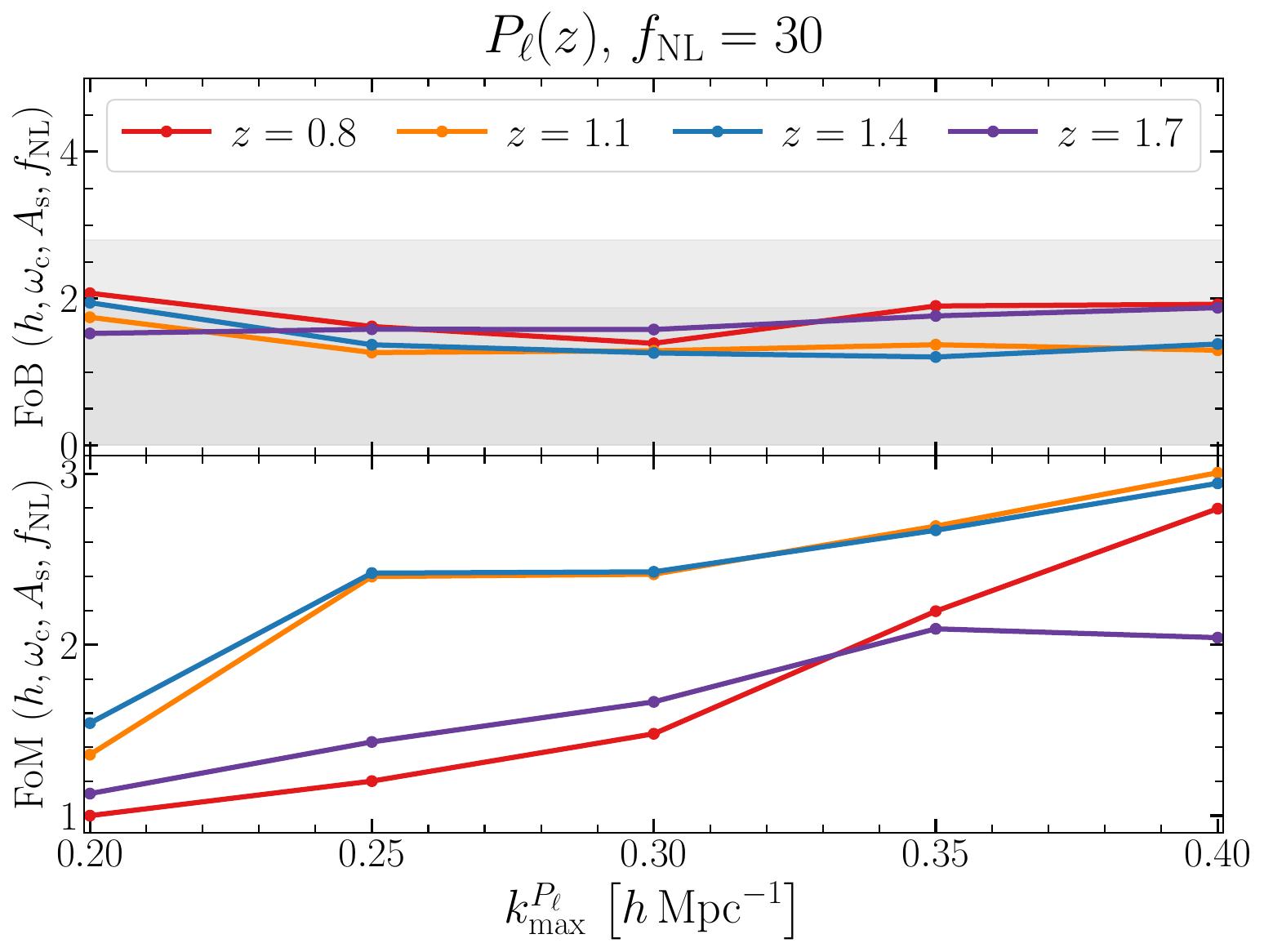}
    \hspace{0.2in}\includegraphics[width=0.9\columnwidth]{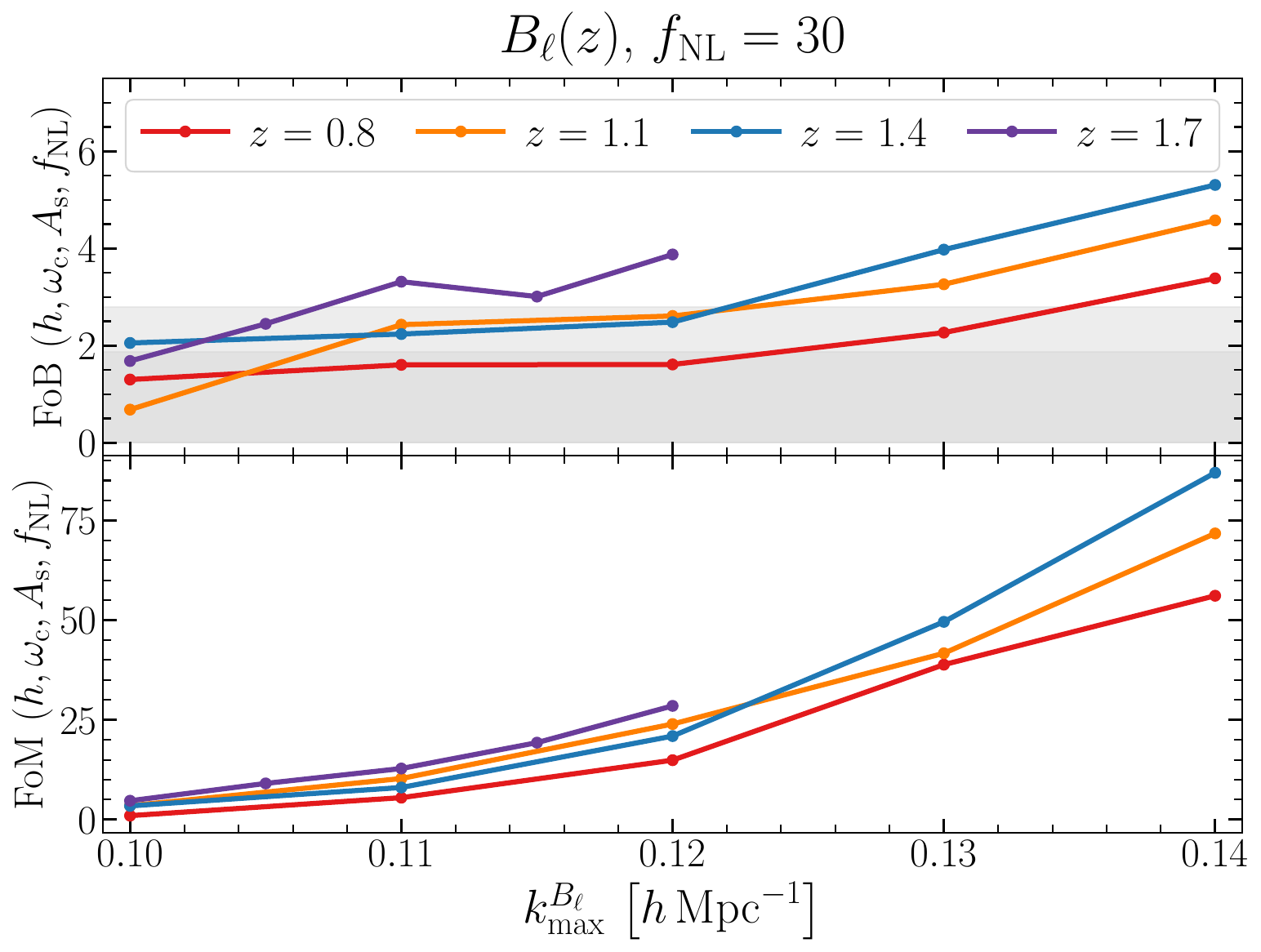}
    \caption{Validity of perturbative model and information content. The plots illustrate the accuracy in terms of FoB (Eq.~\ref{eq:FoB}) and the precision in terms of FoM (Eq.~\ref{eq:FoM}) of various scale cuts across all redshifts for the combination of all the varied cosmological parameters. The FoM is normalised for both observables based on the lowest redshift and scale cut ($z=0.8$, $k_{\rm max}^{P_\ell}=0.2 \,h\,\rm{Mpc}^{-1}$ and $k_{\rm max}^{B_{\ell}}=0.1 \,h\,\rm{Mpc}^{-1}$), indicating the relative increase in information with increasing $k_{\rm max}$. \textit{Left}: results for the powers spectrum. \textit{Right}: results for the bispectrum.}
    \label{fig:FoB_FoM}
\end{figure*}

\begin{figure*}[h!]
    \centering
    \hspace{-0.2in}\includegraphics[width=0.95 \columnwidth]{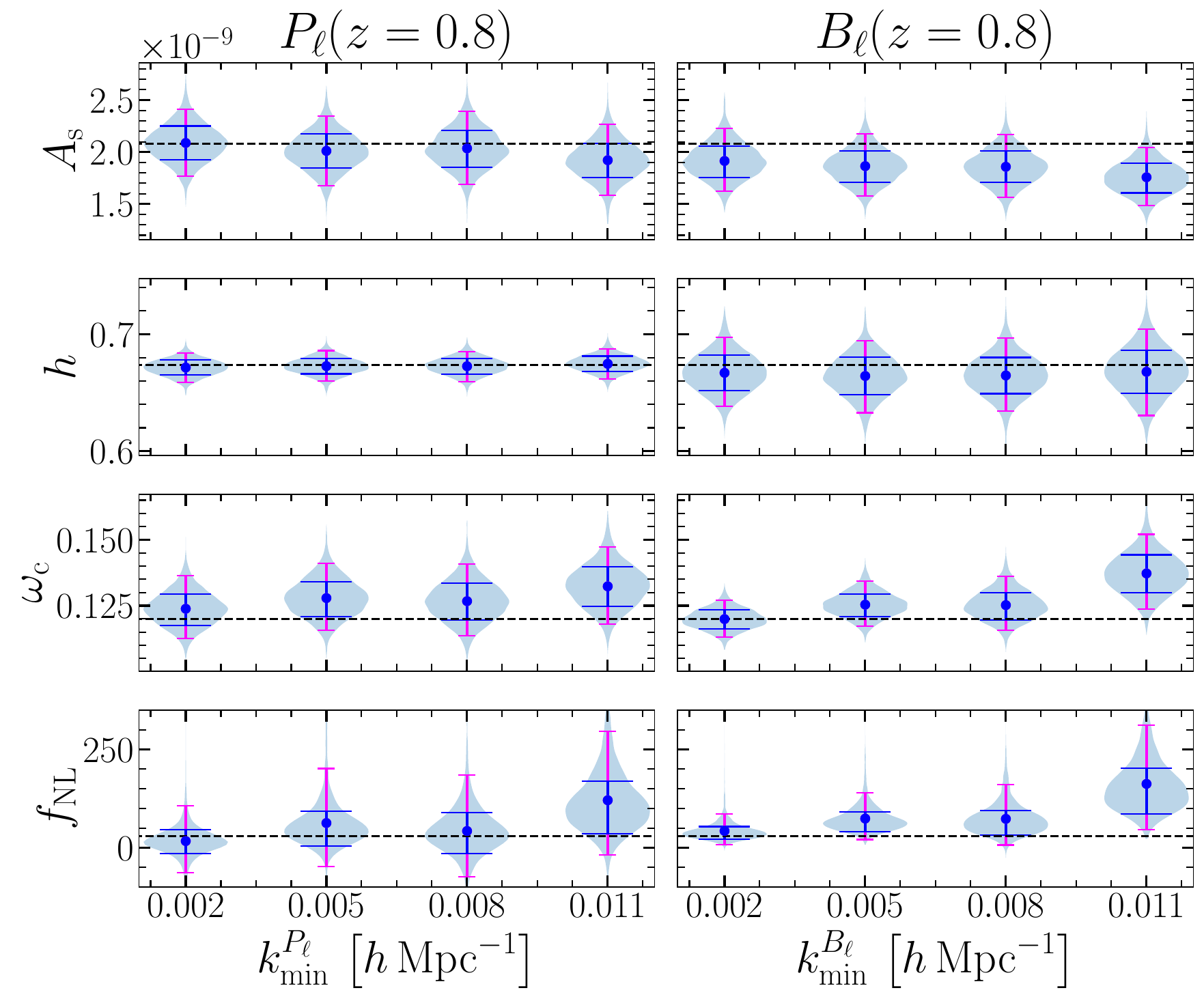}
    \hspace{0.2in}\includegraphics[width=\columnwidth]{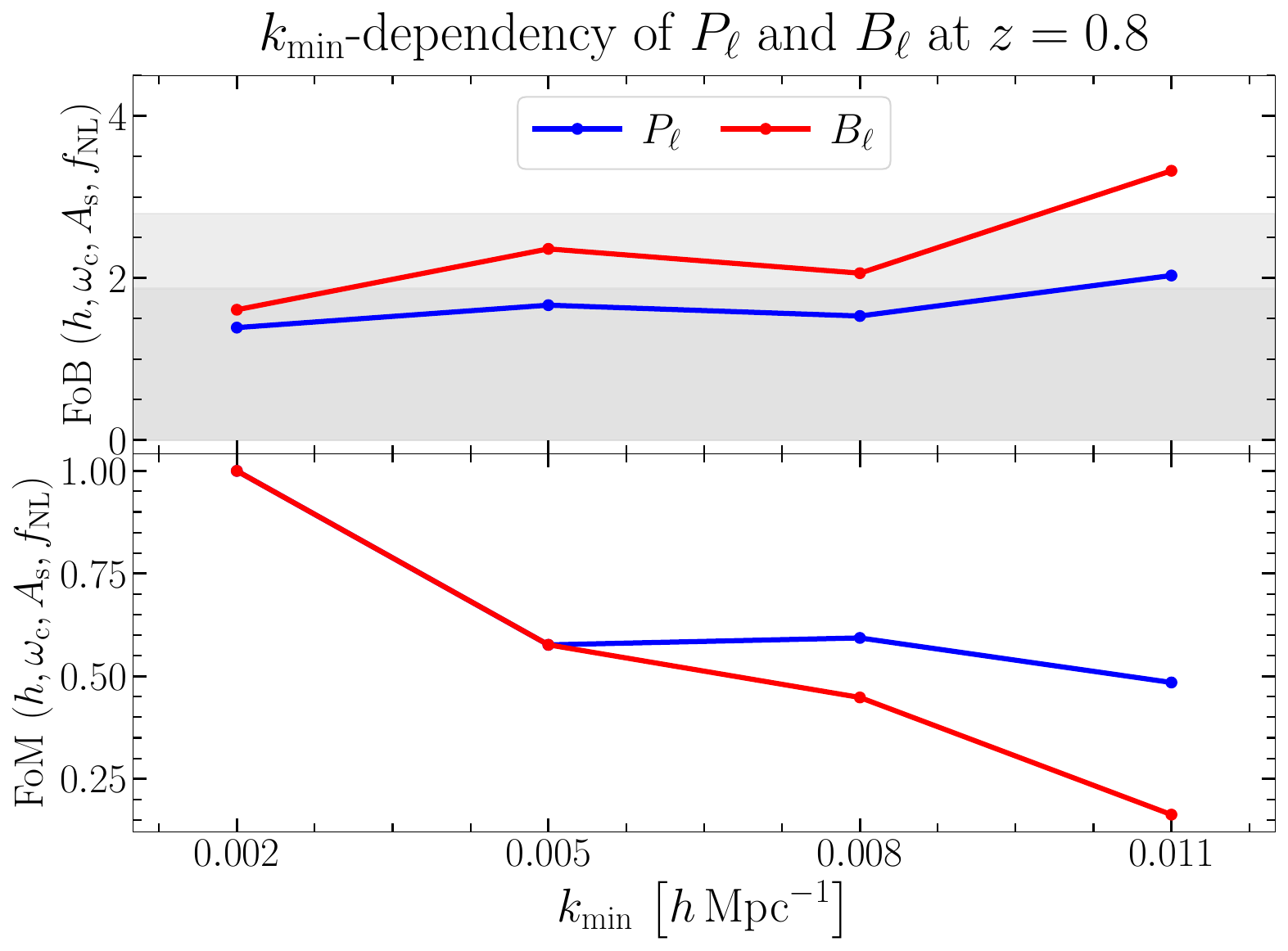}
    \caption{Impact of the exclusion of large-scale modes. \textit{Left}: we present the marginalised $1\sigma$ constraints on cosmological parameters from the power spectrum and bispectrum multipoles at $z=0.8$. \textit{Right}: we show the FoB and FoM for the combined parameters from $P_\ell$ (blue) and $B_\ell$ (red). In this context, $k_{\rm min}$ denotes the lower boundary of the considered bin, which implies excluding each $k_{\rm f}$ bin individually. Moreover, the FoM is normalised to our baseline scale cut of $k_{\rm min}^{P_\ell} = k_{\rm min}^{B_\ell} = k_{\rm f}-\Delta k/2$ (lower edge of the first $k$-bin). The rest of plot styling matches Figs.~\ref{fig:PS_scale_cuts_errors}--\ref{fig:FoB_FoM}.}  
    \label{fig:Scale_cuts_kmin}
\end{figure*}

Figures~\ref{fig:PS_scale_cuts_errors} and \ref{fig:BS_scale_cuts_errors} show $1$-dimensional marginalised posteriors of the cosmological parameters from power-spectrum and bispectrum multipoles across four redshifts and five small-scale cuts. Figure~\ref{fig:FoB_FoM} summarises the two global diagnostics for the combination of all cosmological parameters as a function of the cuts. The FoB quantifies the model accuracy (grey bands mark $1\sigma$ and $2\sigma$), while the FoM quantifies the precision of parameter constraints. We normalised the FoM to the smallest $k_{\rm max}$ to highlight the relative information gain.

For the power spectrum, the one-loop model yields unbiased constraints (within $\sim1\sigma$) for all $\Lambda$CDM parameters at all redshifts over the tested range.\footnote{We note that the analysis of \citet{Camacho2026Euclid} finds a smaller range of validity of EFTofLSS predictions when analysing Flagship 1 simulations. Direct comparison of our results with that paper is not possible due to several reasons: (i) significantly larger volume of Flagship 1 compared to Abacus, $54\ h^{-3}\,{\rm Mpc}^3$ versus $16\ h^{-3}\,{\rm Mpc}^3$, (ii) difference in modelling of stochastic and shot noise contributions, (iii) model parameter space (Gaussian versus non-Gaussian initial conditions).} At a fixed redshift, as we increase $k_{\rm max}$ beyond $0.2\,h\,\mathrm{Mpc}^{-1}$, the constraints on $\omega_{\mathrm{c}}$ improve the most, the gain in $A_{\mathrm{s}}$ is more modest and $h$ remains very stable (see Fig.~\ref{fig:Gbias}). There is a mild drift in $A_{\mathrm{s}}$ at $z=1.7$ at the two highest scale cuts, indicating earlier breaking of the one-loop model, presumably due to larger bias values. Constraints on $f_{\rm NL}$ are essentially unaffected by the choice of $k_{\rm max}$, consistent with the fact that the power spectrum PNG information is governed by the large-scale response. In redshift, the $\fnl$ constraints tighten toward higher $z$ due to the enhancement of the PNG signal from larger $b_\phi$; by contrast, the $A_{\mathrm{s}}$ and $h$ constraints are weakest at the highest redshift bin, plausibly because higher bias and lower number density reduce RSD and amplitude leverage, while $\omega_{\mathrm{c}}$ remains relatively stable across redshifts. The aggregate power spectrum FoB confirms unbiased recovery of the parameter combination (within $1\sigma$--$2\sigma$). The combined FoM increases with $k^{P}_{\max}$, with the exact trend being redshift dependent. Over the scales considered, the FoM grows by at most a factor $\sim3$ beyond $k_{\rm max}=0.2\,h\,\mathrm{Mpc}^{-1}$; at $k_{\rm max}=0.30\,h\,\mathrm{Mpc}^{-1}$ the gain is $\sim1.5$ times for the lowest and highest redshift bins, and $\sim2.5$ times at intermediate redshifts.

For the bispectrum, increasing $k_{\max}^{B_\ell}$ tightens the constraints but eventually induces a bias because the $B_\ell$ model is at tree level. For $z\leq1.4$, posteriors remain consistent within $1$--$2\sigma$ up to $k_{\max}^{B_\ell}=0.12\,h\,\mathrm{Mpc}^{-1}$, whereas at $z=1.7$ the constraints begin to bias beyond $k_{\max}^{B_\ell}\sim0.105\,h\,\mathrm{Mpc}^{-1}$. The first parameter to show tension is typically $\omega_{\mathrm{c}}$, which,  at the lower redshift, shifts low by more than $2\sigma$ once $k_{\max}^{B_\ell} \gtrsim  0.12\,h\,\mathrm{Mpc}^{-1}$. In contrast, $A_{\mathrm{s}}$, $h$, and $f_{\rm NL}$ remain stable over the same range: $f_{\rm NL}$ is consistent within $1\sigma$ up to $k_{\mathrm{max}}^{B_\ell} = 0.14\,h\,\mathrm{Mpc}^{-1}$ at $z\leq1.4$, and within $2\sigma$ at $z=1.7$. Much of the apparent small-scale improvement is enabled by the phenomenological VDG counterterm (Sect.~\ref{sec:th}); replacing it with a standard FoG counterterm triggers an earlier breakdown (around $k_{\max}^{B_\ell} \sim 0.10\,h\,\mathrm{Mpc}^{-1}$ at the lowest redshift). Even so, modestly pushing $k_{\max}^{B_\ell}$ greatly boosts statistical power because each increment adds many triangles that jointly probe large and small scales. As shown in Fig.~\ref{fig:FoB_FoM}, the combined bispectrum FoM improves by a factor of $\sim 10$--$20$, while the combined FoB remains within $2\sigma$. Two caveats are worth noting. First, extending the bispectrum model beyond tree level would likely allow higher $k_{\max}^{B_\ell}$, but at the cost of many additional nuisance parameters; the net gain after marginalising over them must be assessed with a full analysis. Second, the FoM gain for $B$ alone does not map one-to-one into the joint power spectrum-bispectrum FoM (see Sect.~\ref{subsec:Joint}): at the lower redshifts, the power spectrum already carries substantial constraining power, so adding $B$ at higher $k_{\max}^{B_\ell}$ yields a less substantial improvement than suggested by the $B$-only FoM (see Sect.~\ref{subsec:Joint}).

We now assess the impact of removing the largest scale modes on the results from the power spectrum and the bispectrum. We illustrate this explicitly for the lowest-redshift snapshot, $z=0.8$, as a representative and conservative case; the stronger PNG sensitivity at higher redshift is documented separately in Fig.~\ref{fig:Errors_Multipoles_Redshift}. Given that the constraints on local-type $\fnl$ are driven by the large-scale modes (in the power spectrum) and the coupling of large and small-scale modes (in the bispectrum), a crucial concern is the loss of constraining power resulting from our inability to use large-scale modes due to various observational systematic effects. Focussing on the lowest redshift snapshot ($z=0.8$), the two left panels of Fig.~\ref{fig:Scale_cuts_kmin} show the one-dimensional marginalised constraints as a function of the large-scale cut $k_{\min}$ for $P$ and $B$, while the right panel displays the aggregate FoB and FoM for the four cosmological parameters, with the FoM normalised to the baseline with $k_{\min}=k_{\mathrm{f}}-\Delta k/2$. As expected, the $f_{\rm NL}$ errors degrade as $k_{\min}$ increases, more strongly for the bispectrum than for the power spectrum. The sensitivity of $\omega_{\mathrm{c}}$ to $k_{\min}$ is also significant, especially for $B$, whereas the $A_{\mathrm{s}}$ error bars are nearly insensitive to $k_{\min}$. However, for $B$ at the largest $k_{\min}$ considered, removing the lowest modes induces $\sim3\sigma$ biases in all parameters except $h$. Quantitatively (see the right panel), excluding the first three $k$-bins induces less than $1\sigma$ total bias (FoB) for $P$ and about $2\sigma$ for $B$ (with larger biases at the highest $k_{\min}$). Over the scanned range, the FoM drops by $\sim 50\%$ for $P$ and by $\sim 75\%$ for $B$, underscoring the disproportionate importance of the largest scales, particularly for the bispectrum, in constraining $f_{\rm NL}$ and $\omega_{\mathrm{c}}$. 

Guided by the above results, for the rest of our analyses, we adopt rather conservative (and robust) cuts $k_{\max}^{P_0,P_2}=0.30\,h\,\mathrm{Mpc}^{-1}$ and $k_{\max}^{P_4}=0.25\,h\,\mathrm{Mpc}^{-1}$ for the power spectrum across all redshift bins. For the bispectrum, at the lowest three redshifts, we adopt $k_{\max}^{B_0,B_2}=0.12\,h\,\mathrm{Mpc}^{-1}$, while at the highest redshift, we limit ourselves to a lower value of $k_{\max}^{B_0,B_2}=0.105\,h\,\mathrm{Mpc}^{-1}$. We set the large-scale cut to the fundamental mode of the survey $k_{\rm min}^{P_\ell} = k_{\rm min}^{B_\ell} = k_{\rm f}-\Delta k /2$.

\subsubsection{Joint analysis of power spectrum and bispectrum}\label{subsec:Joint}

\begin{figure*}[htbp!]
   \includegraphics[width=0.7\columnwidth, height=2.5in]{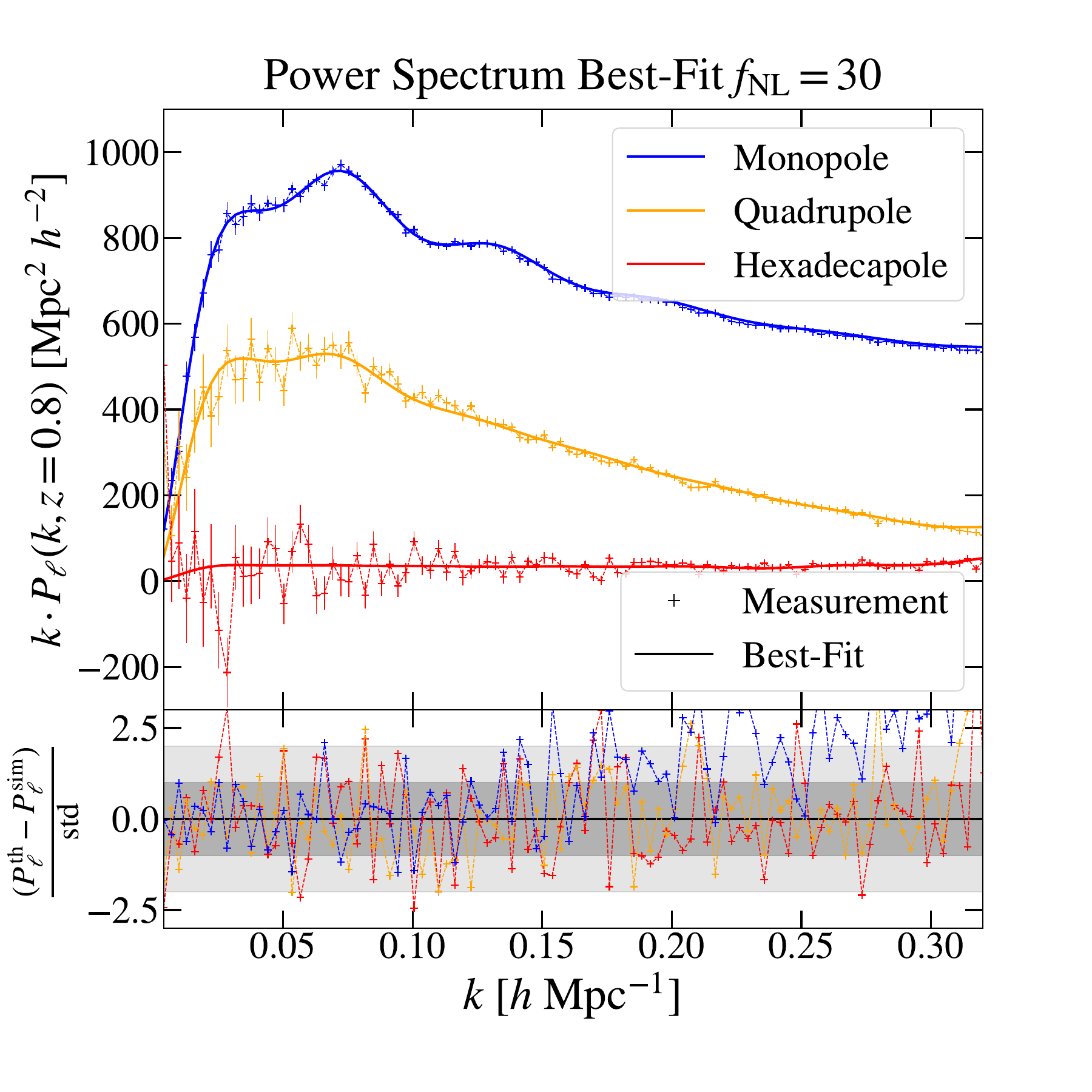}
    \includegraphics[width=0.7\columnwidth, height=2.5in]{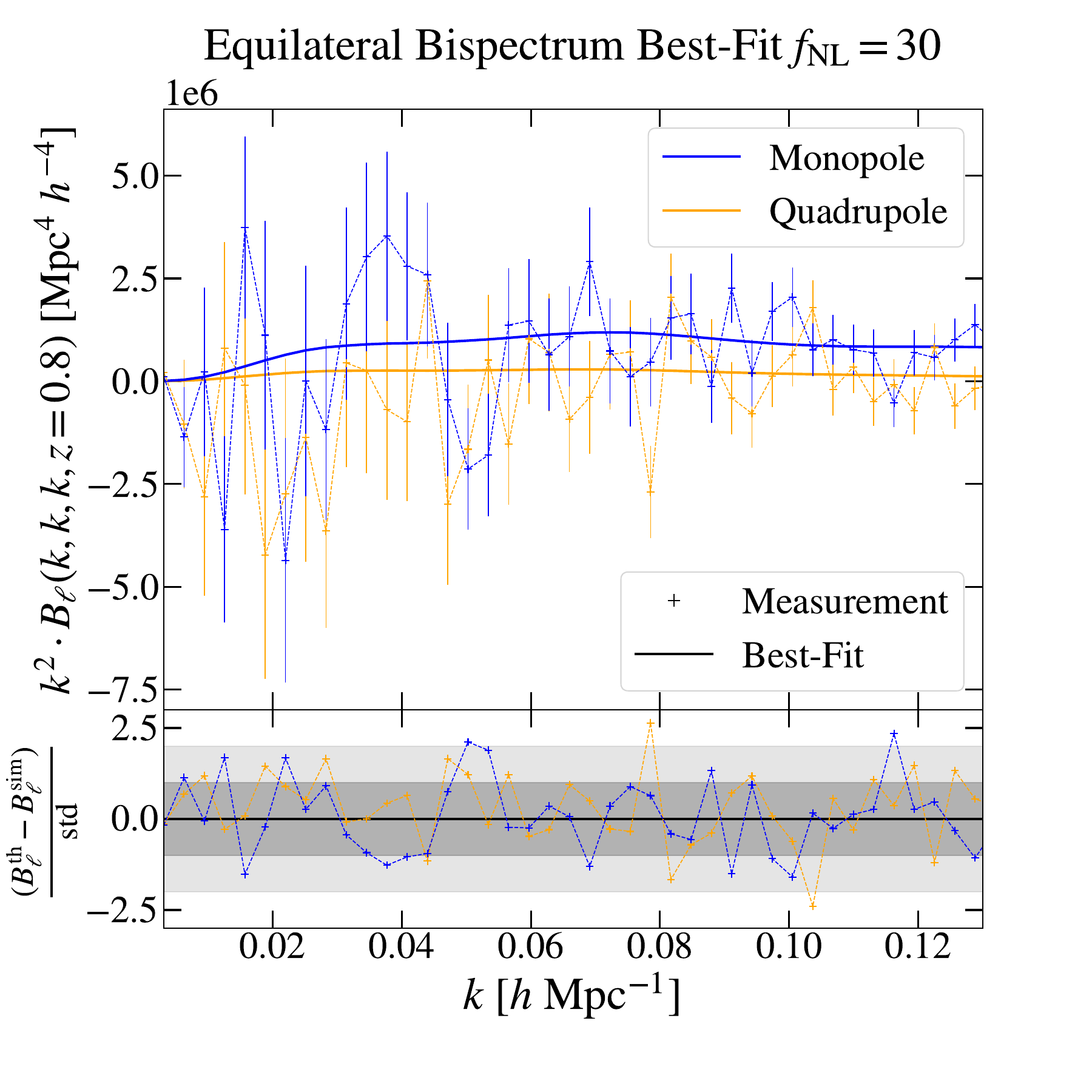}
    \includegraphics[width=0.7\columnwidth, height=2.5in]{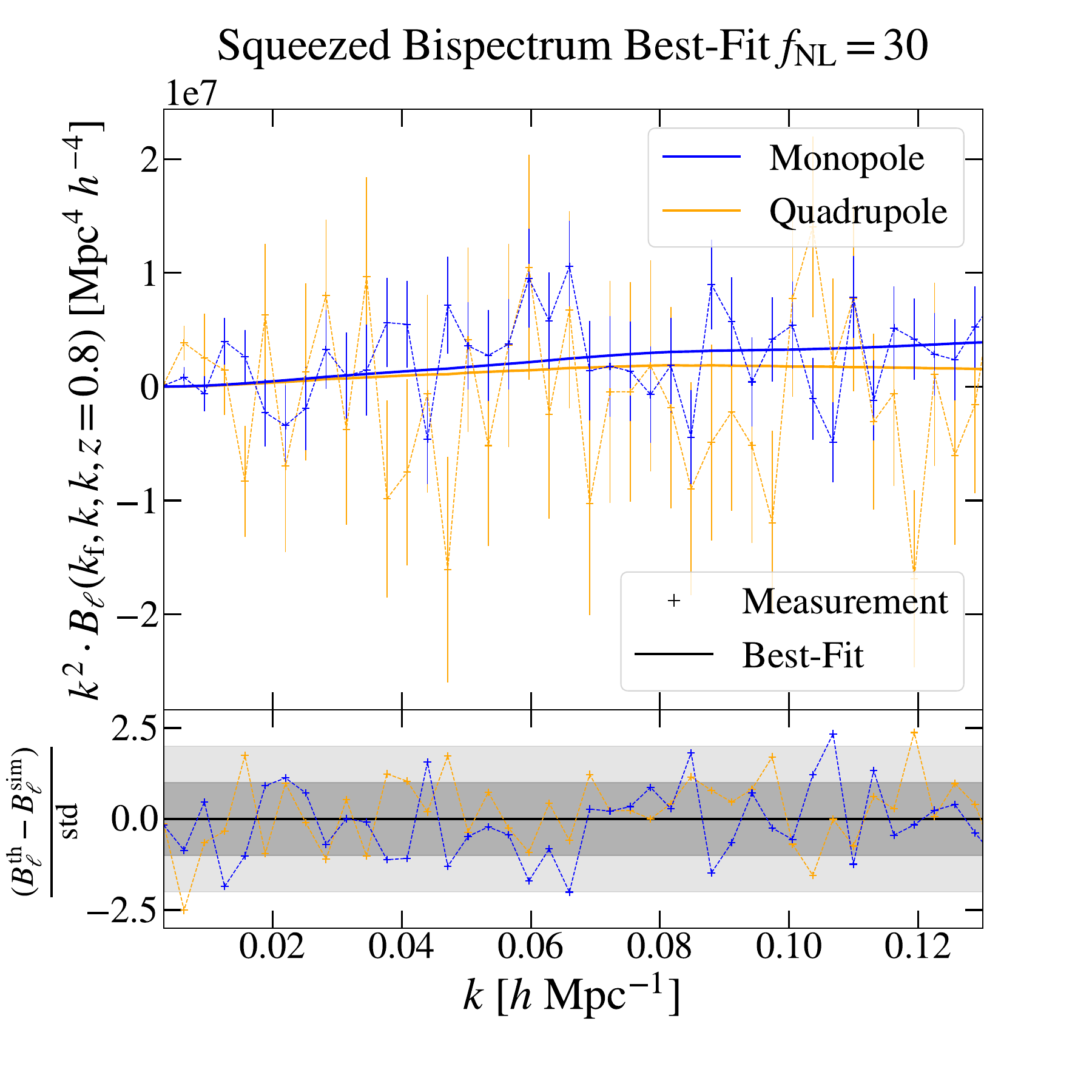}\vspace{-0.1in}
    \caption{Best-fit models from joint power spectrum and bispectrum analysis versus measured spectra at $z=0.8$. \textit{Left}: redshift-space power spectrum multipoles. \textit{Middle}: bispectrum multipoles in equilateral configurations. \textit{Right}: bispectrum multipoles in squeezed configurations. The error bars are determined using the theoretical Gaussian covariance. The bottom panels show the deviation of the model from the measurement, normalised to the measured standard deviation (std). Grey bands show the $1\sigma$ and $2\sigma$ uncertainties. For the bispectrum, we only show the monopole and quadrupole, since the hexadecapole is too noisy.\\}
    \label{fig:PB_Best_fit}
\end{figure*}

\begin{figure*}[htbp!]
    \centering
    \hspace{-0.2in}\includegraphics[width=0.83\columnwidth]{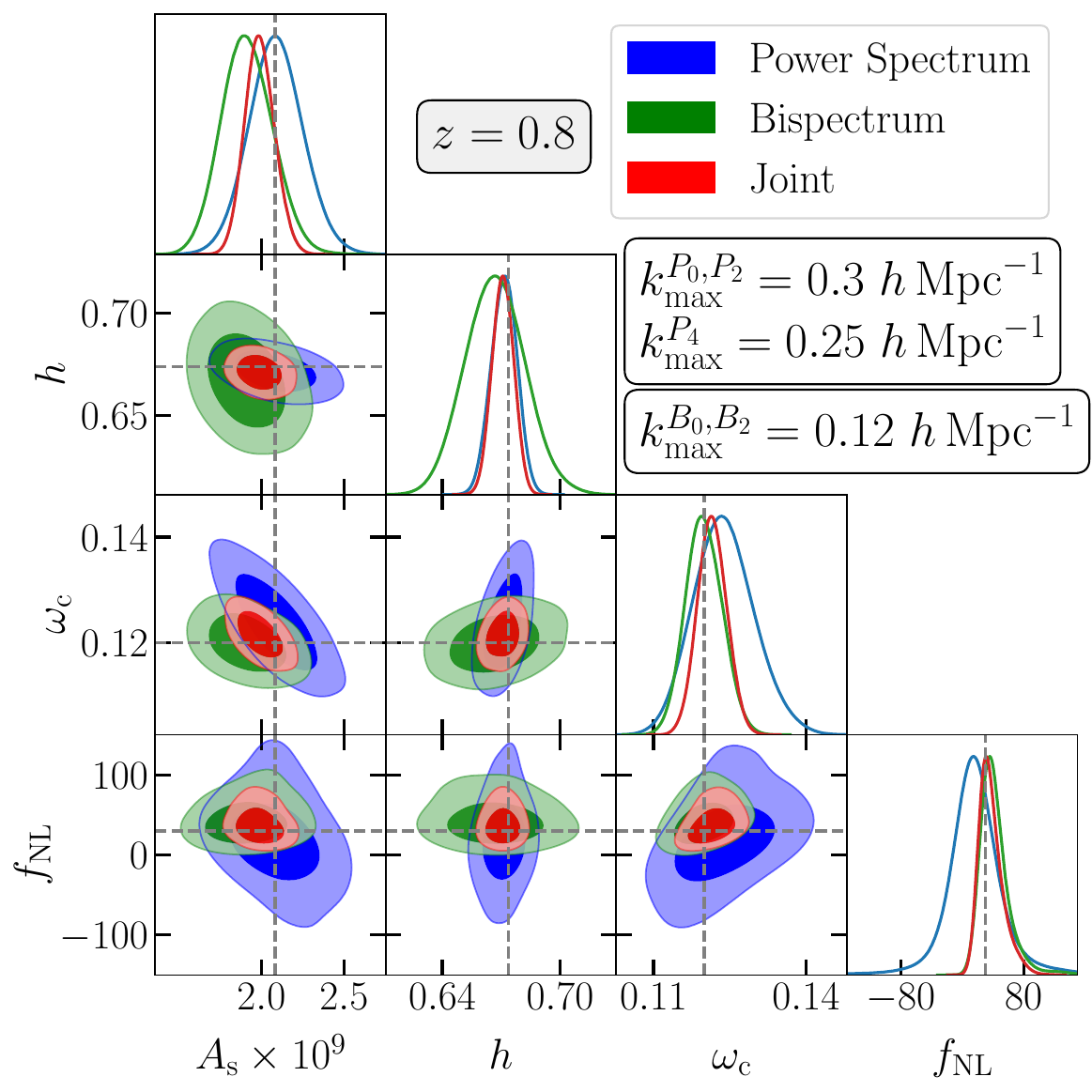}
    \hspace{0.2in}\includegraphics[width=0.83\columnwidth]{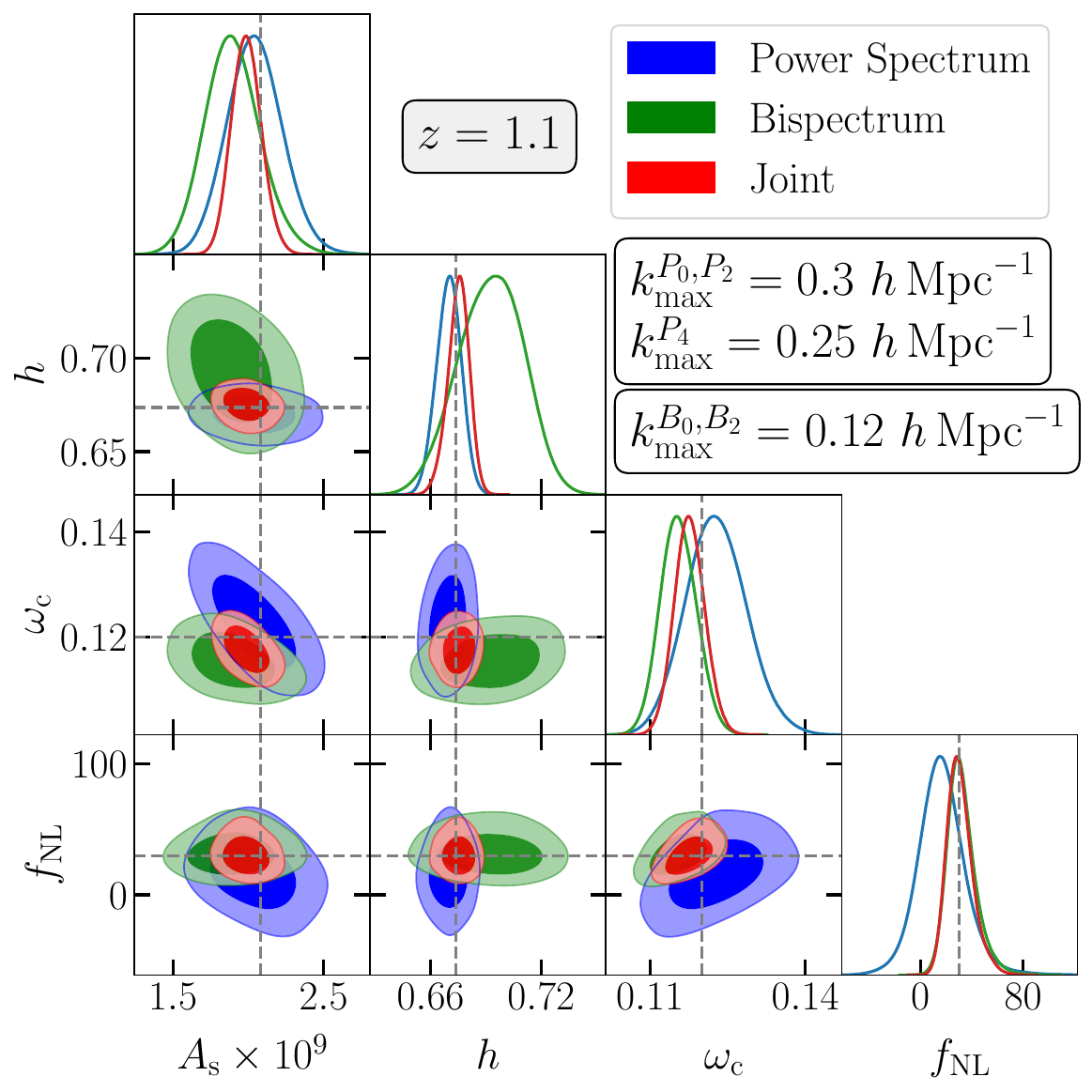}
    \hspace{-0.2in}\includegraphics[width=0.83\columnwidth]{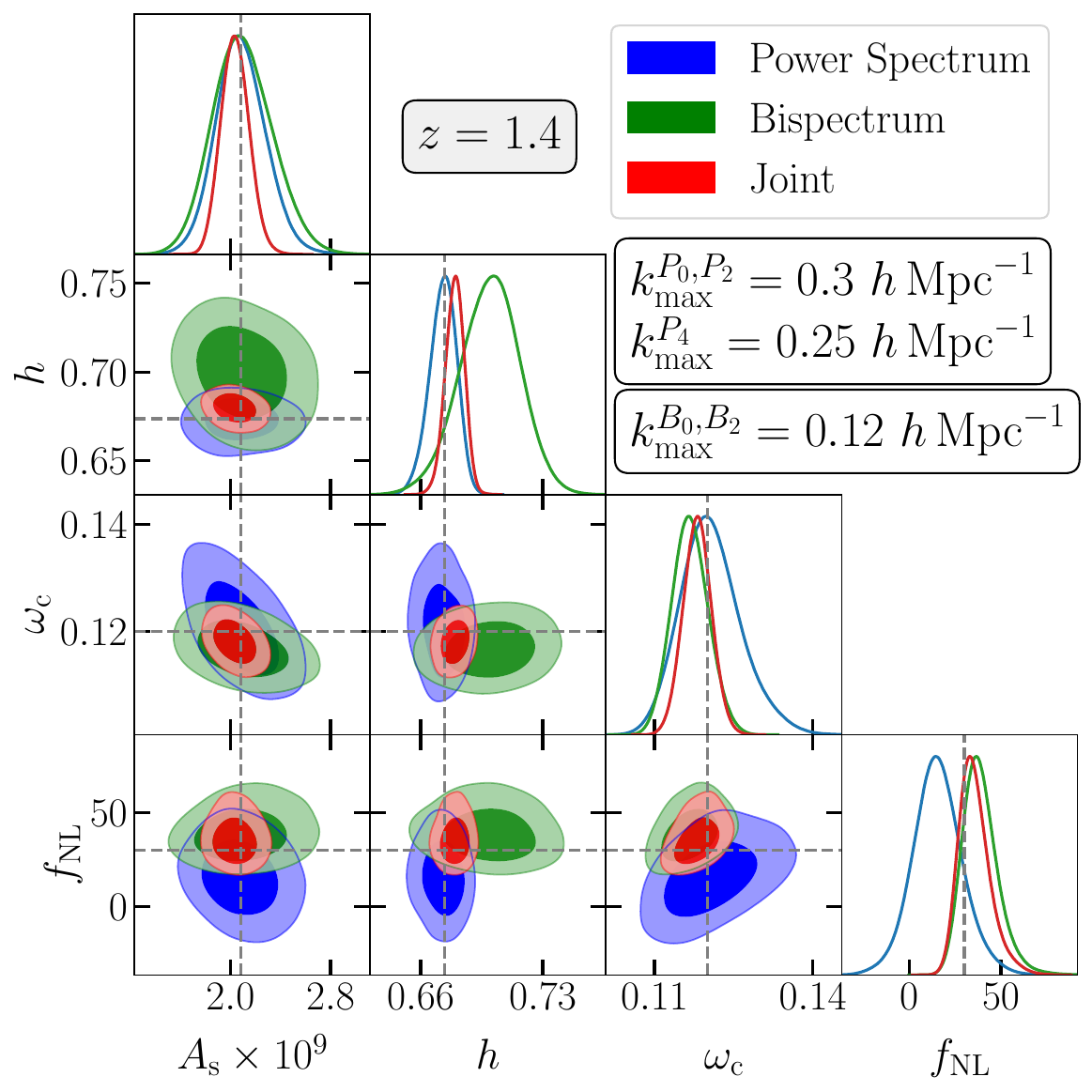}
    \hspace{0.2in}\includegraphics[width=0.83\columnwidth]{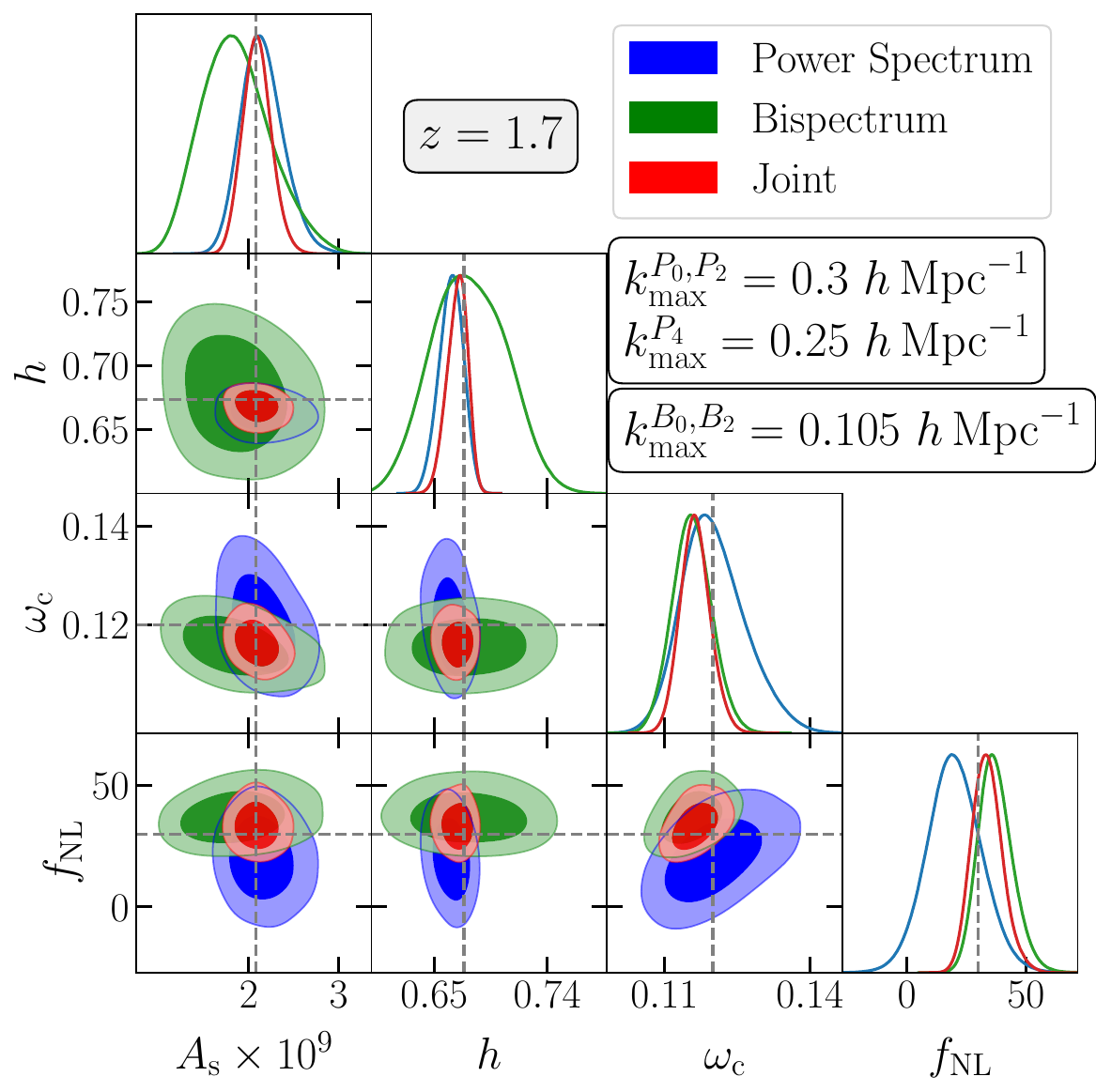}
    \caption{Joint analysis of power spectrum and bispectrum. Marginalised parameter posterior distributions from power spectrum (blue), bispectrum (green), and their combination (red). The dashed lines are fiducial values of simulations. The redshifts and scale cuts are noted on the plot.}
    \label{fig:Joint_Posteriors}
\end{figure*}

\begin{figure}[t]
    \centering
    \hspace{-0.15in}\includegraphics[width=0.92\columnwidth]{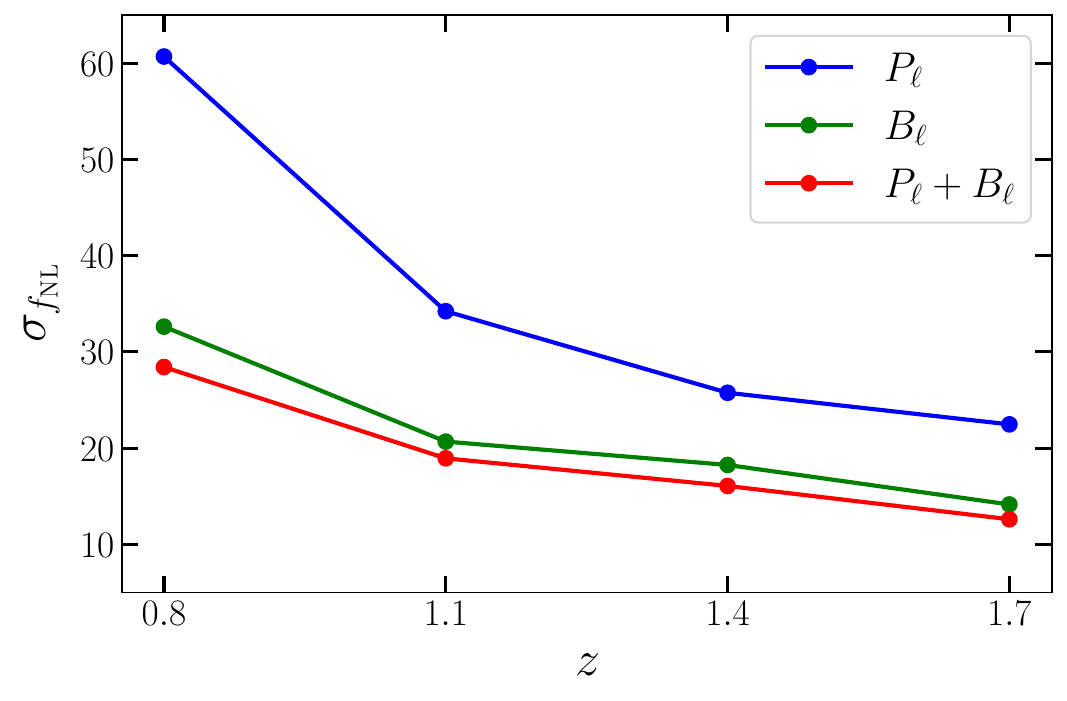}
    \caption{Comparison of constraints from power spectrum and bispectrum alone and combined. Shown are $1\sigma$ error on $\fnl$ for fiducial value of $\fnl=30$ from power spectrum (blue) and bispectrum (green) multipoles, and their combination (red) at four redshift bins considered.}
    \label{fig:Errors_Multipoles_Redshift}
\end{figure}

We now proceed to the joint analysis of the power spectrum and bispectrum multipoles at all four redshift snapshots, with the scale cuts determined above. 

Before comparing the marginalised constraints, as an example of our measurements and performance of the models, we show in Fig.~\ref{fig:PB_Best_fit} the theoretical predictions corresponding to the best-fit values of cosmological and nuisance parameters from the joint analysis in comparison with the measured multipoles at $z=0.8$. The error bars correspond to the standard deviation (std) of the mean estimated from the data. The bottom panels show the relative difference between the model and data, scaled with respect to the standard deviation. Overall, for both the power spectrum and bispectrum, the model agrees with the data within $2\sigma$, even though for the monopole, we observe a larger deviation for some of the $k$-bins on scales $k>0.2\,h\,\mathrm{Mpc}^{-1}$. 

Figure~\ref{fig:Joint_Posteriors} shows the $1$- and $2$-dimensional marginalised posteriors from the power spectrum (blue contours), bispectrum (green contours), and their combination (red contours). Across all four redshifts, the joint posteriors are consistently tighter than either statistic alone. For $\omega_{\mathrm{c}}$, the constraints are largely driven by the bispectrum; adding the power spectrum further reduces the uncertainties thanks to the different degeneracy directions of the two observables. For $\fnl$, the bispectrum yields unbiased constraints that are substantially tighter than those from the power spectrum; combining with $P$ further improves the error bars, though the gain is modest. By contrast, the bispectrum constraints on $h$ are much weaker than those from the power spectrum, and the joint analysis offers only a modest improvement over the power spectrum alone. Overall, the rotated degeneracy directions between the two observables most clearly illustrate their complementarity and the benefits of the joint analysis. Moreover, while the power spectrum and bispectrum alone can show mild bias in some cases, the combined fit is more robust and effectively mitigates prior-volume effects by simultaneously constraining parameters that are poorly determined by either statistic in isolation.

To compare the $\fnl$ constraints from the power spectrum, the bispectrum, and their combination across redshift, Figure~\ref{fig:Errors_Multipoles_Redshift} shows the $1\sigma$ errors on $\fnl$. The errors decrease with redshift for all cases, driven by the increase of $b_\phi$ with $z$ (enhancing the large-scale response and, within the bispectrum, by the growing relative importance of the PNG piece -- especially in $B_2$ -- compared to the Gaussian contribution). At fixed redshift, the bispectrum outperforms the power spectrum, reducing $\sigma_{f_{\rm NL}}$ by
$\sim 46\%$, $40\%$, $30\%$, and $37\%$ at $z\in\{0.8,1.1,1.4,1.7\}$, respectively. Combining $P_\ell$ and $B_\ell$ tightens further over $B_\ell$ alone by
$\sim 13\%$, $8\%$, $12\%$, and $11\%$ at the same redshifts, corresponding to improvements over $P_\ell$ of
$53\%$, $45\%$, $38\%$, and $44\%$. From $z=0.8$ to $z=1.7$ the overall precision improves by a factor of $\sim 2.7$ for $P_\ell$, $\sim 2.3$ for $B_\ell$, and $\sim 2.3$ for the joint case. These trends mirror Fig.~\ref{fig:Joint_Posteriors} and our scale-cut study: $B_\ell$ is the primary driver of the $\fnl$ constraints, with $P_\ell$ providing a non-negligible additional boost when combined.

Across the four redshifts, the joint analysis yields detections ranging from $\sim1.05\sigma$ at $z=0.8$ to $\sim2.35\sigma$ at $z=1.7$. The rise in significance with redshift follows directly from the same two effects noted above: increasing $b_\phi$ amplifies the large-scale response, and within the bispectrum, the PNG fraction of the signal (in particular, in the quadrupole) becomes comparatively larger at higher $z$. Consequently, $\fnl$ constraints from $P_\ell$, from $B_\ell$, and from their combination all tighten toward higher redshift.

\subsubsection{Information content of redshift-space multipoles}

\begin{figure*}[htbp!]
    \centering
    \includegraphics[width=\columnwidth]{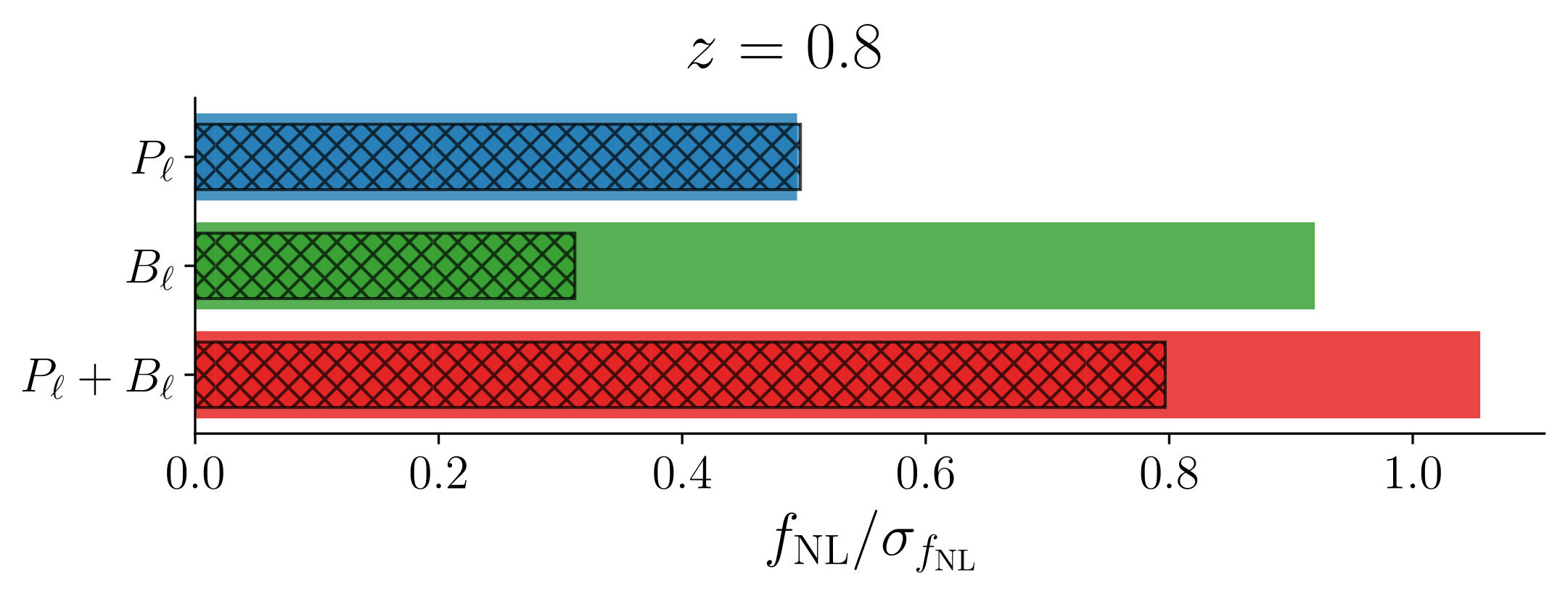}
    \includegraphics[width=\columnwidth]{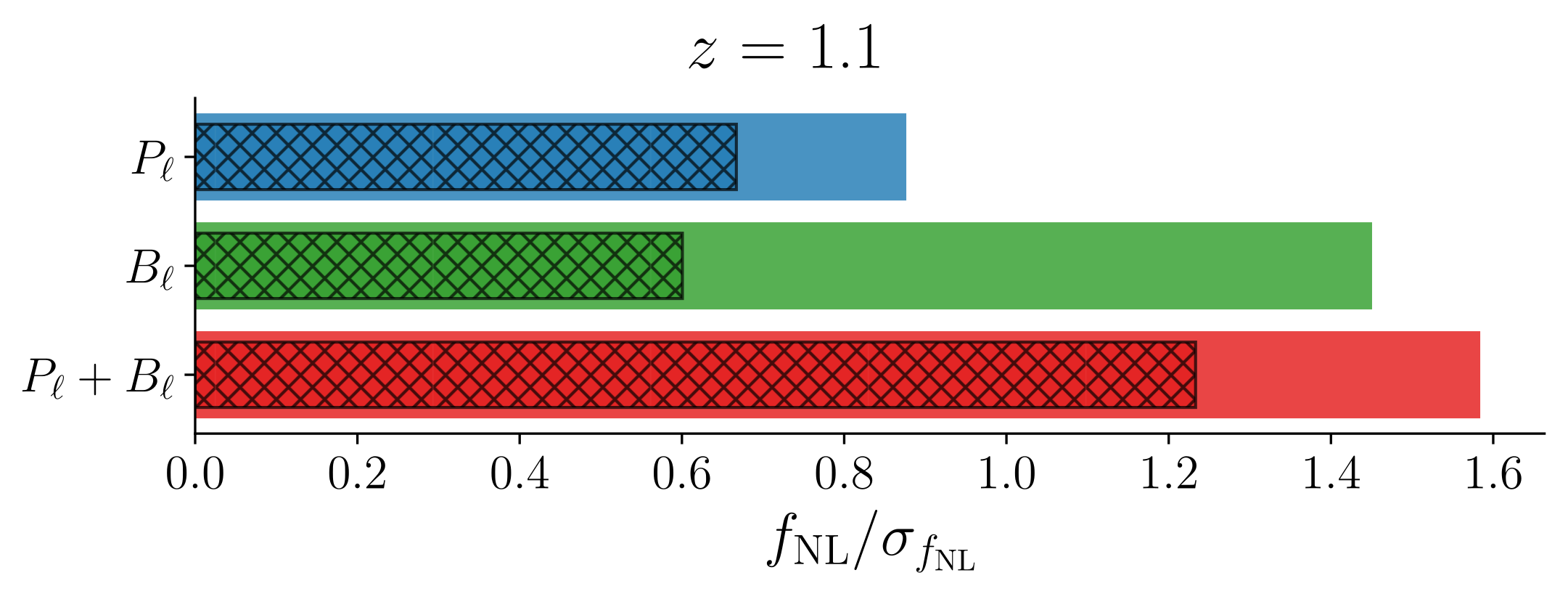}
    \includegraphics[width=\columnwidth]{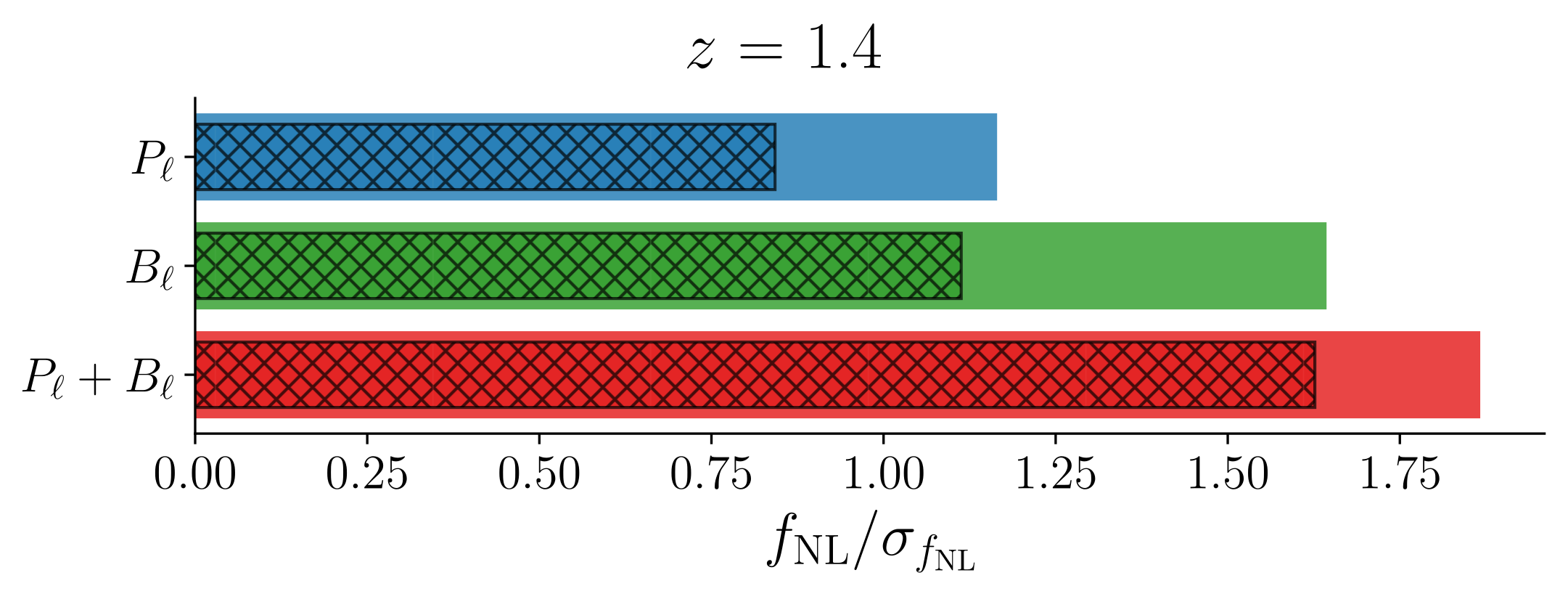}
    \includegraphics[width=\columnwidth]{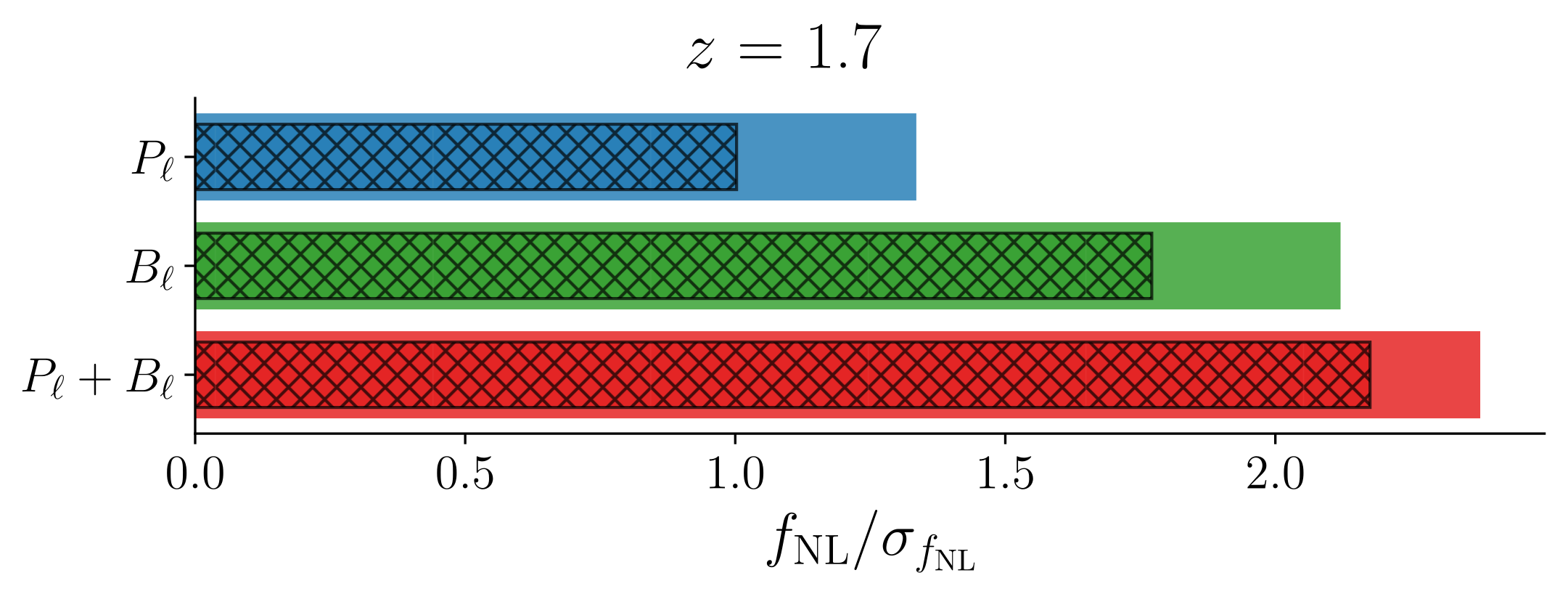}
    \caption{Information content of various multipoles on $\fnl$. The inverse of the relative error on $\fnl$ with respect to the fiducial value of $\fnl = 30$ for power spectrum multipoles (blue), bispectrum multipoles (green), and their combination (red). In each coloured bar, the crossed black bars show the constraints considering only the power spectrum and bispectrum monopoles and their combination. 
    }
    \label{fig:Constraining_Power_Obs}
\end{figure*}

In Fig.~\ref{fig:Constraining_Power_Obs}, we compare the information content of the power-spectrum multipoles (blue contours), bispectrum multipoles (green contours), and their combination (red contours) across the four redshift bins, quantified as $\fnl/\sigma_{\fnl}$. In each panel, the hatched bars show the constraints using only monopoles ($P_0$, $B_0$, and $P_0+B_0$), while the solid bars include all measured multipoles. Across all redshifts, there is a clear gain from adding higher multipoles. For the power spectrum, the improvement ranges from $\sim 0\%$ at $z=0.8$ up to $\sim 30$--$40\%$ for $z=1.1$--$1.7$, reflecting the additional RSD information in $P_2$ (with $P_4$ providing only a small incremental benefit). For the bispectrum, the quadrupole $B_2$ has a much larger relative impact: it enhances the $\fnl$ constraint by $\sim 20\%$ at $z=1.7$ and by up to a factor of $\sim 3$ at $z=0.8$, consistent with the stronger PNG sensitivity of $B_2$ and its distinct parameter degeneracies. The full joint power spectrum and bispectrum analysis then combines these gains.

These trends come with important caveats, which are illustrated in Figs.~\ref{fig:multipoles_zp8} and \ref{fig:multipoles_z1p7}. Here, we summarise the observations: $P_0$ and $B_0$ alone are weakly constraining for our scale cuts, leaving the familiar strong $A_{\mathrm{s}}$--$b_1$ degeneracy. This induces a visible bias in $A_{\mathrm{s}}$ in the monopole-only chains and propagates to $\fnl$: at lower redshifts, where $b_1$ is smaller, the $A_{\mathrm{s}}$--$b_1$ degeneracy makes it easier for $b_1$ to approach $p_{b_\phi}$, projecting $\fnl$ to larger values and artificially tightening the marginalised $\fnl$ error. Adding higher multipoles breaks these effects: $P_2$ largely collapses the $A_{\mathrm{s}}$--$b_1$ degeneracy, and $B_2$ rotates the $\fnl$--$\omega_{\mathrm{c}}$ and $\fnl$--$b_1$ directions. Quantitatively, comparing the full joint power spectrum and bispectrum multipoles ($P_\ell+B_\ell$) to a joint analysis without the bispectrum quadrupole ($P_\ell+B_0$), the $1\sigma$ uncertainty on $\fnl$ increases by $\sim 12\%$ at $z=1.7$ and by up to $\sim 25\%$ at lower redshifts, underscoring the importance of $B_2$.

Beyond $\fnl$, the multipoles carry complementary information about other parameters and galaxy-bias coefficients as shown in Figs.~\ref{fig:multipoles_zp8} and \ref{fig:multipoles_z1p7}. In $P$, $P_2$ is the workhorse: it stabilises $A_{\mathrm{s}}$ and $b_1$, tightens $\omega_{\mathrm{c}}$, and indirectly helps $\fnl$ through reduced $\omega_{\mathrm{c}}$ and $b_1$ degeneracies; $P_4$ adds only modest further shrinkage \citep{Gambardella:2023qrm}. In $B$, $B_0$ already constrains $\fnl$; however, $A_{\mathrm{s}}$ and $b_1$ are poorly constrained and biased, while nonlinear bias parameters ($b_2$, $b_{\mathcal G_2}$) are well-constrained. Adding $B_2$ sharpens all three biases and provides a noticeable extra pull on $\omega_{\mathrm{c}}$. It is worth noting the redshift-dependent trend that at the lowest redshift bin, adding $B_2$ considerably tightens $b_1$, while at the highest redshift bin, $b_1$ is left weakly constrained. The full $P_\ell+B_\ell$ combination yields the tightest and most isotropic posteriors: $P_2$ fixes the amplitude-bias sector, $B_\ell$ sets $\fnl$ and the higher-order/tidal biases, and the joint fit reduces prior-volume effects that appear in single-multipole or monopole-only analyses.

\section{Conclusions}\label{sec:conclusions}

Deciphering the origin of cosmic structure is a central open question in modern cosmology. A detection of local-type PNG -- or, conversely, a high-significance upper bound -- would dramatically sharpen our picture of the very early Universe: a robust non-zero detection of local $\fnl$ would confirm the violation of the single-field consistency relation and points to multi-field dynamics, while pushing $|\fnl|$ below $\mathcal O(1)$ severely restricts broad classes of multi-field models. In preparation for the analysis of the first data release of \Euclid, this work focuses on a detailed investigation of potential theoretical systematic effects in an idealised periodic-box setting with synthetic data, and on quantifying the constraining power of the redshift-space galaxy power spectrum and bispectrum -- highlighting, in particular, the gains from their joint analysis. 

We presented a comprehensive validation of redshift-space power-spectrum and bispectrum analyses, both separately and combined, aimed at constraining local PNG. We built \Euclid-like H$\alpha$ galaxy mocks by applying an HOD model calibrated to the Euclid FS2 simulation (representing \Euclid H$\alpha$ spectroscopic galaxy sample) to the Abacus-PNG simulations with local-type initial conditions and fiducial $f_{\rm NL}=30$. Considering four snapshots of the simulations matching the redshift bins of the \Euclid spectroscopic galaxy sample, we examined the performance of the perturbative model of the redshift-space power spectrum at one-loop order and the tree-level bispectrum, including contributions due to gravitational evolution, galaxy biasing, and primordial non-Gaussianity of the local type. 

A central element of our study is the treatment of PNG bias parameters that appear in the power spectrum and bispectrum models as products with $\fnl$. We assessed whether a prior-agnostic detection of non-zero PNG is feasible and showed that, while the bispectrum contains an $f_{\rm NL}$-only term at tree level, in practice, the dominant PNG response at our scales is $\propto f_{\rm NL} \, b_\phi$. Thus, informative choices for the $b_\phi$ prior are pivotal for robust $f_{\rm NL}$ inference. Furthermore, this implies that the considered data set does not have the sensitivity to constrain $b_{\phi \delta}$; therefore, its value can be safely set to the theoretical prediction (e.g., from UHMF). Guided by literature and internal tests, we adopted a physically motivated parametrisation $b_\phi=2\delta_{\mathrm{c}}\,(b_1-p_{b_\phi})$ and used a Gaussian prior on $p_{b_\phi}$ that brackets the commonly used UHMF ($p_{b_\phi}=1$) and GFM ($p_{b_\phi}=0.55$) prescriptions at the $1\sigma$ level. This prior is agnostic between those models, captures redshift trends through $b_1(z)$, and avoids projection effects that bias $f_{\rm NL}$ when very broad priors are used. We validated this choice a posteriori with SU measurements of $b_\phi$ for our mocks, finding consistency within the prior envelope at all redshifts.

Having demonstrated the ability to recover the fiducial $f_{\rm NL}=30$ in the simulations with PNG, we performed a null test on Gaussian-initial-condition mocks. With the same scale cuts and $p_{b_\phi}$ prior, we recovered $f_{\rm NL}=0$ within $1\sigma$ from $P_\ell$, $B_\ell$, and the combination of $P_\ell$ and $B_\ell$, and the inclusion of the PNG sector left the $\Lambda$CDM parameters essentially unchanged. This confirms that our pipeline and priors do not spuriously detect PNG and do not contaminate standard-parameter inference.

We then mapped the validity range of our perturbative model via scale-cut studies. In an effective volume of $V_{\rm eff} = 16\ h^{-3}\,{\rm Mpc}^3$, for the power spectrum, the one-loop $P_\ell$ model yields unbiased constraints (within $\sim1\sigma$) across all tested redshifts; increasing $k^{P_\ell}_{\rm max}$ chiefly improves $\omega_{\mathrm{c}}$, with $A_{\mathrm{s}}$ and $h$ being more stable and a negligible dependence of $f_{\rm NL}$ on $k^{P_\ell}_{\rm max}$, as expected for a large-scale PNG signal. For the bispectrum, tighter constraints with increasing $k^{B_\ell}_{\rm max}$ are eventually offset by tree-level model breakdown; we found $k^{B_\ell}_{\rm max}\sim0.12\,h\,\mathrm{Mpc}^{-1}$ (for $z\le1.4$) and $\sim0.105\,h\,\mathrm{Mpc}^{-1}$ (for $z=1.7$) to keep posteriors within $1\sigma$--$2\sigma$ of the truth, with $\omega_{\mathrm{c}}$ being the first parameter to show tension when pushing deeper. These choices maximise information while maintaining unbiased recovery according to our FoB and FoM diagnostics. Given that the large-scale systematic effects represent the most significant source of contamination for constraining local PNG, particularly through the scale-dependent bias signature, we also tested the impact of the removal of the largest-scale Fourier modes by varying $k_{\rm min}$. We found that an increase in $k_{\rm min}$ markedly degrades constraints on $\fnl$, and in the extreme case of excluding the first three $k$-bins, the recovered value of $\fnl$ is biased by more than $1\sigma$ in the power spectrum and more than $2\sigma$ in the bispectrum. 

Using these scale cuts, we compared constraints from $P_\ell$, $B_\ell$, and their combination. At fixed redshifts, the bispectrum outperforms the power spectrum by $29$--$46\%$, and the joint power spectrum and bispectrum analysis boosts the constraining power by an additional $\sim8$--$13\%$ while also improving constraints on $\omega_{\mathrm{c}}$ and stabilising $A_{\mathrm{s}}$ and $h$. Overall, with an effective volume of $V=16\,h^{-3} \, \mathrm{Gpc}^3$ (but using the largest-scale mode of half of this volume), the joint analysis yields $1\sigma$--$2.35\sigma$ detections of $f_{\rm NL}=30$ across $z\in\{0.8,1.1,1.4,1.7\}$, with significance increasing toward higher redshift as the PNG response strengthens. Finally, by comparing multipole subsets, we showed that redshift-space information (especially the bispectrum quadrupole) is crucial: it breaks key degeneracies (e.g., among $A_{\mathrm{s}}$, $b_1$, and $f_{\rm NL}$) that limit monopole-only analyses, delivering tighter and more robust constraints. It is important to note that we considered the constraints obtained if the four redshift bins are analysed individually. Combining the information across the whole redshift range, corresponding to the full volume of \Euclid, will further tighten error bars. 

This work is intended as a thorough study of key theoretical systematics affecting PNG inference, and the constraints reported here should not be interpreted as forecasts for the full Euclid Survey. Our analysis does not exploit the entire survey volume, which is the primary reason our bounds are conservative. Substantial improvements can be achieved by using the full volume and implementing optimal redshift weighting rather than hard redshift binning \citep{eBOSS:2019sma}, as well as by jointly analysing photometric and spectroscopic samples -- leveraging the photometric sample’s sensitivity to ultra-large scales (via its higher number density) and the spectroscopic sample’s bispectrum shape information over a broader range of scales \citep[see, e.g.,][]{SPHEREx:2014bgr}. Further gains are expected from cross-correlations with external tracers (e.g., CMB lensing: \citealp{DESI:2023duv,Bermejo-Climent:2024bcb,Fabbian:2025fdk}, and reconstructed velocities from the kinetic Sunyaev--Zeldovich effect: \citealp{Lague:2024czc,Hotinli:2025tul}), and more generally from multi-tracer analyses \citep{Seljak:2008xr} -- including near–zero-bias tracers \citep{Castorina:2018zfk} -- to suppress sample variance and tighten PNG constraints.

Beyond the aspects studied here (the validity range of perturbative models, priors on PNG biases, and the information content of the observables), robust and precise $f_{\rm NL}$ constraints from Stage-IV spectroscopic surveys will require further modelling advances, including extending the bispectrum beyond tree level, accounting for general-relativistic projection effects, incorporating non-Gaussian covariance (particularly in the squeezed limit), and moving from periodic boxes to realistic survey geometries with full observational systematics.

To prepare specifically for upcoming \Euclid data, our immediate next steps focus on observational systematics and survey geometry. Unlike DESI -- where imaging-driven target-selection systematics dominate the largest modes \citep[see, e.g.,][]{Chaussidon:2024qni} -- \Euclid\/’s slitless spectroscopy largely sidesteps that issue but introduces its own large-scale risks: slowly varying completeness and contamination fields (from spectral overlap or zodiacal background) and line-identification/redshift-failure systematics that can imprint broad angular and radial modes. To quantify the impact of these \Euclid-specific effects, we will utilise the \Euclid lightcone mocks, which incorporate the relevant observational systematics and survey mask, to test their impact on PNG constraints from joint power-spectrum–bispectrum analyses across the full survey volume.

\begin{acknowledgements} 
\AckEC
We thank Giovanni Cabass, Marko Simonovic, and Zvonomir Vlah for helpful discussions. The research of AMD and MC is supported by the {\it Agence Nationale de la Recherche} (ANR), grant ANR-23-CPJ1-0160-01. The development of the code and the validation tests were carried out on the Baobab Cluster of the University of Geneva and on the High Performance Computing facility of the University of Parma, Italy. For running MCMC chains, this project was provided with computing and storage resources by GENCI at IDRIS, thanks to grant 2025-AD010416529 on the supercomputer Jean Zay's CSL partition. This work has made use of CosmoHub, developed by PIC (maintained by IFAE and CIEMAT) in collaboration with ICE-CSIC. It received funding from the Spanish government (MCIN/AEI/10.13039/501100011033), the EU NextGeneration/PRTR (PRTR-C17.I1), and the Generalitat de Catalunya.
\end{acknowledgements}

%
%

\bibliography{main.bib}
%

\appendix

\section{Theory}\label{app:theory}
\subsection{Infrared resummation}\label{app:IR}

Large-scale bulk flows displace matter and galaxies by ${\sim}10\,{\rm Mpc}$, which -- while locally unobservable by the equivalence principle \citep{Peloso:2013zw,Kehagias:2013yd,Creminelli:2013mca} -- smooth the oscillatory BAO feature with little impact on the broadband. Fixed-order standard (Eulerian) perturbation theory (SPT) treats these displacements perturbatively \citep{Eisenstein:2006nj,Crocce:2007dt,Sugiyama:2013gza} and, given their size, underestimates the BAO damping. In redshift space, in addition to resumming the long-wavelength displacements, one should also account for the LoS displacement induced by long-wavelength velocities through the RSD mapping (i.e., the anisotropic bulk-flow contribution to BAO damping). The damping can be reliably modelled by resumming the long-wavelength bulk flows in LPT \citep{Matsubara:2007wj,Carlson:2012bu,Porto:2013qua,Vlah:2015sea}, in SPT-based resummations \citep{Baldauf:2015xfa,Vlah:2015zda}, and in hybrid LPT-SPT/EFTofLSS schemes \citep{Senatore:2014via,Senatore:2017pbn,Lewandowski:2018ywf}, or via time-sliced perturbation theory (TSPT) \citep{Blas:2016sfa,Ivanov:2018gjr}. 

We follow the effective implementation of IR resummation based on TSPT, in which the linear power spectrum is split into smooth $P_{\text{nw}}$ and wiggly parts $P_{\text{w}}$, 
\be
P_0(k)= P_{\text{nw}}(k)+P_{\text{w}}(k)\,,
\ee
and the leading-order IR resummation is modified by applying a damping factor on the wiggly part, 
\be\label{eq:PO_real}
P_{\text{LO}}(k) \equiv P_{\text{nw}}(k)+\e^{-k^2\,\Sigma^2}P_{\text{w}}(k)\,.
\ee
In real space, the damping is isotropic, with
\be
\Sigma^2=\frac{4\pi}{3}\int_0^{k_{\mathrm{s}}}\diff q\,P_{\rm nw}(q)\,\left[1-j_0\,\left(\frac{q}{k_{\rm osc}}\right)+2\,j_2\,\left(\frac{q}{k_{\rm osc}}\right)\right]\,,
\ee
where $k_{\rm osc}\sim(110\,h^{-1}\,{\rm Mpc})^{-1}$, $k_{\mathrm{s}}$ is the separation scale for the resummed modes, and $j_n$ are spherical Bessel functions. In principle, $k_{\mathrm{s}}$ is arbitrary, and any residual dependence is a theoretical error. At next-to-leading order we use Eq.~\eqref{eq:PO_real} inside the one-loop integrals,
\begin{align}
P_{\rm NLO}(k)=&\;P_{\rm nw}(k)+\e^{-k^2\,\Sigma^2}P_{\rm w}(k)\,\left(1+k^2\,\Sigma^2\right)\nn \\
&+P_{\rm 1\text{-}loop} \left[P_0(k) \rightarrow P_{\rm nw}(k)+\e^{-k^2\,\Sigma^2}\,P_{\rm w}(k)\right]\,,
\end{align}
treating $P_{\rm 1\text{-}loop}$ as a functional of the linear spectrum. In redshift space, the damping exponent becomes angle-dependent,
\be
\Sigma_{ s}^2(\mu)=\left[1+f\,(f+2)\,\mu^2\right]\,\Sigma^2,
\ee
so that the IR-resummed one-loop power spectrum reads
\begin{align}
P^s_{\rm NLO}(k,\mu) =&\;P^s_{\rm nw}(k,\mu)+\e^{-k^2\,\Sigma_{s}^2(\mu)}\,P_{\rm w}(k,\mu)\left(1+k^2\,\Sigma_{s}^2(\mu)\right) \nn\\
&+P^s_{\rm 1\text{-}loop}\Bigl[P_0(k,\mu) \rightarrow P_{\rm nw}(k,\mu) \nn \\
&+\e^{-k^2\,\Sigma_{s}^2(\mu)} \, P_{\rm w}(k,\mu)\Bigr]\,,
\end{align}
with $P_0(k,\mu)=\left(1+f\,\mu^2\right)^2 P_0(k)$ for matter, and in the loop calculation we perform the usual replacement of the linear power spectrum with the leading order IR-resummed power spectrum \citep[see, e.g.,][]{Chen2020JCAP}.

For the bispectrum, we implement IR resummation by the straightforward substitution $P_0\rightarrow P_{\rm LO}$ in both the tree-level expression and the corresponding stochastic term \citep{IvanovEtal2022B}. Over the scales considered, IR resummation typically has a smaller quantitative impact on the bispectrum than on the power spectrum \citep{Alkhanishvili:2021pvy}.

\subsection{Alcock--Paczynski effect}\label{app:AP}
Mapping observed angles and redshifts to $3$-dimensional positions requires adopting a fiducial cosmology. If the fiducial and true cosmologies differ, the inferred clustering is geometrically distorted and must be accounted for in the theoretical model of the power spectrum and bispectrum -- the Alcock--Paczynski effect \citep{Alcock:1979mp}. We parametrise these distortions with the standard dilation factors
\be
\alpha_{\parallel}(z)\equiv \frac{H_{\rm true}(z)}{H_{\rm fid}(z)}\,,
\qquad
\alpha_{\perp}(z)\equiv \frac{D_{{\rm A, fid}}(z)}{D_{{\rm A, true}}(z)}\,,
\ee
so that wavenumbers measured in the fiducial cosmology, $(k,\mu)$, map to the underlying true coordinates, $(q,\nu)$, as
\begin{gather}
q = k \,\sqrt{\alpha_{\parallel}^{-2}\,\mu^2+\alpha_{\perp}^{-2}\,(1-\mu^2)}\,,\label{eq:APq}\\ 
\nu = \frac{\mu}{\alpha_{\parallel} \,\sqrt{\alpha_{\parallel}^{-2}\,\mu^2+\alpha_{\perp}^{-2}\,(1-\mu^2)}}\,. \label{eq:APnu}
\end{gather}
The Jacobian of this transformation yields the familiar overall factor $(\alpha_{\parallel}\,\alpha_{\perp}^2)^{-1}$ for the power spectrum and $(\alpha_{\parallel}^2\,\alpha_{\perp}^4)^{-1}$ for the bispectrum. The corresponding multipoles in the fiducial coordinates are then
\begin{align}
    P_\ell(k) =&\, \frac{2\ell + 1}{2 \,\alpha_{\parallel}\, \alpha_{\perp}^2}\,\int_{-1}^1\diff\mu \, \cL_\ell(\mu) \, P_{\mathrm{g}}\left(q[k,\mu],\nu[\mu]\right),  \\   
    B_\ell(k_1,k_2,k_3) =&\, \frac{2\ell + 1}{4\pi \, \alpha_{\parallel}^2 \, \alpha_{\perp}^4}\,\int_{0}^{2\pi}\diff\phi \,\int_{-1}^1\diff\mu_3 \, \cL_\ell(\mu_3) \nn \\  
    & \times  B_{\mathrm{g}}\left(q_1,q_2,q_3,\nu_1,\nu_2, \nu_3\right).
\end{align}
where $(q_i,\nu_i)$ are obtained from $(k_i,\mu_i)$ through the same AP mapping (Eqs.~\ref{eq:APq} and \ref{eq:APnu}). For the bispectrum, $\mu_1$ and $\mu_2$ are functions of $(\mu_3,\phi;k_1,k_2,k_3)$ determined by triangle geometry (law of cosines) and the line-of-sight projection.

\section{Measurements}\label{app:measurment-all}
\subsection{Estimators}\label{app:measurment}

For power spectrum and bispectrum measurements, we use \texttt{PBI4},\footnote{\url{https://github.com/matteobiagetti/pbi4}} an FFT-based Python package that upgrades the Fortran code \texttt{PowerI4}.\footnote{\url{https://github.com/sefusatti/PowerI4}} It computes the power spectrum and bispectrum for particle catalogues in periodic boxes, in both real and redshift space, implements grid assignment up to the fourth order, and employs interlacing to suppress aliasing from density-field sampling \citep{Sefusatti:2015aex}. We turn on interlacing to minimise aliasing at large wavenumbers. We use a grid of $512$ for the power spectrum and of $256$ for the bispectrum, yielding a Nyquist frequency of $\sim 0.8 \ h \ \mathrm{Mpc}^{-1}$ and $\sim 0.4 \ h \ \mathrm{Mpc}^{-1}$, respectively. These values are far beyond the maximum scale we will use in the analysis that follows.

The estimators in redshift space are specified as 
\begin{align}
&\hat{P}_\ell(k)=\frac{(2\ell+1)}{V\,N_k}\,\sum_{\bq \in k}|\delta_{s}(\bq)|^2\, \cL_\ell(\mu) \, , \\
&\hat{B}_\ell(k_1,k_2,k_3)=\frac{(2\ell+1)\,k_{\rm f}^6}{N_{\mathrm{B}}} \nn \\
&\hspace{0.3in}\times\sum_{\bq_1 \in k_1}\sum_{\bq_2 \in k_2}\sum_{\bq_3 \in k_3}\,\delta_{s}(\bq_1)\,\delta_{s}(\bq_2)\,\delta_{s}(\bq_3)\,\cL_\ell(\mu_1)\,\delta_{\mathrm{K}}(\bq_{123}) \, ,
\end{align}
where $V=L^3=\left(2\pi \,k_{\rm f}^{-1}\right)^3$ represents the volume of the simulation. The sum over $\bq \in k$ includes all wavevectors $\bq$ falling in the $k$-bin of size $\Delta k=k_{\rm f}$, with $|\bq|\in[k-\Delta k/2,\,k+\Delta k/2]$. The parameters $N_k$ and $N_{\mathrm{B}}$ denote the counts of wavenumbers and triangles, respectively, within a specified bin, $\delta_{\rm K}$ is the Kronecker delta, and $\bq_{123} = \bq_1+\bq_2+\bq_3$. The term $\delta_{s}$ represents the displaced density fields along the LoS, which is chosen to align with the $\hat z$-axis without loss of generality.

\subsection{Binning effects}\label{app:binnig}

Rather than evaluating the forward model at every mode/triangle inside a bin and then averaging, we evaluate the theory once per bin at an {\it effective} point that reproduces the bin average. For the power spectrum, we use the effective wavenumber
\be
k_{\rm eff} \equiv \frac{1}{N_k}\sum_{\bq \in k}\bq\,,
\ee
while for the bispectrum the effective triangle $(k_{1,\rm eff},k_{2,\rm eff},k_{3,\rm eff})$ is defined following \citet{Oddo:2021iwq} as the average over all triangles in the bin, 
\begin{gather}
    k_{1,\rm eff}(k_1,k_2,k_3)\equiv \frac{1}{N_{\mathrm{B}}}\sum_{\bq_1 \in k_1}\sum_{\bq_2 \in k_2}\sum_{\bq_3 \in k_3}\delta_{\mathrm{K}}(\bq_{123})\;{\rm max}(q_1,q_2,q_3) \,, \\
    k_{2,\rm eff}(k_1,k_2,k_3)\equiv \frac{1}{N_{\mathrm{B}}}\sum_{\bq_1 \in k_1}\sum_{\bq_2 \in k_2}\sum_{\bq_3 \in k_3}\delta_{\mathrm{K}}(\bq_{123})\;{\rm med}(q_1,q_2,q_3)\,, \\
    k_{3,\rm eff}(k_1,k_2,k_3)\equiv \frac{1}{N_{\mathrm{B}}}\sum_{\bq_1 \in k_1}\sum_{\bq_2 \in k_2}\sum_{\bq_3 \in k_3}\delta_{\mathrm{K}}(\bq_{123})\;{\rm min}(q_1,q_2,q_3)\,.
\end{gather}

\section{Inference}\label{app:inference}
\subsection{Analytic covariance matrices}\label{app:cov}

For the covariance of the power spectrum (between the $i$-th and $j$-th $k$-bins) and the bispectrum (between the $i$-th and $j$-th triangles), we use \citep{Grieb:2015bia,Sugiyama:2019ike,Rizzo:2022lmh,Ivanov:2023qzb}\footnote{If the theoretical Gaussian covariance accurately describes the covariance of a single realisation, then the covariance of the sample mean over $N_{\mathrm{R}}$ independent realisations is $\tens{C}\,N_{\mathrm{R}}^{-1}$. By contrast, when estimating the full (non-diagonal) covariance from $N_{\mathrm{R}}$ mocks, the unbiased sample-covariance estimator uses the Bessel correction, i.e., a normalisation $(N_{\mathrm{R}}-1)^{-1}$ rather than $N_{\mathrm{R}}^{-1}$.}
\begin{align}
C_{\mathrm{G},ij}^{P_{\ell_1}, P_{\ell_2}} =&\;\frac{2 \;\delta_{ij}}{N_{\mathrm{R}}\,N_{\mathrm{P}}(k_i)}(2\ell_1 +1)\,(2\ell_2 +1) \notag \\
&\times \int_{-1}^1 \diff\mu \ \cL_{\ell_1}(\mu) \, \cL_{\ell_2}(\mu) \, \left[P_{\mathrm{g}}(k,\mu)\right]^2 \, , \\
C_{\mathrm{G},ij}^{B_{\ell_1}, B_{\ell_2}}  =&\;  \frac{s_{\mathrm{B}}\,\;\delta_{ij}}{N_{\mathrm{R}}\,N_{\mathrm{B}}([k_1,k_2,k_3]_i)\,k_{\rm f}^3} (2\ell_1 +1)\,(2\ell_2 +1) \nn\\ 
& \times \int_{-1}^1 \diff\mu_1 \int_{-1}^1 \diff\mu_2 \int_0^{2\pi} \diff \varphi_2 \, \cL_{\ell_1}(\mu_1) \, \cL_{\ell_1}(\mu_2) \notag \\
&\times P_{\mathrm{g}}(k_1,\mu_1) \, P_{\mathrm{g}}(k_2,\mu_2) \, P_{\mathrm{g}}(k_3,\mu_3) + 5 \, {\rm perms}. \, ,
\end{align}
where $\delta_{ij}$ is the $2$-dimensional Kronecker delta, $s_{\mathrm{B}}$ is a symmetry factor defined such that $s_{\mathrm{B}} = 6, 2, 1$ for equilateral, isosceles, and general triangles, respectively, and $N_{\mathrm{P}}$ and $N_{\mathrm{B}}$ correspond to the number of modes and triangles in that particular bin, which in thin-shell approximation are given by
\begin{align}
    N_{\mathrm{P}}(k_i)&\simeq 4\pi\,k_i^2\,\Delta k \, k_{\rm f}^{-3}\,,\\
    N_{\mathrm{B}}([k_1, k_2, k_3]_i) &\simeq 8 \pi^2 k_1 \, k_2  \, k_3 ~\Delta k^3 \,k_{\rm f}^{-6}\,.
\end{align}
While the thin-shell approximation typically provides precision at the per cent level and is particularly accurate for smaller scales, we employ the exact mode and triangle counts used in the measurements themselves.

Since we use the measured multipoles to build the covariance matrices, we rewrite the Gaussian covariances directly in multipole space. For the power spectrum, the derivation follows \citet{Grieb:2015bia},\footnote{We correct a typo in the final expression of \citet{Grieb:2015bia}: a factor of $2\,(2\ell+1)$ was missing.} and for the bispectrum we adopt \citet{Rizzo:2022lmh}. We quote the final results below,
\begin{align}
&C_{\mathrm{G},ij}^{P_{\ell_1}, P_{\ell_2}} =  \frac{2k_{\rm f}^3\,\delta_{ij} \, (2\ell_1 +1)\,(2\ell_2 +1)}{N_{\mathrm{R}}\,N_{\mathrm{P}}(k_i)}\nn \\
&\hspace{0.18in} \times\sum_{\ell_3=0}^{\infty}\sum_{\ell_4=0}^{\ell_3}\,\left[P_{\ell_4}(k_i)+\frac{1}{\bar n}\,\delta_{\ell_4,0}\right]\,\left[P_{\ell_4}(k)+\frac{1}{\bar n}\,\delta_{\ell_4,0}\right]
 \nn \\
&\hspace{0.18in} \times \sum_{\ell={\rm max}(|\ell_1-\ell_2|,|2\ell_4-\ell_3|)}^{{\rm min}(\ell_1+\ell_2,\ell_3)}(2 \ell +1) \,\tj{\ell_1}{\ell_2}{\ell}{0}{0}{0}^2 \tj{\ell_4}{\ell_3-\ell_4}{\ell}{0}{0}{0}^2\,,
\end{align}
\begin{align}
    &C_{\mathrm{G},ij}^{B_{\ell_1}, B_{\ell_2}} = \delta_{ij}\,\frac{(2\ell_1+1)\,(2\ell_2+1)}{N_{\mathrm{B}, i}\,k_{\rm f}^3} \nn \\
    &\hspace{0.25in}\times\sum_{\ell_3, \ell_4, \ell_5} \sum_{\bq_1 \in k_1} \sum_{\bq_2 \in k_2} \sum_{\bq_3 \in k_3} 
    \delta_{\mathrm{K}}(\bq_{123}) \, P_{\ell_3}(q_1) \, P_{\ell_4}(q_2) \, P_{\ell_5}(q_3) \nn \\
    &\hspace{0.25in}\times \cL_{\ell_3}(\mu_1) \, \cL_{\ell_4}(\mu_2) \, \cL_{\ell_5}(\mu_3)\,\Bigg[ \Big(1+\delta_{\mathrm{K}}(\bk_{23})\Big) \, \cL_{\ell_1}(\mu_1) \, \cL_{\ell_2}(-\mu_1) \nn \\
    &\hspace{0.25in} + \Big(\delta_{\mathrm{K}}(\bk_{12}) + \delta_{\mathrm{K}}(\bk_{12}) \, \delta_{\mathrm{K}}(\bk_{23})\Big) \, \cL_{\ell_1}(\mu_1) \, \cL_{\ell_2}(-\mu_2) \nn \\ 
    &\hspace{0.25in} + \Big(\delta_{\mathrm{K}}(\bk_{13}) + \delta_{\mathrm{K}}(\bk_{12}) \, \delta_{\mathrm{K}}(\bk_{23})\Big) \, \cL_{\ell_1}(\mu_1) \, \cL_{\ell_2}(-\mu_3) \Bigg] \, ,
\end{align}
where we substituted $N_{\mathrm{B}}^2([k_1, k_2, k_3]_i)$ with $N_{\mathrm{B}, i}$ for brevity. 

By separating the bin-averaging quantities and applying the thin-shell approximation, we obtain
\begin{align}
    C_{\mathrm{G},ij}^{B_{\ell_1}, B_{\ell_2}} \equiv &\; \delta_{ij}\frac{(2\ell_1+1)\,(2\ell_2+1)}{N_{\mathrm{B}, i}\,k_{\rm f}^3} \nn \\
    &\times \sum_{\ell_3, \ell_4, \ell_5} P_{\ell_3}(k_1) \, P_{\ell_4}(k_2) \, P_{\ell_5}(k_3) \, R_{\ell_1,\ell_2;\ell_3,\ell_4,\ell_5}(k_1,k_2,k_3)\,,
\end{align}
where $R_{\ell_1,\ell_2;\ell_3,\ell_4,\ell_5}(k_1,k_2,k_3)$ is defined in Eq. (4.6) of \citet{Rizzo:2022lmh}, and can be further simplified within the continuous limit\footnote{Where we are replacing the bin-averaging of the triangles and the corresponding angles by a continuous integral over the shells 
\begin{gather*} 
\frac{1}{N_{\mathrm{B}}} \sum_{\bq_1 \in k_1} \sum_{\bq_2 \in k_2} \sum_{\bq_3 \in k_3} 
    \delta_{\mathrm{K}}(\bq_{123}) \\ \simeq \frac{1}{N_{\mathrm{B}}\,k_{\rm f}^6}\int_{k_1}\int_{k_2}\int_{k_3} \diff^3q_1\,\diff^3q_2\,\diff^3q_3 \;\delta_{\mathrm{D}}(\bq_{123}) = \frac{1}{4\pi}\int_{0}^{2\pi}\diff\varphi_1\int_{-1}^1 \diff\mu_1\,.
\end{gather*}} 
of the bin-average to the following expression

\begin{align} 
    &R_{\ell_1,\ell_2;\ell_3,\ell_4,\ell_5}(k_1,k_2,k_3) \simeq \frac{1}{512} \int_0^{2\pi}\diff\varphi_1 \int_{-1}^1 \diff \mu_1 \nn \\
    &\hspace{0.05in}\Bigg\{ \left[8P_0(k_1)+4\,\left(3\mu_1^2-1\right)\,P_2(k_1) + \left(3 - 30 \mu_1^2+35\mu_1^4\right) \, P_4(k_1) \right]\nn\\ 
    &\hspace{0.05in}\times  \left[8P_0(k_2)+4\,\left(3\mu_2^2-1\right)\,P_2(k_2) + \left(3 - 30 \mu_2^2+35\mu_2^4\right) \, P_4(k_2) \right] \Bigg\}\nn\\ 
    &\hspace{0.05in} \times \Bigg\{ \left[1+\delta_{\mathrm{K}}(\bk_{23})\right]\,\cL_{\ell_1}(\mu_1) \, \cL_{\ell_2}(-\mu_1)\nn\\
    &\hspace{0.05in} + \left[\delta_{\mathrm{K}}(\bk_{12}) + \delta_{\mathrm{K}}(\bk_{12}) \, \delta_{\mathrm{K}}(\bk_{23})\right] \, \cL_{\ell_1}(\mu_1) \, \cL_{\ell_2}(-\mu_2) \nn \\ 
    &\hspace{0.05in} + \left[\delta_{\mathrm{K}}(\bk_{13}) + \delta_{\mathrm{K}}(\bk_{12}) \, \delta_{\mathrm{K}}(\bk_{23})\right] \, \cL_{\ell_1}(\mu_1) \, \cL_{\ell_2}(-\mu_3) \Bigg\} \, . 
\end{align}

\subsection{Analytical marginalisation}\label{app:analytic_marg}

Analytically marginalising nuisance parameters can substantially improve the sampling efficiency in our high-dimensional power-spectrum and bispectrum model \citep{Bridle:2001zv,Taylor:2010pi,DAmico:2019fhj,Philcox:2020zyp}. For generic (nonlinear) parameters, this can be done approximately (e.g., via a local Gaussian/Laplace approximation or a first-order Taylor expansion). In contrast, for parameters that enter the model linearly -- and under a Gaussian likelihood with Gaussian priors -- the marginalisation is exact.

Let the parameter vector be split into linear $\{\theta_{\mathrm{lin},i}\}$ and nonlinear parameters $\{\theta_{\mathrm{nl},\alpha}\}$. We denote the data vector by $D_{\ell,k}$ with the covariance $C_{\ell\ell',kk'}$, where $\ell$ labels the multipole of the summary statistic (e.g., $\ell\in\{0,2,4\}$ for $P_\ell$ and $B_\ell$) and $k$ labels the wavenumber bin. With a Gaussian prior on the linear parameters (with prior covariance $\sigma_{ij}$), the Gaussian likelihood can be written in quadratic form as
\begin{multline}
\mathcal{L}(\theta_{\mathrm{lin},i},\theta_{\mathrm{nl},\alpha}) \\
\propto \exp \left\{
-\frac{1}{2}\,\left[
\theta_{\mathrm{lin},i}\,F_{2,ij}\,\theta_{\mathrm{lin},j}
-2\,\theta_{\mathrm{lin},i}\,F_{1,i}
+F_{0}(\theta_{\mathrm{lin},i}=0,\theta_{\mathrm{nl},\alpha})
\right]\right\} \, ,
\label{eq:marg_Likelihood}
\end{multline}
where repeated indices are summed and
\begin{align}
F_{2,ij}&=P^{\,i}_{\mathrm{lin};\ell,k}\,(C^{-1})_{\ell\ell',kk'}\,P^{\,j}_{\mathrm{lin};\ell',k'}+\sigma^{-1}_{ij}\,,\\
F_{1,i}&=P^{\,i}_{\mathrm{lin};\ell,k}\,(C^{-1})_{\ell\ell',kk'}\big[P_{\mathrm{nl};\ell',k'}(\theta_{\mathrm{nl},\alpha})-D_{\ell',k'}\big]\,,\\
F_{0}&=\big[P_{\mathrm{nl};\ell,k}(\theta_{\mathrm{nl},\alpha})-D_{\ell,k}\big]\,(C^{-1})_{\ell\ell',kk'}\,\big[P_{\mathrm{nl};\ell',k'}(\theta_{\mathrm{nl},\alpha})-D_{\ell',k'}\big]\,.
\end{align}
Here $P_\mathrm{lin}^{\,i}$ is the template multiplying the $i$-th linear parameter, $P_{\mathrm{nl}}$ depends only on $\theta_{\mathrm{nl},\alpha}$, $\sigma_{ij}$ is the prior covariance of the linear parameters, $C_{\ell\ell',kk'}$ is the data covariance, and $D_{\ell,k}$ are the data components. Integrating over the linear parameters yields the marginalised likelihood
\begin{align}
\mathcal{L}_{\rm marg}(\theta_{\mathrm{nl},\alpha})
&= \int \diff^{n_\mathrm{lin}}\theta_\mathrm{lin} \,\mathcal{L}(\theta_{\mathrm{lin},i},\theta_{\mathrm{nl},\alpha}) \nn \\
& \propto \exp\left\{-\frac{1}{2}\,F_0(\theta_{\mathrm{lin},i}=0,\theta_{\mathrm{nl},\alpha}) + \frac{1}{2}\,F_{1,i}\,(F_{2}^{-1})_{ij}\,F_{1,j} \right. \nn \\
 & \hspace{0.44in} \left.
-\frac{1}{2}\, \ln\left[\det\left(\frac{{\bf F}_2}{2\pi}\right)\right]\right\} \, .
\end{align}

To inspect degeneracies with $\theta_{\mathrm{nl},\alpha}$ and assess the impact of different prior choices on the nonlinear parameters, it is useful to reconstruct the posteriors of the analytically marginalised parameters. This is straightforward because the linear parameters enter the log-likelihood quadratically; their conditional posterior at fixed $\theta_{\mathrm{nl},\alpha}$ is Gaussian. At each MCMC step (for the current $\theta_{\mathrm{nl},\alpha}$), the optimal linear amplitudes are
\begin{align}
\hat\theta_{\mathrm{lin},i}(\theta_{\mathrm{nl},\alpha},D_{\ell,k}) = (F^{-1}_{2})_{ij}\,F_{1,j}(\theta_{\mathrm{nl},\alpha},D_{\ell,k})\,,
\end{align}
which are the maximum-a-posteriori (and conditional mean) values, with conditional covariance ${\bf F}_2^{-1}$ (cf., Eq.~\ref{eq:marg_Likelihood}). If one wishes to reconstruct the posteriors of the analytically marginalised parameters, one can draw them at each step
\begin{align}
\theta_{\mathrm{lin},i}\sim\mathcal{N}\Big(\hat\theta_{\mathrm{lin},i}(\theta_{\mathrm{nl},\alpha},D_{\ell,k}),\,{\bf F}_2^{-1}\Big)\,, \label{eq:resample}
\end{align}
store these samples alongside the chain and thereby recover their marginalised distributions. In Eq.~\eqref{eq:resample} $\mathcal{N}$ identifies a multivariate normal distribution.
 
In practice, our analysis demonstrated that analytic marginalisation reduced the effective sampling dimensionality and noticeably accelerated convergence, improving runtime and chain stability for our data vector. Although the parameter space is smaller and experiences fewer degeneracies in the Gaussian case, analytical marginalisation results in requiring $2$ to $5$ times fewer steps for the corresponding convergence, depending on the specific sampler used. By extending the parameter space to additionally include the PNG parameters and their inherent degeneracies, analytical marginalisation not only reduces the steps by decreasing the sampling dimensionality but also improves the convergence of the chains by enforcing them to follow the steepest descent path within Gaussian parameters.

\subsection{Parameter priors}\label{app:priortable}

We list our priors in Table~\ref{tab:Priors}. See Sect.~\ref{sec:prior} for further discussion about PNG priors. 

\begin{table}[htbp!]
  \renewcommand{\arraystretch}{1.5}
  \centering
  \caption{hoices for priors on cosmological and nuisance parameters.}
    \begin{tabular}{ccc}
      \toprule
       & Parameter & Prior\\
       \midrule
      \multirow{4}{*}{Cosmology} & $h$ & $\mathcal{U}\left[0.6,0.8\right]$\\
      \noalign{\vskip 2pt} \cline{2-3} \noalign{\vskip 1pt}
      & $\omega_\mathrm{c}$ & $\mathcal{U}[0.085,0.165]$ \\
      \noalign{\vskip 2pt} \cline{2-3} \noalign{\vskip 1pt}
      & $10^9\times A_{\mathrm{s}}$ & $\mathcal{U}[1,3.2]$ \\
      \noalign{\vskip 2pt} \cline{2-3} \noalign{\vskip 1pt}
      & $\fnl$ & $\mathcal{U}[-500,500]$ \\
      \hline
      \multirow{4}{*}{Bias} & $b_1$ & $\mathcal{U}[0.9,5]$\\
      \noalign{\vskip 2pt} \cline{2-3} \noalign{\vskip 1pt}
      & $b_2$ & $\mathcal{U}[-20,20]$\\
      \noalign{\vskip 2pt} \cline{2-3} \noalign{\vskip 1pt}
      & $b_{\mathcal{G}_2}$ & $\mathcal{U}[-20,20]$\\
      \noalign{\vskip 2pt} \cline{2-3} \noalign{\vskip 1pt}
      & $b_{\Gamma_3}$ & $\mathcal{U}[-20,20]$\\
      \midrule
      \multirow{5}{*}{Counter-Term} & $c_0\,\left[h^{-2}\,{\rm Mpc}^2\right]$ & $\mathcal{N}\left(0,100^2\right)$\\    
      \noalign{\vskip 2pt} \cline{2-3} \noalign{\vskip 1pt}
      & $c_2\,\left[h^{-2}\,{\rm Mpc}^2\right]$ & $\mathcal{N}\left(0,100^2\right)$\\    
      \noalign{\vskip 2pt} \cline{2-3} \noalign{\vskip 1pt}
      & $c_4\,\left[h^{-2}\,{\rm Mpc}^2\right]$ & $\mathcal{N}\left(0,100^2\right)$\\    
      \noalign{\vskip 2pt} \cline{2-3} \noalign{\vskip 1pt}
      & $c_6\,\left[h^{-4}\,{\rm Mpc}^4\right]$ & $\mathcal{N}\left(0,100^2\right)$\\    
      \noalign{\vskip 2pt} \cline{2-3} \noalign{\vskip 1pt}
      & $c_{\rm VDG}^{B}$ & $\mathcal{U}[-50,50]$\\    
      \midrule
      \multirow{7}{*}{Shot-Noise} & $s_0$ & $\mathcal{N}\left(0,2^2\right)$ \\
      \noalign{\vskip 2pt} \cline{2-3} \noalign{\vskip 1pt}
      & $s_{02}\,\left[h^{-2}\,{\rm Mpc}^2\right]$ & $\mathcal{N}\left(0,10^2\right)$\\
      \noalign{\vskip 2pt} \cline{2-3} \noalign{\vskip 1pt}
      & $s_{2}\,\left[h^{-2}\,{\rm Mpc}^2\right]$ & $\mathcal{N}\left(0,10^2\right)$\\
      \noalign{\vskip 2pt} \cline{2-3} \noalign{\vskip 1pt}
      & $s_{4}\,\left[h^{-2}\,{\rm Mpc}^2\right]$ & $\mathcal{N}\left(0,50^2\right)$\\
      \noalign{\vskip 2pt} \cline{2-3} \noalign{\vskip 1pt}
      & $\tilde{s}_1$ & $\mathcal{N}\left(0,2^2\right)$\\
      \noalign{\vskip 2pt} \cline{2-3} \noalign{\vskip 1pt}
      & $\tilde{s}_2$ & $\mathcal{N}\left(0,2^2\right)$\\
      \noalign{\vskip 2pt} \cline{2-3} \noalign{\vskip 1pt}
      & $\tilde{s}_3$ & $\mathcal{N}\left(0,2^2\right)$\\
      \midrule
      \multirow{6}{*}{PNG Counter-Terms} & $c^{\phi}_0\,\left[h^{-2}\,{\rm Mpc}^2\right]$ & $\mathcal{N}\left(0,5000^2\right)$\\    
      \noalign{\vskip 2pt} \cline{2-3} \noalign{\vskip 1pt}
      & $c^{\phi}_2\,\left[h^{-2}\,{\rm Mpc}^2\right]$ & $\mathcal{N}\left(0,5000^2\right)$\\    
      \noalign{\vskip 2pt} \cline{2-3} \noalign{\vskip 1pt}
      & $c^{\phi}_4\,\left[h^{-2}\,{\rm Mpc}^2\right]$ & $\mathcal{N}\left(0,5000^2\right)$\\    
      \noalign{\vskip 2pt} \cline{2-3} \noalign{\vskip 1pt}
      & $c^{12}_0\,\left[h^{-4}\,{\rm Mpc}^4\right]$ & $\mathcal{N}\left(0,50^2\right)$\\    
      \noalign{\vskip 2pt} \cline{2-3} \noalign{\vskip 1pt}
      & $c^{12}_2\,\left[h^{-4}\,{\rm Mpc}^4\right]$ & $\mathcal{N}\left(0,50^2\right)$\\    
      \noalign{\vskip 2pt} \cline{2-3} \noalign{\vskip 1pt}
      & $c^{12}_4\,\left[h^{-4}\,{\rm Mpc}^4\right]$ & $\mathcal{N}\left(0,50^2\right)$\\    
      \bottomrule
    \end{tabular}
    \tablefoot{The symbol $\mathcal{U}$ represents a uniform distribution, with specified lower and upper limits in brackets, while the symbol $\mathcal{N}$ denotes a Gaussian distribution, with the mean and variance provided in the brackets.}
    \label{tab:Priors}
\end{table}

\section{Additional results from individual and combined statistics}\label{app:add_joint}

For the same analysis settings as in Sect.~\ref{subsec:Joint}, we present additional plots, which complement those shown earlier.

\begin{figure*}[htbp!]
    \centering
    \includegraphics[width=0.7\columnwidth]{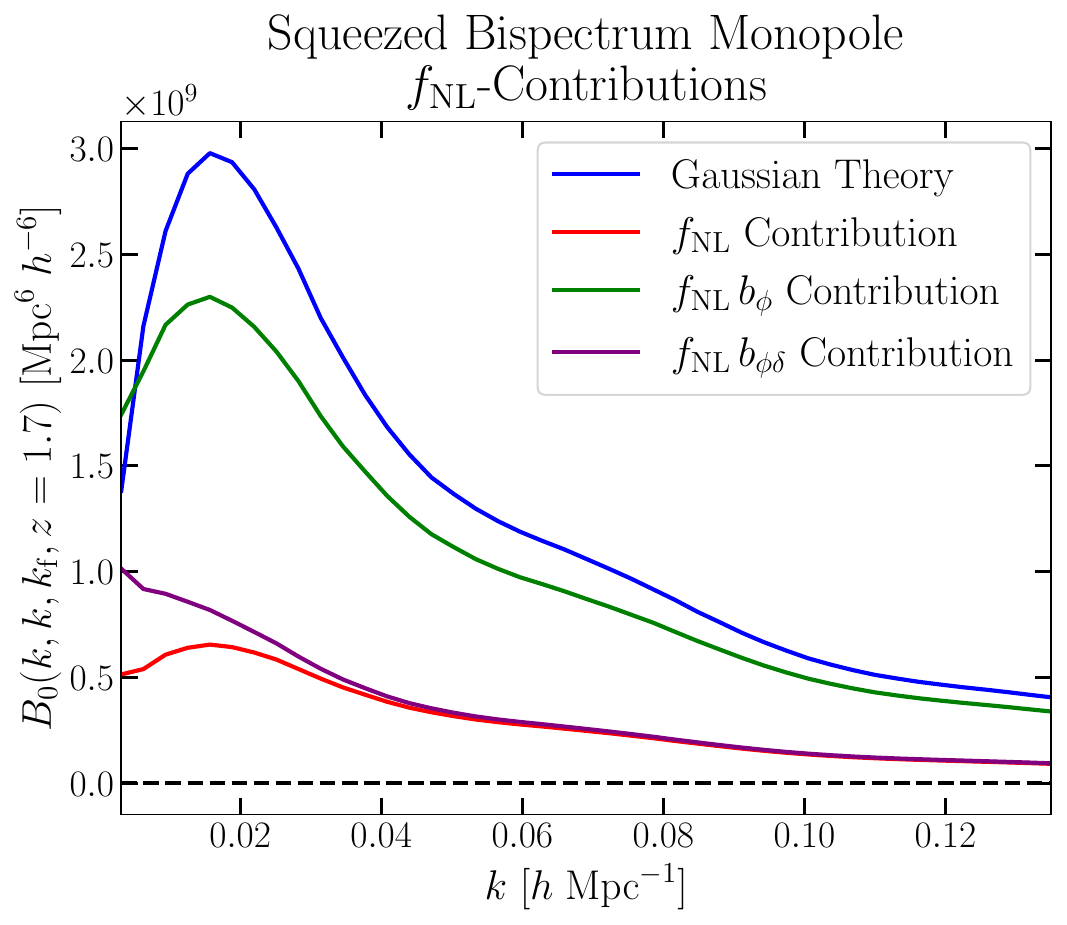}
    \includegraphics[width=0.7\columnwidth]{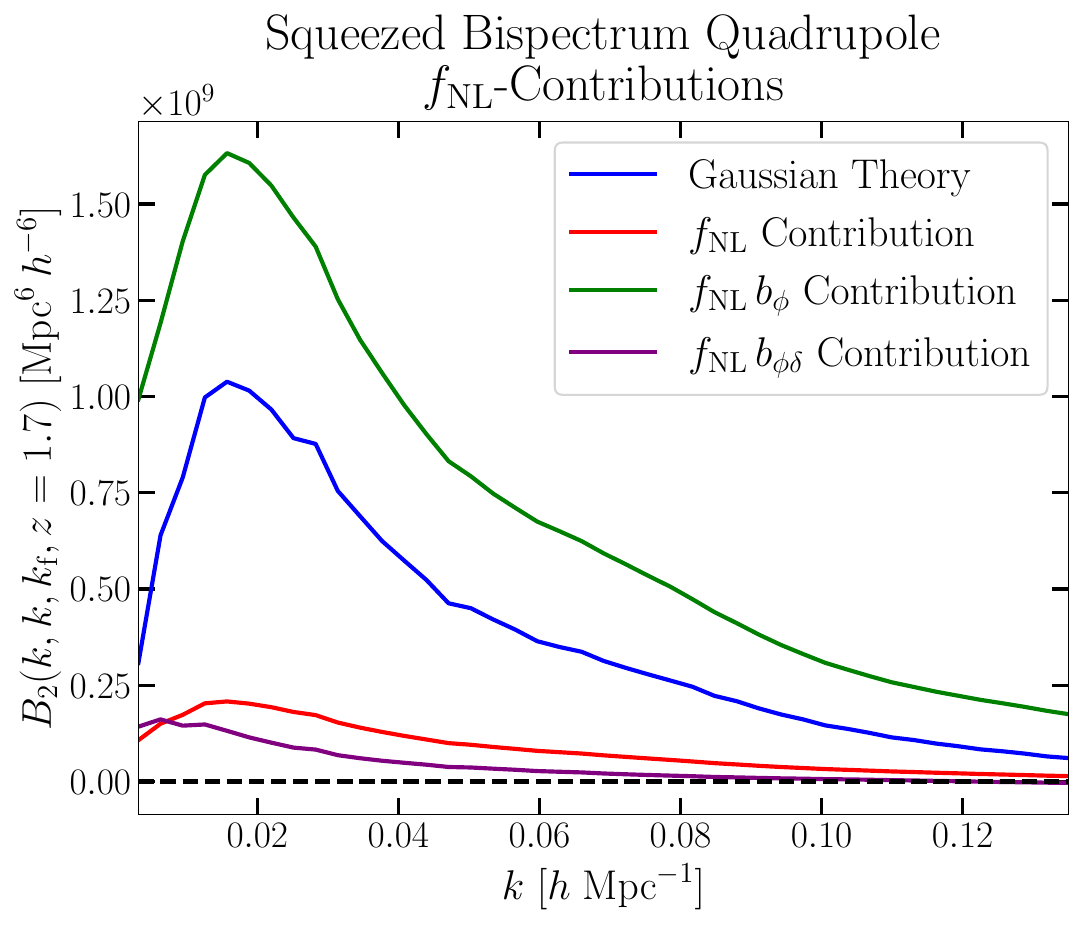}
    \includegraphics[width=0.7\columnwidth]{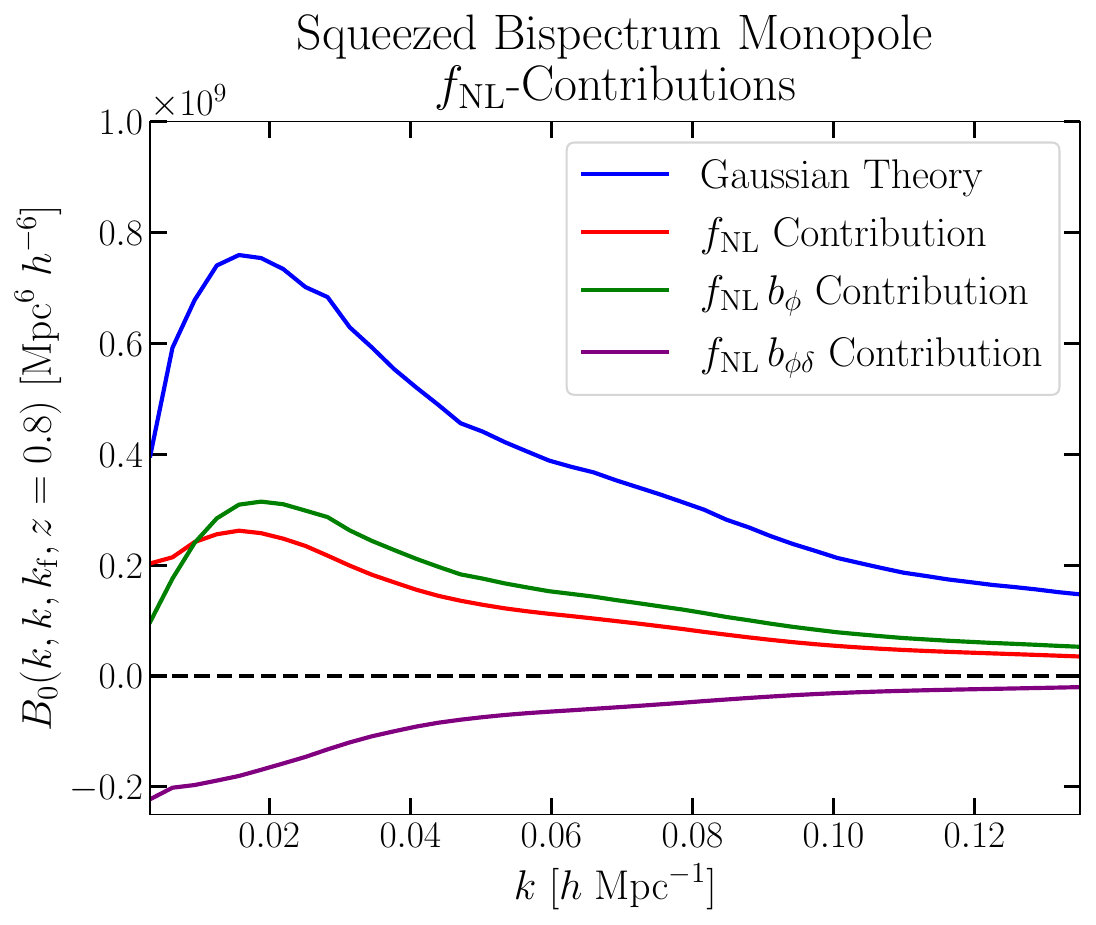}
    \includegraphics[width=0.7\columnwidth]{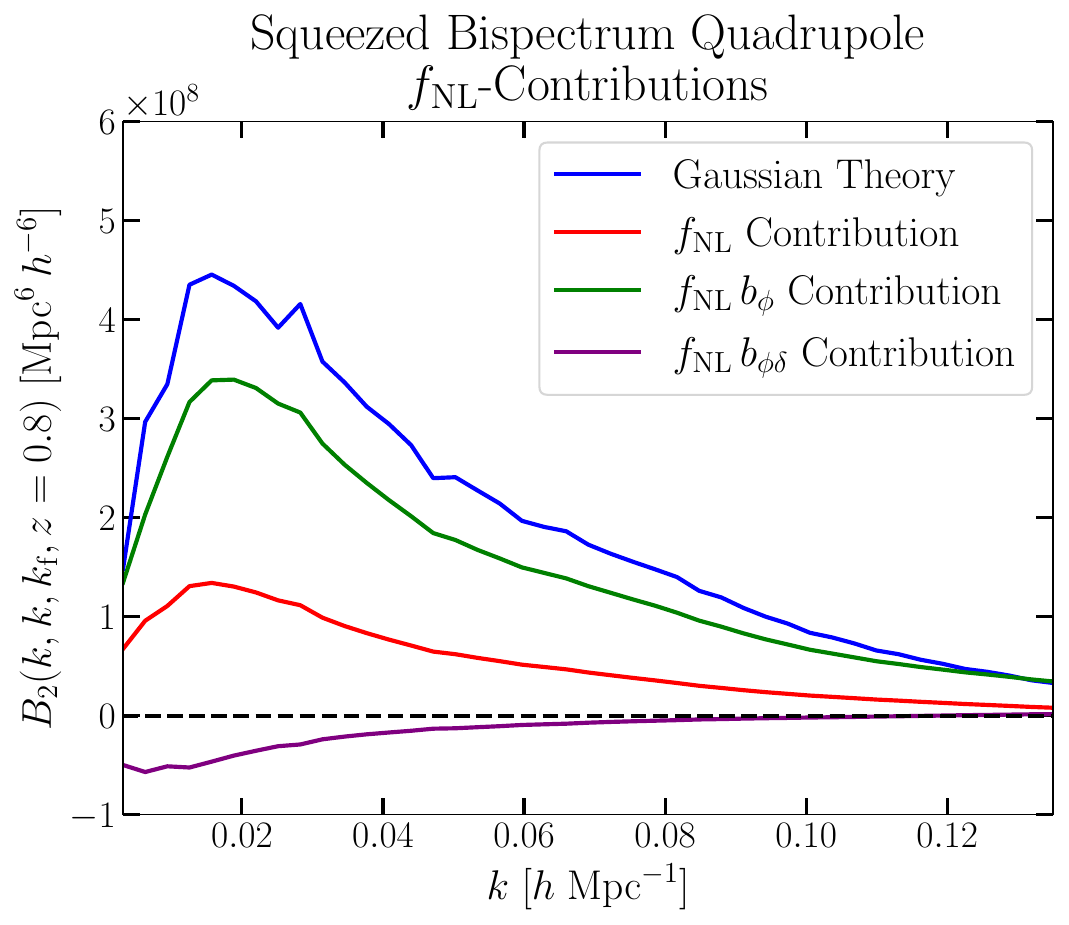}
    \caption{Relative size of PNG contributions to squeezed-limit bispectrum at $z=1.7$ (top) and $z=0.8$ (bottom). Assuming the best-fit values of model parameters from joint power spectrum-bispectrum fit with our baseline scale cuts, the left panels show the monopole, while the right ones show the quadrupole. The non-primordial bispectrum is shown in blue. We fix one of the three sides to the fundamental mode $k_3 = k_{\rm f} = 0.00314 \, h\ \mathrm{Mpc}^{-1}$ and vary $k_1 = k_2 \equiv k$ across different scales. 
    \\}    \label{fig:Bispectrum_contributions_z1p7}
\end{figure*}

Figure~\ref{fig:Bispectrum_contributions_z1p7} shows the relative size of the three PNG bispectrum components in a squeezed configuration where one of the three sides is fixed to the fundamental mode $k_3 = k_{\rm f} = 0.00314 \, h\ \mathrm{Mpc}^{-1}$ and the other two sides $k_1 = k_2 \equiv k$ are varied across different scales. The top rows are results at $z=1.7$, while the bottom ones are at $z=0.8$. The non-primordial bispectrum contribution is shown in blue for comparison. The left (right) panels show the monopole (quadrupole). The values of model parameters are set to the best-fit values from the power spectrum-bispectrum joint analysis with our baseline scale cuts in Sect.~\ref{subsec:Joint}. While for the monopole, PNG contributions are always subdominant to non-PNG ones, at $z=1.7$, the PNG component dominates the quadrupole. 

\begin{figure}[t!]
    \centering
    \includegraphics[width=0.8\columnwidth]{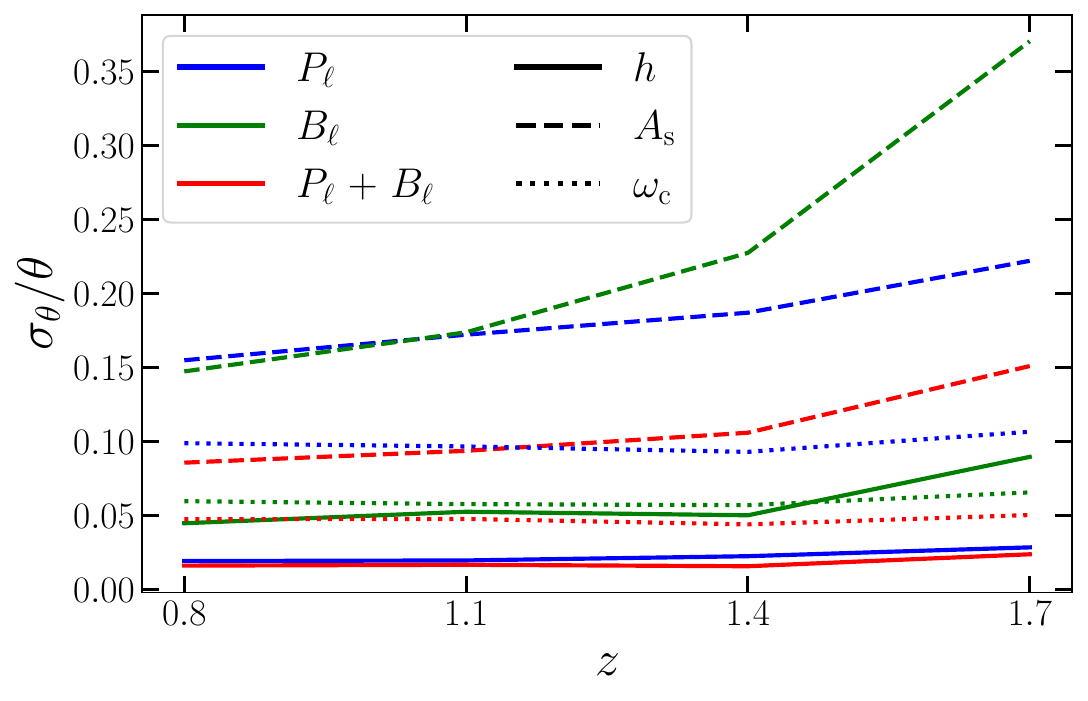}
    \caption{$\Lambda$CDM constraints from individual and joint data vectors. We show the 1$\sigma$ errors scaled with the values of three $\Lambda$CDM parameters as a function of redshift from multipoles of the power spectrum (blue), bispectrum (green) and combined (red).}\label{fig:1sigma_lcdm}
\end{figure}

Figure~\ref{fig:1sigma_lcdm} shows fractional $1\sigma$ errors $\sigma_\theta/\theta$ for three $\Lambda$CDM parameters from the power spectrum multipoles (blue), the bispectrum multipoles (green), and their combination (red). For $h$, the bispectrum is markedly less informative than the power spectrum, and the joint result offers only a marginal improvement over $P_\ell$ alone. For $A_{\mathrm{s}}$, $P_\ell$ and $B_\ell$ are comparable at low $z$ but $B_\ell$ degrades quickly with redshift; the combined fit remains substantially tighter than either observable, effectively stabilising the constraint at high $z$. For $\omega_{\mathrm{c}}$, $B_\ell$ outperforms $P_\ell$ at all redshifts, and the joint analysis delivers the best (and comparatively flat-with-$z$) precision. In both $A_{\mathrm{s}}$ and $\omega_{\mathrm{c}}$, the rotated degeneracy directions in $P_\ell$ and $B_\ell$ drive the gains in the combined constraints, whereas for $h$ the bispectrum adds little beyond $P_\ell$.

\begin{figure*}[htbp!]
    \centering
    \includegraphics[width=0.61\linewidth]{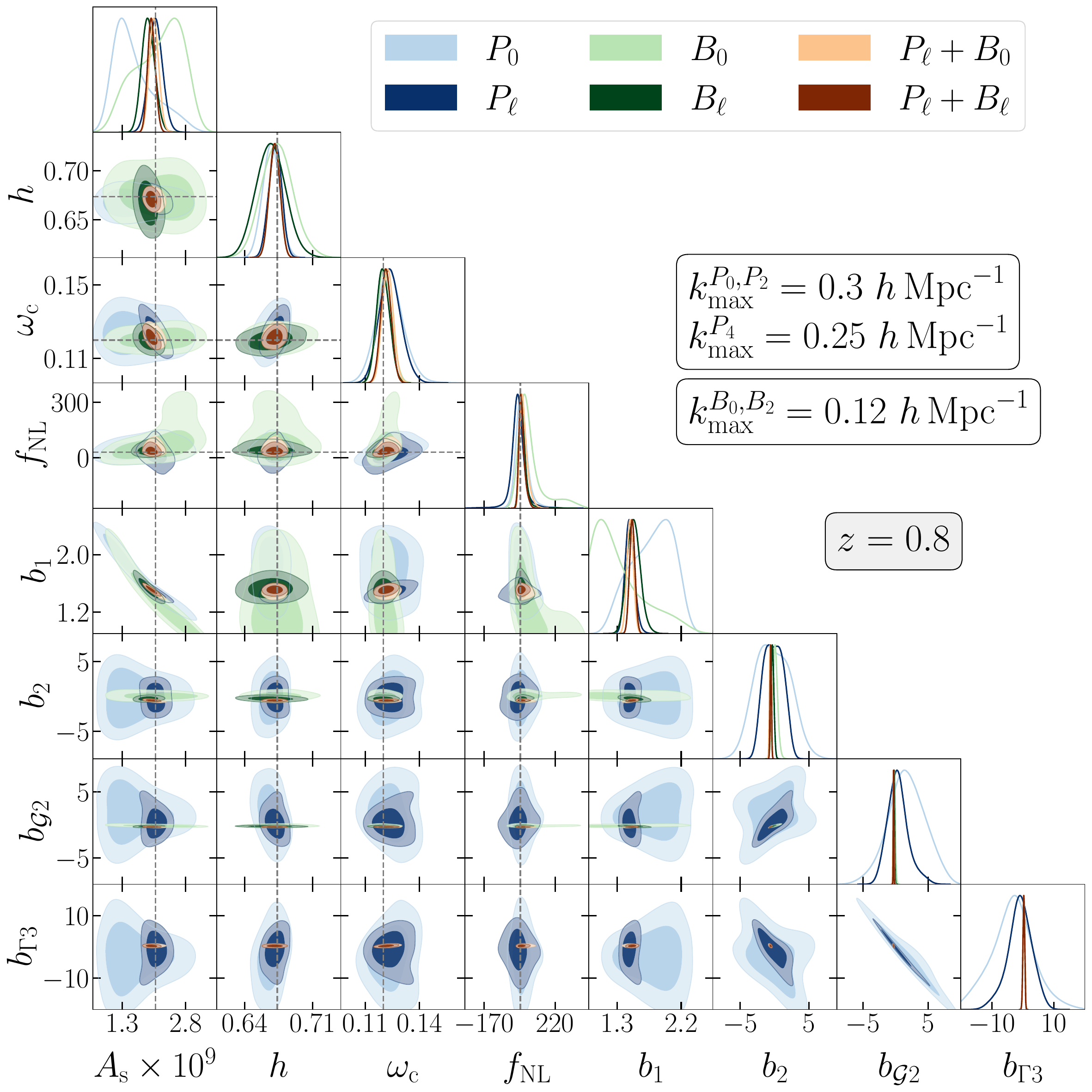}
    \caption{Information content of redshift-space multipoles on cosmological parameters and galaxy biases at $z=0.8$. The marginalised parameter posteriors from different combinations of power spectrum and bispectrum multipoles for the noted choice of scale cuts.}
    \label{fig:multipoles_zp8}
\end{figure*}

\begin{figure*}[htbp!]
    \centering
    \includegraphics[width=0.61\linewidth]{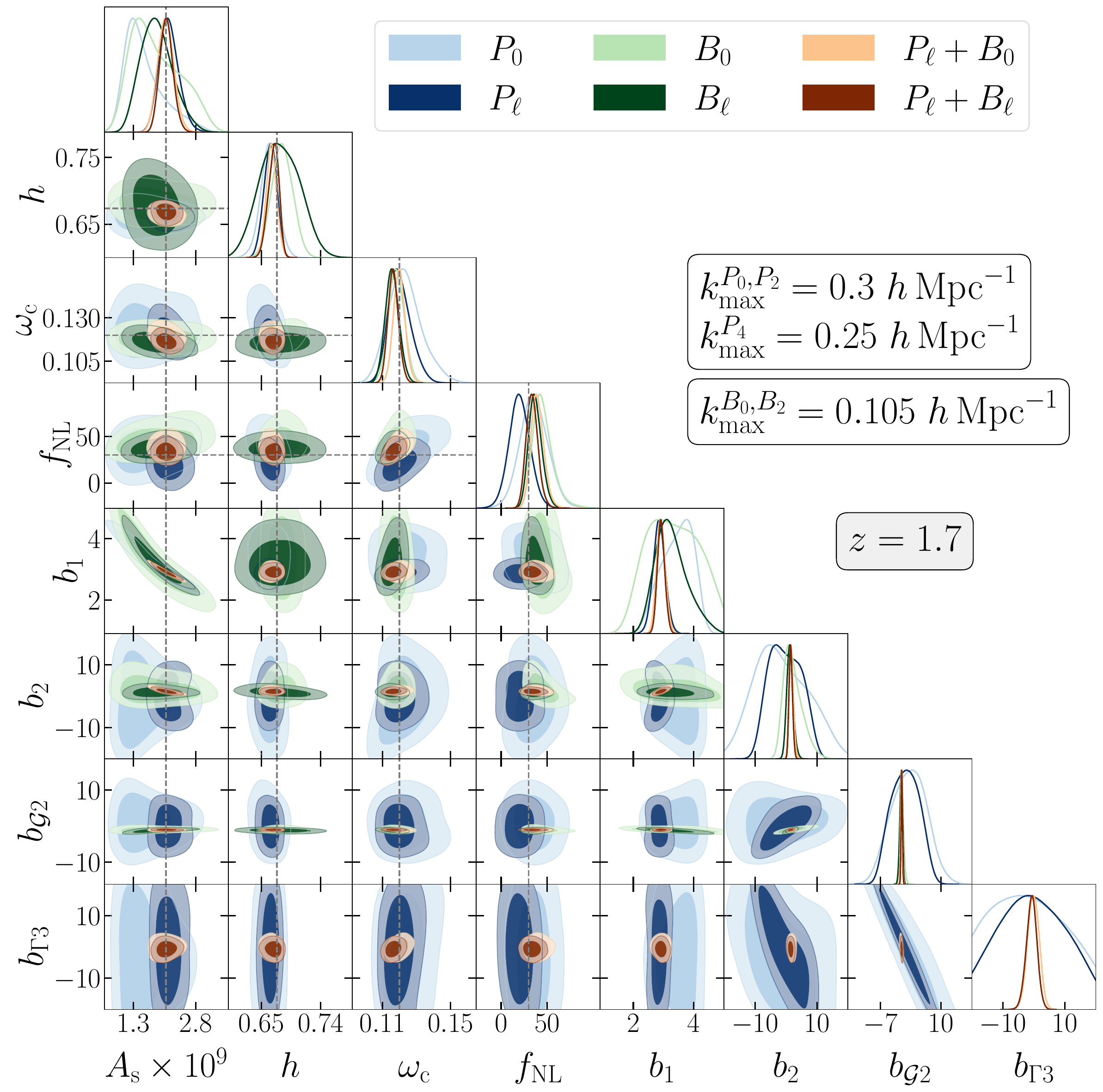}
    \caption{Same as Fig.~\ref{fig:multipoles_zp8} but at redshift $z=1.7$.}
    \label{fig:multipoles_z1p7}
\end{figure*}

Lastly, to illustrate the role of multipoles in the inferred constraints, Figs.~\ref{fig:multipoles_zp8} and \ref{fig:multipoles_z1p7} show the $1$- and $2$-dimensional marginalised posteriors on cosmological parameters and galaxy biases from different combinations of power spectrum and bispectrum multipoles at two redshifts of $z\in\{0.8,1.7\}$. Detailed description of the observed trends is provided in Sect.~\ref{subsec:Joint}. 

\end{document}